\documentclass[review]{elsarticle}
\usepackage{graphicx}
\usepackage{bm}
\usepackage{lineno}
\usepackage{amsmath}
\usepackage{amssymb}
\usepackage{color}

\newcommand{\araa}{{ARA\&A, }}          
\newcommand{\apj}{{ApJ, }}         
\newcommand{\apjl}{{ApJ, }}          
\newcommand{\apjs}{{ApJS, }}          
\newcommand{\aap}{{A\&A, }}          
\newcommand{\mnras}{{MNRAS, }}          
\newcommand{\nat}{Natur, }       
\newcommand{\physrep}{PhR, } 
\newcommand{\prc}{PhRvC, }
\newcommand{\prd}{PhRvD, }
\newcommand{\prl}{PhRvL, }
\newcommand{\prx}{PhRvX, }
\newcommand{\nphysa}{NuPhA, }         
\newcommand{\npb}{NuPhB, }            
\newcommand{\ssr}{SSRv, }         
\newcommand{\pasj}{PASJ, }         
\newcommand{\rmp}{RvMP} 
\newcommand{\ijmpd}{IJMPD, } 
\newcommand{\lrr}{LRR, } 

\newcommand{\ep}{\varepsilon}
\newcommand{\ptrans}{P_{\rm trans}}
\newcommand{\ntrans}{\rho_{\rm trans}}
\newcommand{\etrans}{\varepsilon_{\rm trans}}

\newcommand{\cQMsq}{c^2_{\rm QM}}

\title{Neutron star equation of state: QMF modeling and applications}

\author[XMU]{A. Li}
\author[XMU,ITP]{Z.-Y. Zhu}
\author[AEI]{E.-P. Zhou}
\author[IMP]{J.-M. Dong}
\author[NKU,RIKEN]{J.-N. Hu}
\author[NB,TOKAI]{C.-J. Xia}

\address[XMU]{Department of Astronomy, Xiamen University, Xiamen 361005, China; liang@xmu.edu.cn}
\address[ITP]{Institute for Theoretical Physics, Frankfurt am Main 60438, Germany}
\address[AEI]{Max Planck Institute for Gravitational Physics (Albert Einstein Institute), Am M{\"u}hlenberg 1, Potsdam-Golm, 14476, Germany}
\address[IMP]{Institute of Modern Physics, Chinese Academy of Sciences, Lanzhou 730000, China}
\address[NKU]{School of Physics, Nankai University, Tianjin 300071, China}
\address[RIKEN]{Strangeness Nuclear Physics Laboratory, RIKEN Nishina Center, Wako, 351-0198, Japan}
\address[NB]{School of Information Science and Engineering, Zhejiang University Ningbo Institute of Technology, Ningbo 315100, China}
\address[TOKAI]{Advanced Science Research Center, Japan Atomic Energy Research Institute, Tokai, Ibaraki 319-1195, Japan}

\begin{document}

\begin{abstract}
\noindent
Because of the development of many-body theories of nuclear matter, the long-standing, open problem of the equation of state (EOS) of dense matter may be understood in the near future through the confrontation of theoretical calculations with laboratory measurements of nuclear properties \& reactions and increasingly accurate observations in astronomy. In this review, we focus on the following six aspects: 1) providing a survey of the quark mean-field (QMF) model, which consistently describes a nucleon and many-body nucleonic system from a quark potential; 2) applying QMF to both nuclear matter and neutron stars; 3) extending QMF formalism to the description of hypernuclei and hyperon matter, as well as hyperon stars; 4) exploring the hadron-quark phase transition and hybrid stars by combining the QMF model with the quark matter model characterized by the sound speed; 5) constraining interquark interactions through both the gravitational wave signals and electromagnetic signals of binary merger event GW170817; and 6) discussing further opportunities to study dense matter EOS from compact objects, such as neutron star cooling and pulsar glitches.

\end{abstract}

\maketitle

\section{Introduction}

The equation of state (EOS) of dense stellar matter is a problem for both nuclear physics and relativistic astrophysics and has been greatly promoted by the detection of gravitational waves from the GW170817 binary neutron star (NS) merger event~\citep{2017PhRvL.119p1101A}\footnote{arXiv page: http://blogs.cornell.edu/arxiv/2017/10/16/gw170817/}.
Multimessenger observations of NS mergers~\citep{2017ApJ...848L..12A} can provide information for determining the EOS of supranuclear matter~\citep{2017CQGra..34d4001A,2019PrPNP.10903714B,2017ApJ...837...67C} and that can possibly constrain the phase diagram of the quantum chromodynamics (QCD)~\citep{2008RvMP...80.1455A,2018RPPh...81e6902B,2018PhRvD..97h4038P,2019PhRvD.100j3022H}.

In NSs, nuclear matter is present in beta equilibrium from very low density to several times the saturation density ($\rho\approx0.16~\rm fm^{-3}$) and is extremely neutron-rich~\citep{1983bhwd.book.....S,2004Sci...304..536L,2012RPPh...75b6301B,2017IJMPD..2630015G}.
One assumes that there is one theoretical model that can correctly explain the nuclear matter data of different physical situations obtained in both laboratory nuclear experiments~\citep[e.g.,][]{2016RvMP...88c5009M,2002Sci...298.1592D,2012ChPhC..36....3M,2013ADNDT..99...69A,2015RPPh...78i6301F} and astronomical observations~\citep[e.g.,][]{2006Sci...311.1901H,2010Natur.467.1081D,2013Sci...340..448A,2016ApJ...832..167F,2018ApJS..235...37A,2020NatAs...4...72C,2016ARA&A..54..401O,2019ApJ...887L..24M,2019ApJ...887L..21R,2019PhRvX...9a1001A,2020CQGra..37d5006A}. However, this is a demanding task. It not only requires the theoretical models to extrapolate from lower density/temperature/isospin to unknown regions at high density/temperature/isospin~\citep{2001LNP...578..364P,2018ASSL..457..255F} but also depends on the relevant degrees of freedom of the problem, from nucleons to exotic particles~\citep{2005PrPNP..54..193W,2008IJMPE..17.1635L,2016RvMP...88c5004G,2017RvMP...89a5007O,2020arXiv200209223T}, even dark matter particles \citep[e.g.,][]{2012APh....37...70L,2020arXiv200200594D}.

In this paper, we follow a widely used relativistic mean-field (RMF) approach~\citep{2011PrPNP..66..519N} based on an effective Lagrangian with meson fields mediating strong interactions between quarks, which we call the quark mean-field (QMF) model~\citep{2000PhRvC..61d5205S,1998PhRvC..58.3749T}.
It self-consistently relates the internal quark structure of a nucleon and a hyperon to the RMFs arising in nuclear and hyperonic matter, respectively, and has been employed extensively in the calculations of finite (hyperon-)nuclei and infinite dense matter~\citep{2000PhRvC..61d5205S,2019PhRvC..99b5804Z,2002NuPhA.707..469S,2014PTEP.2014a3D02H,2014PhRvC..89b5802H,2017PhRvC..96e4304H,2016PhRvC..94d4308X,2017PhRvC..95e4310X,2018PhRvC..97c5805Z,2018ApJ...862...98Z}.
We focus on the EOS that have been developed so far, testing the QMF predictions concerning the constraints from experiments.
We also illustrate the developments of this approach for applications to open questions in the present multiscale multimessenger gravitational wave era of astronomy.
Another complementary approach for nuclear matter is the {\it ab initio} approach, such as the Brueckner theory \citep[e.g.,][]{1999IRvNP...8.....B,2015A&A...584A.103S}, the chiral effective field theory \citep[e.g.,][]{2013ApJ...773...11H,2019EPJA...55...97T}, the quantum Monte Carlo method \citep[e.g.,][]{2015PhRvL.114i2301L,2020arXiv200101374G}, and the variational method \citep[e.g.,][]{1998PhRvC..58.1804A}, which starts from microscopic nucleon-nucleon potentials explicitly including many-body forces.
As a comparison, we include some results based on these {\it ab initio} many-body approaches.

The paper is organized as follows. In Sec. 2, we introduce QMF models by introducing the confinement potential of the constituent quarks for a nucleon. Sec. 3 is then devoted to the NS properties based on the QMF EOSs. In Sec. 4, we demonstrate how strange baryons, e.g., hyperons, are incorporated in the QMF model and discuss the hyperon puzzle with the obtained hyperon star maximum mass. We also discuss hybrid stars and strange quark stars (QSs) by introducing quark matter models. This is followed by the discussions of the NS binary in Sec. 5. Other opportunities for studying EOS are given in Sec. 6, including NS cooling and pulsar glitches.
Sec. 7 contains the main conclusions and future perspectives of this review.

\section{EOS models from the quark level within QMF}

In 1988, Guichon~\citep{1988PhLB..200..235G} developed a novel model for nuclear matter to treat the changes in the nucleon properties of nuclear matter, i.e., the European Muon Collaboration (EMC) effects.
This model is similar to the RMF model, but the scalar and the vector meson fields couple not with the nucleons but directly with the quarks.
Then, the nucleon properties change according to the strengths of the mean fields acting on the quarks, and the nucleon is dealt with in terms of the MIT bag model~\citep{1975PhRvD..12.2060D}.
The Guichon model was extended by Thomas and his collaborators under the name of the quark-meson coupling (QMC) model. Excellent reviews on the QMC model can be found in the literature~\citep{2018PrPNP.100..262G,2007PrPNP..58....1S}; see also, e.g., \citep{2016IJMPE..2550007B,2016PhRvL.116i2501S,2019ApJ...878..159M,2020PhLB..80235266M} for some of the latest improvements.
Taking an alternative model for the nucleon, the quark potential model~\citep{1978PhRvD..18.4187I}, Toki and his collaborators constructed the QMF model~\citep{1998PhRvC..58.3749T}.
For a more detailed comparison of these two models, we refer to~\citep{2000PhRvC..61d5205S,2019PhRvC..99b5804Z}.
Briefly, the bag model assumes the nucleon is constituted by bare quarks in the perturbative vacuum, i.e., current quarks, with a bag constant to account for the energy difference between the perturbative vacuum and the nonperturbative vacuum, while in the potential model, the nucleon is described in terms of the constituent quarks, which couple with the mesons and gluons.
We shall first introduce the potential model and then introduce the QMF formalism.

\subsection{Quark potential model}

In the MIT bag model, the quarks inside the nucleon are confined by a bag, which ensures that the quarks can only move freely and independently inside the nucleon through an infinite potential well.
In the potential model,
quarks are confined by a phenomenological confinement potential, where the polynomial forms are widely used.
A harmonic oscillator potential is usually adopted, with which the Dirac equation can be solved analytically,
\begin{eqnarray}
U(r)=\frac{1}{2}(1+\gamma^0)(ar^2+V_0),
\end{eqnarray}
where the scalar-vector form of the Dirac structure is chosen for the quark confinement
potential and the parameters $a$ and $V_0$ are determined from the vacuum nucleon properties.
When the effect of the nuclear medium is considered, the quark field $\psi_{q}(\vec{r})$ satisfies the following Dirac equation:
\begin{eqnarray}
[\gamma^{0}(\epsilon_{q}-g_{\omega q}\omega-\tau_{3q}g_{\rho q}\rho-\vec{\gamma}\cdot\vec{p} \nonumber \\
-(m_{q}-g_{\sigma q}\sigma)-U(r)]\psi_{q}(\vec{r})=0,
\end{eqnarray}
where $\sigma$, $\omega$, and $\rho$ are the classic meson fields. $g_{\sigma q}$, $g_{\omega q}$, and $g_{\rho q}$ are the coupling constants of $\sigma, ~\omega$ and $\rho$ mesons with quarks, respectively. $\tau_{3q}$ is the third component of the isospin matrix, and $m_q$ is the constitute quark mass at approximately 300 MeV.
The nucleon mass in the nuclear medium can be expressed as the binding energy of three quarks,
defined by the zeroth-order term after solving the Dirac equation $E_N^0=\sum_q\epsilon_q^\ast$.
The quarks are simply confined in a two-body confinement potential. Three corrections
are taken into account in the zeroth-order nucleon mass in the nuclear medium, including
the contribution of the center-of-mass (c.m.) correction $\epsilon_{\rm c.m.}$, pionic correction $\delta M_N^\pi$ and gluonic correction $(\Delta E_N)_g$.
The pion correction is generated by the chiral symmetry of QCD theory and the gluon
correction by the short-range exchange interaction of quarks.
Finally, the mass of the nucleon in the nuclear medium becomes
\begin{eqnarray}
M^\ast_N=E^{0}_N-\epsilon_{\rm c.m.}+\delta M_N^\pi+(\Delta E_N)_g.
\end{eqnarray}
The nucleon radius is written as
\begin{eqnarray}
\langle r_N^2\rangle = \frac{\mathop{11\epsilon'_q + m'_q}}{\mathop{(3\epsilon'_q + m'_q)(\epsilon'^2_q-m'^2_q)}}.
\end{eqnarray}
where $\mathop{\epsilon'_q}=\epsilon_q^\ast-V_0/2,\ \mathop{m'_q}=m_q^\ast+V_0/2$. The effective single quark energy is given by $\epsilon_q^*=\epsilon_{q}-g_{q\omega}\omega-\tau_{3q}g_{q\rho}\rho$, and the effective quark mass is given by $m_q^\ast = m_q-g_{\sigma q}\sigma$.
By reproducing the nucleon mass and radius $(M_N, r_N)$ in free space, we determine the potential parameters ($a$ and $V_0$) in Eq.~(1). We obtain $V_0=-62.257187$ MeV and $a=0.534296$ fm$^{-3}$ with $m_q=300$ MeV by fitting $M_N = 939$ MeV and $r_N = 0.87$ fm~\citep{2016RvMP...88c5009M}.

\subsection{Nuclear matter from an RMF Lagrangian}

In the above section, we construct the nucleon at the quark level with the confinement potential and the pion and gluon corrections. Next, we would like to connect such nucleons in a nuclear medium with nuclear objects, such as nuclear matter and systems of finite nuclei.
A good bridge is the RMF model at the hadron level, which is developed based on the one-boson exchange potential between two nucleons. The effective nucleon mass from the quark model is inserted into the RMF Lagrangian.
The nucleon and meson fields are solved self-consistently, and then, the properties of the nuclear many-body system are obtained.
We mention here that the nucleons are treated as point-like particles even though a quark model is used to describe the structure of the nucleon.
In many-body calculations, the structure of the nucleon only modifies the effective mass of a nucleon, i.e., Eq.~(3).

 We consider the $\sigma,~\omega$ and $\rho$ mesons exchanging in the Lagrangian~\citep{2019PhRvC..99b5804Z,2018PhRvC..97c5805Z,2018ApJ...862...98Z}, and the cross-coupling from the $\omega$ meson and $\rho$ meson is introduced to achieve a reasonable slope of symmetry energy (see Sec.~\ref{sec:sym})~\citep{2001PhRvL..86.5647H},
\begin{eqnarray}
\mathcal{L}& = & \overline{\psi}\left(i\gamma_\mu \partial^\mu - M_N^\ast - g_{\omega N}\omega\gamma^0 - g_{\rho N}\rho\tau_{3}\gamma^0\right)\psi  \nonumber \\
&& -\frac{1}{2}(\nabla\sigma)^2 - \frac{1}{2}m_\sigma^2 \sigma^2 - \frac{1}{3} g_2\sigma^3 - \frac{1}{4}g_3\sigma^4 \nonumber \\
& & + \frac{1}{2}(\nabla\rho)^2 + \frac{1}{2}m_\rho^2\rho^2 + \frac{1}{2}g_{\rho N}^2\rho^2 \Lambda_v g_{\omega N}^2\omega^2 \nonumber \\
&& + \frac{1}{2}(\nabla\omega)^2 + \frac{1}{2}m_\omega^2\omega^2 ,
\end{eqnarray}
where $g_{\omega N}$ and $g_{\rho N}$ are the nucleon coupling constants for $\omega$ and $\rho$ mesons. From the simple quark counting rule, we obtain $g_{\omega N}=3g_{\omega q}$ and $g_{\rho N}=g_{\rho q}$. The calculation of the confined quarks gives the relation of the effective nucleon mass $M_N^*$ as a function of the $\sigma$ field, $g_{\sigma N} = -\partial M_N^\ast/\partial \sigma$, which defines the $\sigma$ coupling with nucleons (depending on the parameter $g_{\sigma q}$). $m_{\sigma} = 510~\rm{MeV}$,~$m_{\omega}=783~\rm{MeV}$, and $m_{\rho}=770~\rm{MeV}$ are the meson masses.
In this Lagrangian, we already consider the static approximation on the mesons so that their time components are neglected. The spatial part of the $\omega$ meson disappears for the time reversal symmetry.
The infinite nuclear matter has translational invariance, which further removes the partial part of the coordinate space.

The equations of motion of nucleons and mesons can be generated by the Euler-Lagrangian equation from the Lagrangian,
\begin{eqnarray}
&&(i\gamma^{\mu}\partial_\mu-M_{N}^{\ast}-g_\omega\omega\gamma^0-g_\rho\rho\tau_3\gamma^0)\psi=0,\\
&&m_{\sigma}^2\sigma+g_2\sigma^2+g_3\sigma^3=-\frac{\partial M_N^*}{\partial\sigma}\langle\bar{\psi}\psi\rangle,\\
&&m_{\omega}^2\omega+\Lambda_vg_{\omega N}^2g_{\rho N}^2\omega \rho^2=g_{\omega N}\langle\bar{\psi}\gamma^0\psi\rangle,\\
&&m_{\rho}^2\rho+\Lambda_vg_{\rho N}^2g_{\omega N}^2\rho \omega^2=g_{\rho N}\langle\bar{\psi}\tau_3\gamma^0\psi\rangle.
\end{eqnarray}
where
\begin{eqnarray}
\rho_S & = & \langle\bar{\psi}\psi\rangle = \frac{1}{\pi^2}\sum_{i=n,p}\int_0^{p_F^i}dpp_i^2\frac{M^\ast_N}{\sqrt{M^{\ast2}_N+p_i^2}},
\end{eqnarray}
\begin{eqnarray}
E_F^i=\sqrt{M_N^{\ast2}+(p_F^i)^2} ,
\end{eqnarray}
\begin{eqnarray}
m_\omega^{\ast2}=m_\omega^2+\Lambda_vg_{\omega N}^2g_{\rho N}^2\rho^2, ~~ m_\rho^{\ast2}=m_\rho^2+\Lambda_vg_{\rho N}^2g_{\omega N}^2\omega^2.
\end{eqnarray}
$p_F^n~(p_F^p$) is the Fermi momentum for a neutron~(proton), $\rho= \langle\bar{\psi}\gamma^0\psi\rangle = \rho_p+\rho_n $, and $\rho_3 =\langle\bar{\psi}\tau_3\gamma^0\psi\rangle = \rho_p-\rho_n$, which equals $0$ in symmetric nuclear matter.
Then, the energy density and pressure, with arbitrary isospin asymmetry $\beta = (\rho_n-\rho_p)/\rho$, can be generated by the energy-momentum tensor,
\begin{eqnarray}
&\varepsilon & =\frac{1}{\pi^2}\sum_{i=n,p}\int^{k^i_F}_0\sqrt{k^2+M_N^{\ast2}}k^2dk \nonumber \\
& &+\frac{1}{2}m^2_\sigma\sigma^2+\frac{1}{3}g_2\sigma^3+\frac{1}{4}g_3\sigma^4  \nonumber \\
& &
+\frac{1}{2}m^2_\omega\omega^2+\frac{1}{2}m^2_\rho\rho^2+ \frac{3}{2}\Lambda_vg_{\rho N}^2g_{\omega N}^2\rho^2\omega^2,
\end{eqnarray}
\begin{eqnarray}
&P & =  \frac{1}{3\pi^2}\sum_{i=n,p}\int_0^{k_F^i}\frac{k^4}{\sqrt{k^2+M_N^{\ast2}}}dk \nonumber \\
& &- \frac{1}{2}m_\sigma^2\sigma^2-\frac{1}{3}g_2\sigma^3-\frac{1}{4}g_3\sigma^4  \nonumber \\
& &
 + \frac{1}{2}m_\omega^2\omega^2 + \frac{1}{2}m_\rho^2\rho^2 + \frac{1}{2}\Lambda_vg_{\rho N}^2g_{\omega N}^2\rho^2\omega^2.
\end{eqnarray}
where we have written the meson field with their mean-field values denoted by $\sigma$, $\omega$, and $\rho$.

\subsection{Symmetry energy}\label{sec:sym}

We subtract the nucleon mass from the energy density (Eq.~(13)) to study the binding energy per nucleon, $E/A = \varepsilon/\rho- M_N$.
The parabolic approximation is usually applicable, and the energy per nucleon can be written as
\begin{eqnarray}
E/A (\rho, \beta) =  E/A (\rho, \beta=0) + E_{\rm sym}(n)\beta^2 + ...
\label{eq:snm}
\end{eqnarray}
and it is sufficient for performing the calculations only for symmetric nuclear matter and
pure neutron matter.
$E/A (\rho, \beta=0)$ can be expanded around the saturation density,
\begin{eqnarray}
E/A (\rho,0) = E/A (\rho_0) +\frac{1}{18} K\frac{\rho-\rho_0}{\rho_0}  + ...
\end{eqnarray}
where $K$ is the incompressibility at the saturation point.
The symmetry energy $E_{\rm sym}(\rho)$ can be expressed in terms
of the difference between the energies per particle of pure neutrons ($\beta=1$) and symmetric ($\beta=0$) matter, $E_{\rm sym}(\rho)\approx E/A (\rho, 1) -E/A (\rho, 0)$. To characterize
its density dependence, $E_{\rm sym}(\rho)$ can be expanded around the saturation density $\rho_0$ as follows:
\begin{eqnarray}
E_{\rm sym}(\rho) &=& E_{\rm sym}(\rho_0) \nonumber \\
&& + \frac{dE_{\rm sym}}{d\rho}(\rho-\rho_0)
 + \frac{1}{2}\frac{d^2E_{\rm sym}}{d\rho^2}(\rho-\rho_0)^2  + ...
 \label{eq:esym}
\end{eqnarray}
and the following parameters can be defined, where all have an energy dimension (MeV),
\begin{eqnarray}
E_{\rm sym} &=& E_{\rm sym}(\rho_0),  \\
L &=& 3\rho_0 (\frac{dE_{\rm sym}}{d\rho})_{\rho_0},\\
K_{\rm sym} &=& 9\rho_0^2 (\frac{d^2E_{\rm sym}}{d\rho^2})_{\rho_0}.
\end{eqnarray}
$E_{\rm sym}(\rho)$ can also be written as
\begin{eqnarray}
E_{\rm sym}(\rho) = E_{\rm sym} + \frac{1}{3} L\frac{\rho-\rho_0}{\rho_0} + \frac{1}{18} K_{\rm sym} (\frac{\rho-\rho_0}{\rho_0})^2 + ...
\end{eqnarray}

\begin{table*}
\tabcolsep 1pt
\caption{Properties of nuclear matter at saturation predicted by the EOSs employed in this study, in a comparison with the empirical ranges. The BCPM EoS, named after the Barcelona-Catania-Paris-Madrid energy density functional~\citep{2015A&A...584A.103S}, is based on the microscopic Brueckner-Hartree-Fock (BHF) theory~\citep{1999IRvNP...8.....B}. The BSk20 and BSk21 EoS belong to the family of Skyrme nuclear effective forces derived by the Brussels-Montreal group~\citep{2013A&A...560A..48P}. The high-density part of the BSk20 EoS is adjusted to fit the result of the neutron matter APR EOS~\citep{1998PhRvC..58.1804A}, whereas the high-density part of the BSk21 EOS is adjusted to the result of the BHF calculations using the Argonne v18 potential plus a microscopic nucleonic three-body force. The TM1 EOS is based on a phenomenological
nuclear RMF model with the TM1 parameter set~\citep{1998NuPhA.637..435S}, as well as the GM1 EOS, which uses a different parameter set~\citep{1991PhRvL..67.2414G}.
The number density $n_0$ is in fm$^{-3}$. The energy per baryon $E/A$ and the compressibility $K$, as well as the symmetry energy $E_{\rm sym}$ and its slope $L$ at saturation, are in MeV. The empirical values are taken from \citep{2012ChPhC..36....3M,2013ADNDT..99...69A,2017RvMP...89a5007O,2006EPJA...30...23S}.} \vspace*{-12pt}
\begin{center}
\def\temptablewidth{0.9999\textwidth}
{\rule{\temptablewidth}{0.5pt}}
\begin{tabular*}{\temptablewidth}{@{\extracolsep{\fill}}cccccc}
\hline
~~&~$\rho_0$~&~$E/A$~&~$K$~&~$E_{\rm sym}$~&~$L$
\\
~EoS~&(fm$^{-3}$)&(MeV)&(MeV)&(MeV)&(MeV)
\\ \hline
 QMF & 0.16 & -16.00 & 240.00 & 31.00 & 40.0
\\
 BCPM & 0.16 & -16.00 & 213.75 & 31.92 & 53.0
\\
TM1 & 0.145 & -16.26 & 281.14 & 36.89 & 110.8
\\
BSk20 & 0.159 & -16.08 & 241.4 & 30.0 & 37.4
\\
BSk21 & 0.158 & -16.05 & 245.8 & 30.0 & 46.6
\\
 APR  & 0.16 & -16.00 & 247.3  & 33.9 & 53.8
 \\
 GM1 & 0.153 & -16.32 & 299.2 & 32.4 & 93.9
  \\ \hline
 Empirical& $0.16\pm0.01$ & $-16.0\pm0.1$  & $240\pm20$ & $31.7 \pm 3.2$ & $58.7 \pm 28.1$
\\ \hline
\end{tabular*}
 {\rule{\temptablewidth}{0.5pt}}
\end{center} \label{tab:sat}
\end{table*}

In laboratory experiments, the symmetry energy $E_{\rm sym}(\rho)$ can be studied by analyzing the neutron skin~\citep[e.g.,][]{2020arXiv200306168T}, the different isovector nuclear excitations~\citep[e.g.,][]{2014NuPhA.922....1D}, and the data on heavy-ion collisions such as isospin diffusion and the isotopic distribution in multifragmentation processes~\citep[e.g.,][]{2008PhR...464..113L}. The large amount of
novel exotic nuclei produced in the laboratory and the development of radioactive ion beams have greatly stimulated new research projects on symmetry energy~\citep{2017RvMP...89a5007O,2014EPJA...50....9L,2016PrPNP..91..203B,2020arXiv200210884Z}.
We mention here that in the following discussion, we only discuss up to the second expansion terms in both the binding energy (Eq.~\ref{eq:snm}) and the symmetry energy (Eq.~\ref{eq:esym}); see, e.g., \citep{2012PhRvC..85c5201D,2014PhRvC..90e5203D} for detailed discussions on the higher order terms and the suitability of a nuclear EOS for up to the high density matter possible in NSs.
Some of the latest constraints on higher order terms are also discussed in, e.g., \citep{2018PhRvC..98c5804M,2019ApJ...879...99Z,2019EPJA...55...39Z,2020arXiv200203210Z}.

\subsection{Results and discussion}

There are six parameters ($g_{\sigma q}, g_{\omega q}, g_{\rho q}, g_3, c_3, \Lambda_v$) in this Lagrangian~(Eq.~(1)) to be determined by fitting the saturation density $\rho_0$ and the corresponding values at the saturation point of the binding energy $E/A$, the incompressibility $K$, the symmetry energy $E_{\rm sym}$, the symmetry energy slope $L$ and the effective (Landau) mass $M_N^\ast$($\approx0.74 M_N$).
In particular, we use the most preferred values for $(K, E_{\rm sym}, L)$ as recently suggested by~\citep{2017RvMP...89a5007O,2006EPJA...30...23S}, namely, $K = 240 \pm 20$ MeV, $E_{\rm sym} = 31.7 \pm 3.2$ MeV, and $L = 58.7 \pm 28.1 $ MeV.
A recent fitting of finite nuclei data in the same model yielded $K = 328$ MeV~\citep{2016PhRvC..94d4308X}, and we choose this case as well for a comparison.
To study the effect of $r_N$, we varied this parameter from the intermediate value $0.87$ fm~\citep{2016RvMP...88c5009M} by approximately $10 \%$ according to our model capability: $r_N$ = 0.80 fm, 0.87 fm, and 1.00 fm. This covers both of the most recent experimental analyses of the $rms-$radius of the proton charge distribution: 0.879 $\pm$ 0.009 fm~\citep{2015JPCRD..44c1204A} from electron-proton scattering and 0.8409 $\pm$ 0.0004 fm~\citep{2010Natur.466..213P} from the Lamb shift measurement in muonic hydrogen.
For each nucleon radius, we first determine the potential parameters ($a$ and $V_0$) by reproducing $(m_N, r_N)$ and then determine QMF many-body parameters by reproducing the saturation properties of nuclear matter ($\rho_0, E/A, E_{\rm sym}, K, L, M_N^*/M_N$), which is shown in the first line of Table~\ref{tab:sat}. Six EOS models from other theoretical frameworks are also listed, together with the empirical ranges in the last row.

\begin{figure}
\vspace{-0.3cm}
{\centering
\resizebox*{0.9\textwidth}{0.4\textheight}
{\includegraphics{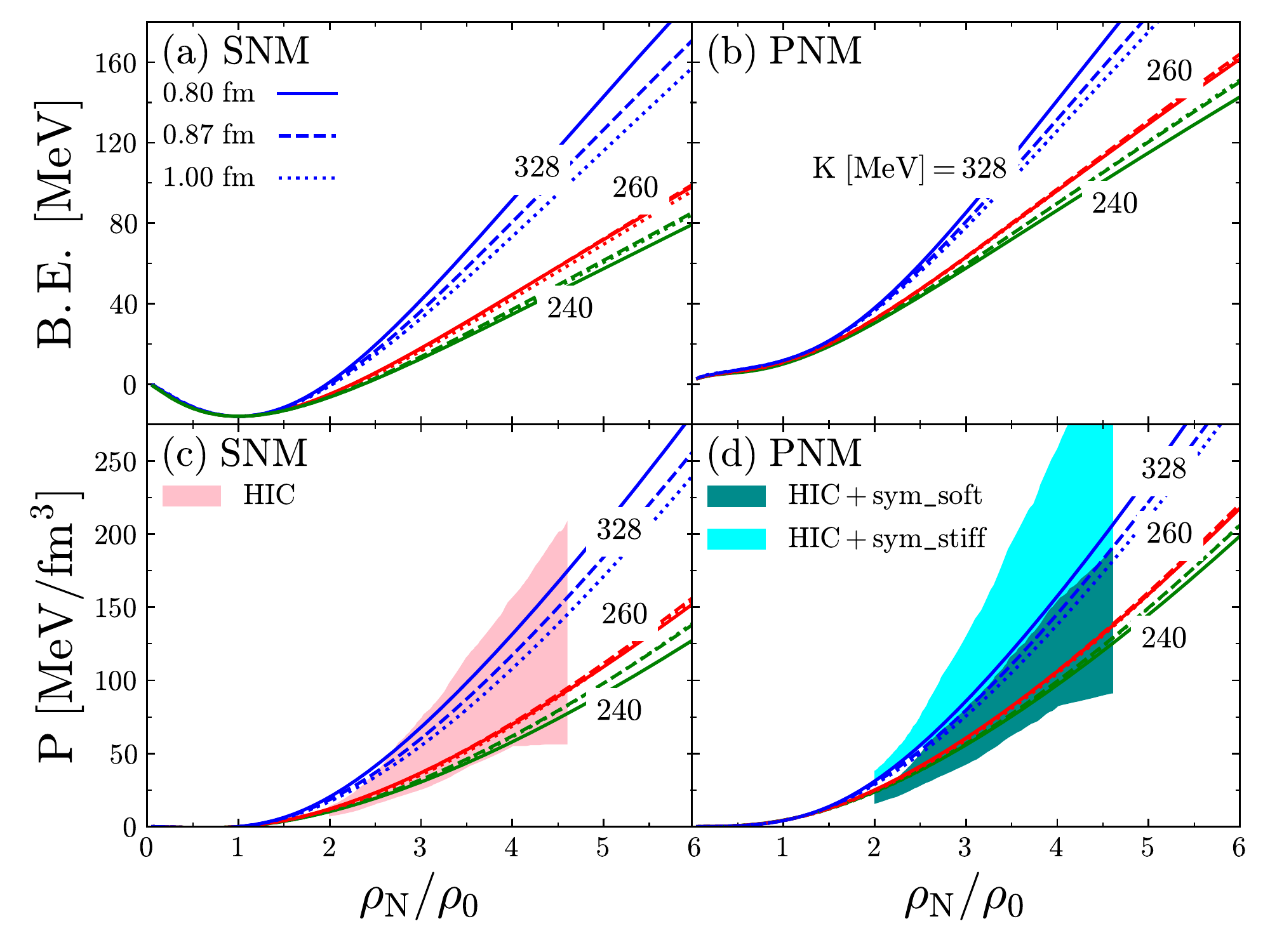}}}
\caption{(Color online) Binding energy (B.E.) and pressure as a function of the number density for symmetric nuclear matter (SNM) and pure neutron matter (PNM). The calculations are performed for fixed symmetry parameters $E_{\rm sym}$ = 31 and $L = 60$ MeV and different cases of incompressibility $K$ at saturation: $K = 240, 260, 328$ MeV. The results with different nucleon radii of 0.80, 0.87, and 1.00 fm, chosen from the CODATA values and two recent experiments~\citep{2016RvMP...88c5009M,2015JPCRD..44c1204A,2010Natur.466..213P}, are shown by the solid, dashed, and dotted curves, respectively.
Heavy-ion collisions (HIC) are expected to go through a quark¨Cgluon plasma (QGP) phase, where matter is strongly interacting, resulting in the development of collective motion.
The EOS results for SNM and PNM lie inside the boundaries obtained from the analysis of the collective flow in HIC~\citep{2002Sci...298.1592D}, which are shown with two density-dependent cases of symmetry energy (light blue for the stiff case and dark blue for the soft case). The radius of the nucleon is shown to have limited effects on the nuclear matter EOSs even at high density. Taken from~\citet{2018PhRvC..97c5805Z}. }\label{fig:rN}
\vspace{-0.3cm}
\end{figure}

The binding energy and pressure from the QMF are displayed in Fig.~\ref{fig:rN} for symmetric nuclear matter and pure neutron matter with different nucleon radii.
The EOS results within the QMF fulfill the flow constraints from heavy-ion collisions for both symmetric nuclear matter and pure neutron matter.
The nucleon radius has a weak effect on the nuclear matter even at high density.

We address other important aspects before closing this section:
\begin{itemize}
\item \relax $Temperature$: The above discussions are only for the zero-temperature case, below $\sim$1 MeV for cold NSs, lower than the characteristic nuclear Fermi energy, while dense matter is usually hot in heavy-ion collisions and proto-neutron stars, with a temperature as high as $\sim50$ MeV. Although the matter is expected to cool down on timescales of $10^{-22}-10^{-24}$ seconds and $1-10$ seconds, respectively, the thermal effects cannot be ignored, especially in the study of  dynamic processes~\citep[e.g.,][]{2013A&A...555A.129L,2015APh....62..115L}. However, for the equilibrium configurations of cold NSs, the EOSs are not affected much by finite temperature. For example, the temperature influence on the maximum mass is very limited, and there is an increase in the NS radius for a fixed amount of gravitational mass~\citep[e.g.,][]{2010PhRvC..81b5806L}.
\item \relax $Meson$-$coupling~parameters$: The present calculations are structured to be renormalizable to fix the coupling constants and the mass parameters by the empirical properties of nuclear matter at saturation. They can also be determined by fitting the ground-state properties of closed-shell nuclei. In the latter case, a substantial stiff EOS with an extremely high incompressibility is usually obtained, $\sim328$ MeV, which is not consistent with recent experimental results~\citep{2006EPJA...30...23S} (as seen in Table~\ref{tab:sat}). An alternatively low compressibility usually cannot describe the finite nuclei with a proper spin-orbit coupling.
\item \relax $~Beyond~mean$-$field$: As a starting point, we choose the mean-field approximation, which should be reasonably good at very high densities (a few times the nuclear matter density). There have been studies that demonstrate that the isoscalar Fock terms could be important for the prediction of NS properties (see, e.g.,~\citet{2016PhRvC..94d5803Z} for a study based on relativistic Hartree-Fock theory). In such models, the Lorentz covariant structure is kept in full rigor, which guarantees all well-conserved relativistic symmetries.
Additionally, the attractive Fock term introduced in the framework of QMC could effectively decrease the incompressibility at the saturation point~\citep{2007NuPhA.792..341R}.
\end{itemize}

\section{Neutron star}

NSs with typical masses $M\approx 1-3~M_{\odot}$ (where $M_{\odot}$ is the mass of the sun, $M_{\odot}=1.99\times10^{33}$ g) and radii on the order of $R\approx10$ km have many extreme features that are unique in the universe~\citep{1983bhwd.book.....S,2004Sci...304..536L,2012RPPh...75b6301B,2017IJMPD..2630015G} and lie outside the realm of terrestrial laboratories, such as rapid rotation, extremely strong magnetic fields, superstrong gravitation, interior superfluidity and superconductivity, and superprecise spin period.
These intriguing features have aroused much interest from researchers of many branches of contemporary physics as well as astronomy because of their importance to fundamental physics.
However, information regarding the NS interior has not yet been sufficiently revealed through the current observations due to the complexity of the NS system and many uncertain factors~\citep{2007PhR...442..109L}.
It is time to combine the efforts from different communities and discuss mutual interests and problems.
In this section, we introduce the basic insights into NSs, in particular the global properties such as the mass, radius, and tidal deformability of the star, which have a one-to-one correspondence to its underlying EOS and are usually used as a tool to connect nuclear physics to astrophysics for the study of dense matter above the nuclear saturation density.
~\citep[e.g.,][]{2018ApJ...862...98Z,2020arXiv200203210Z,2019ApJ...878..159M,2012ARNPS..62..485L,2017ApJ...850L..34B,2018PhRvL.120q2703A,2018PhRvL.121f2701L,2018PhRvL.120z1103M,2018ApJ...857...12N,2018PhRvD..98f3020Z,2018PhRvD..97h3015Z,2019PhRvD..99h3014H,2018ApJ...857L..23R,2019PhLB..796....1T,2019arXiv191202312W,2019ApJ...886...52Z,2020PhRvD.101f3007E,2020PhRvD.101d3021F,2020arXiv200110259G,2020PhLB..80335306L,2020EPJA...56...32L,2020arXiv200208951T}.

A wide range of matter density from $\sim0.1$ g cm$^{-3}$ in the star atmosphere to values larger than $\sim$ $10^{14}$ g cm$^{-3}$ in the star core is encountered in these objects.
Theoretically, the global properties are studied by using the overall EOSs as basic input and ignoring their thin atmosphere ($\sim$ $0.1-10$) cm, where hot X-rays originate.
The observations of massive NSs~\citep{2010Natur.467.1081D,2013Sci...340..448A,2016ApJ...832..167F,2018ApJS..235...37A,2020NatAs...4...72C} have already ruled out soft EOSs that cannot reach $2M_{\odot}$. Here, this serves as a criterion for the selection of the NS (core) EOSs.
The saturation properties of the employed core EOSs are collected in Table \ref{tab:sat}, with the empirical ranges listed in the last row.
The determination of the EOS above the saturation density represents one of the main problems in NS study because first principle QCD calculations are difficult to perform in such a many-body system.
In most of the model calculations available in the literature, a central density as high as $(2-10) \rho_0$ is found for the maximum mass, and one or more types of strangeness-driven phase transitions (hyperons, kaons, Delta isobars or quarks)
may take place in the NSs' innermost parts, e.g., ~\citep{2014PhRvC..89b5802H,2010PhRvC..81b5806L,2016PhRvC..94d5803Z,2007ChPhy..16.1934L,2011PhRvC..83b5804B,2004PhRvC..70e5802Z,2006PhRvC..74e5801L,2008ChPhL..25.4233L,2008PhRvC..77f5807P,2009ChPhC..33...61L,2015PhRvC..91c5803L}.
NSs with exotic phases are discussed in Sec.~\ref{sec:exotic}.
In this section, we restrict ourselves to normal nuclear matter.

\subsection{Neutron star crust}

In the outer crust, at densities below $\sim$ $10^7$ g cm$^{-3}$, nuclei arrange themselves in a Coulomb lattice mainly populated by $^{56}$Fe nuclei. At higher densities ($10^7$ g cm$^{-3}-4\times 10^{11}$ g cm$^{-3}$), the nuclei are stabilized against beta decay by the filled Fermi sea of electrons and become increasingly neutron-rich. The composition of the outer crust is mainly determined by the nuclear masses, which are experimentally measured close to stability, whereas the masses of the very neutron-rich nuclei are not known, and they have to be calculated using nuclear models.

The inner crust is a nonuniform system of more exotic neutron-rich nuclei, degenerate electrons, and superfluid neutrons. The density range extends from $\sim4\times 10^{11}$ g cm$^{-3}$ to the nuclear saturation density $2.8 \times10^{14}$ g cm$^{-3}$, at which point the nuclei begin to dissolve and merge together. Nonspherical nuclear structures, generically known as nuclear "pasta'', may appear at the bottom layers of the inner crust. In fact, one of NSs' irregular behaviors, the $glitch$, is closely related to the inner crust EOS and the crust-core transition properties~\citep[e.g.,][]{2008LRR....11...10C,2014PhRvC..90a5803P,2015ChPhL..32g9701L,2016ApJS..223...16L}.
The crust is also crucial for NS cooling \citep{2008LRR....11...10C}.

\begin{figure*}
{\centering
\resizebox*{0.47\textwidth}{0.265\textheight}
{\includegraphics{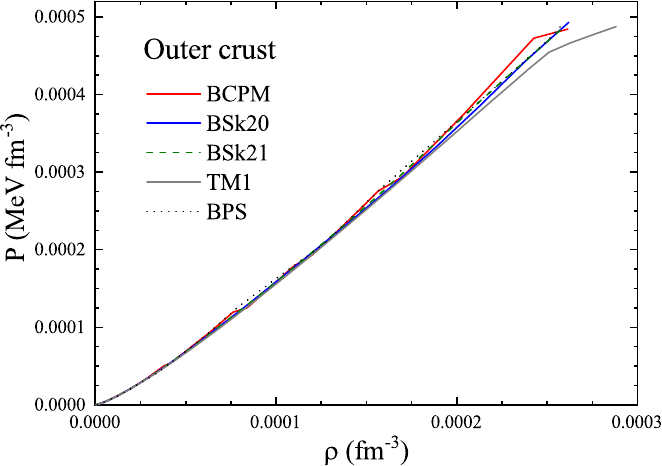}}
\resizebox*{0.47\textwidth}{0.27\textheight}
{\includegraphics{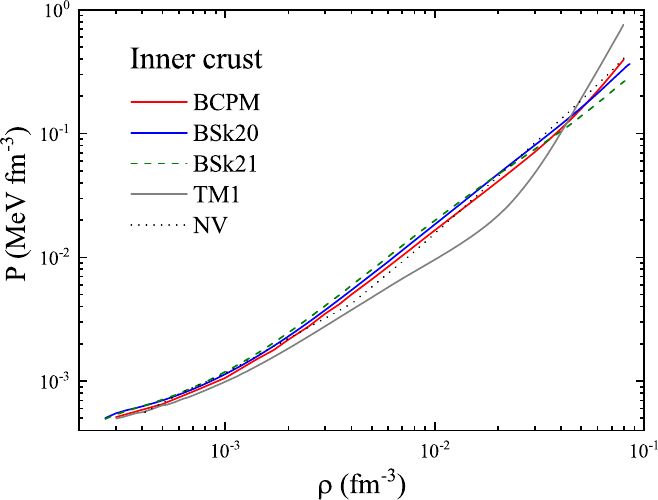}}}
\caption{Various EOSs for the outer crust (left) and inner crust (right). Among them, BCPM, TM1, BSk20, and BSk21 are unified NS EOSs, namely, all EOS segments (outer
crust, inner crust, liquid core) are calculated using the same nuclear interaction. The BPS (NV) EOS for the outer (inner) crust part is indicated by the black dotted line.
The BPS outer crust EOS is based on a semi-empirical mass formula for matter from $10^7$ g cm$^{-3}$ to 3.4 $\times 10^{11}$ g cm$^{-3}$~\citep{1971ApJ...170..299B}, whereas the NV inner crust EOS is based on quantal Hartree-Fock calculations for spherical Wigner-Seitz cells~\citep{1973NuPhA.207..298N}. }\label{fig:crust}
\vspace{-0.3cm}
\end{figure*}

It may be necessary to calculate all EOS segments (outer crust, inner crust, and liquid core) using the same nuclear interaction, the so-called ''unified'' EOS~\citep[e.g.,][]{2013A&A...559A.128F,2013A&A...560A..48P,2015A&A...584A.103S,2019arXiv191203815A}, since matching problems in nonunified EOS could cause nontrivial conflicts in the predictions of the stars' properties~\citep{2016PhRvC..94c5804F}.
Fig.~\ref{fig:crust} shows the crust EOS for the different theoretical approaches in Table~\ref{tab:sat}. We observe that all outer crust EOSs display a similar pattern, with some differences around the densities where the composition changes from one nucleus to the next one. Only the TM1 EOS, based on an RMF model, shows a slightly different trend due to the semiclassic-type mass calculations, in which $A$ and $Z$ vary in a continuous way, without jumps at the densities associated with a change in the nucleus in the crust.
$A$ is the number of nucleons in the nucleus, and $Z$ is the atomic number.
On the other hand, the energy in the inner crust is largely determined by the properties of the neutron gas; hence, the neutron matter EOS plays an important role. Moreover, the treatment of complicated nuclear shapes, in a range of average baryon densities between the crust and the core, produces some uncertainties in the EOS of the inner crust, where some differences are visible.

\begin{figure}
\vspace{0.3cm}
\begin{center}
\resizebox*{0.65\textwidth}{0.32\textheight}
{\includegraphics{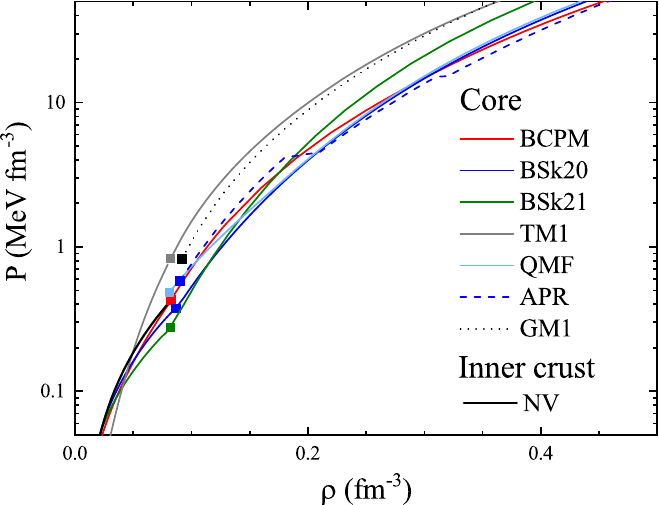}}
\end{center}
\caption{Various EOSs for the NS core from the low-density inner crust indicated with symbols.
In addition to the models in Fig.~\ref{fig:crust}, we include another model within the RMF, the GM1 EOS~\citep{1991PhRvL..67.2414G}, as well as the present QMF model. The inner crust EOS of NV is also included. }\label{fig:crustcore}
\vspace{-0.3cm}
\end{figure}

In Fig.~\ref{fig:crustcore}, we show the above discussed EOSs, with the full symbols indicating the transition point from the inner crust to the core for each chosen EOS.
The APR and GM1 EOSs have to be matched with an inner crust EOS, which is at variance with the unified EOSs (BCPM, TM1, BSk20, and BSk21), and we achieve this by imposing that the pressure is an increasing function of the energy density. It is evident that the matching of the GM1 core (dotted black line) to the TM1 crust (solid gray line) shows nonsmooth behavior in the $dP/d\rho$ slope compared to the matching to the BCPM and NV crust. %
Since the crust effects were shown to be more important for distorted fast-rotating stars than for static stars~\citep{2003LRR.....6....3S}, later in Sec.~\ref{sec:rns} on rotating NSs, we discuss three widely used crust EOSs (TM1, BCPM, NV + BPS) that are matched with one core EOS (GM1).

Note that the above crust is based on the ground state approximation for zero-temperature matter, which can only be applied to an isolated NS born in a core-collapse supernova explosion. It is assumed that during the process of cooling and crystallization, the plasma maintains nuclear equilibrium. Consequently, when the matter becomes strongly degenerate, the structure and EOS of the crust can be approximated well via cold-catalyzed matter.
For an NS crust formed by accreted plasma from the companion star in a low-mass X-ray binary, the outermost layer of the accreted plasma undergoes thermonuclear flashes, observed as X-ray bursts, during the active stages. The layers deeper than a few meters are at $T < 5\times10^8 \rm K$, becoming increasingly
neutron-rich due to electron capture and neutrino emissions
and finally dissolving in the liquid core. After the fully accreted crust is formed, the layered structure of the crust ceases to evolve and becomes quasistationary, with matter elements moving inwards due to compression and undergoing exothermic nuclear transformations~\citep{2018MNRAS.475.5010F}.
There is a microscopic model for a fully accreted crust~\citep{2008A&A...480..459H} that calculates the EOS and distribution of deep crustal heating sources by following the nuclear evolution
of an element of matter consisting initially of X-ray ashes under quasistatic compression from $10^7~\rm g~cm^{-3}$ to $10^{14}\rm~g~cm^{-3}$ (crust-core interface).

\subsection{Mass-radius relation}

To study the structure of NSs, we have to calculate the composition and EOS of cold, neutrino-free, catalyzed matter. We require that the NS contains charge-neutral matter consisting of neutrons, protons, and leptons ($e^-$, $\mu^-$) in beta equilibrium.
Additionally, since we are looking at NSs after neutrinos have escaped, we set the neutrino chemical potentials equal to zero.
The energy density of NS matter can be written as a function of the different partial densities,
\begin{eqnarray}
\varepsilon(\rho_n,\rho_p,\rho_e,\rho_\mu) &=& \rho M_N +  \rho E\left(\rho_n,\rho_p\right)/A \nonumber  \\
&& +\rho_\mu m_\mu + \frac{1}{2m_{\mu}} \frac{(3\pi^2\rho_\mu)^{5/3}}{5\pi^2}  + \frac{(3\pi^2\rho_e)^{4/3}}{4\pi^2} \nonumber  \\
\end{eqnarray}
where we use ultrarelativistic and nonrelativistic approximations for the electrons and muons, respectively, from textbooks~\citep{1983bhwd.book.....S}.
Then, the various chemical potentials $\mu_i$ of the species ($i=n,p,e,\mu$) can be computed,
\begin{eqnarray}
\mu_i = \partial \varepsilon/\partial \rho_i,
\end{eqnarray}
which fulfills beta-equilibrium,
\begin{eqnarray}
\mu_i = b_i \mu_n -q_i\mu_e
\end{eqnarray}
($b_i$ and $q_i$ denote the baryon number and charge of species $i$).
Supplemented with the charge neutrality condition,
\begin{eqnarray}
\sum_i \rho_iq_i=0
\end{eqnarray}
the equilibrium composition $\rho_i(\rho)$ can be determined at the given baryon density $\rho$,
and finally, the EOS is
\begin{eqnarray}
P(\rho)=\rho^2\frac{d(\varepsilon/\rho)}{d\rho}=\rho\frac{d\varepsilon}{d\rho}-\varepsilon=\rho\mu_n-\varepsilon
\end{eqnarray}
for the interior of NSs.

The NS stable configuration in hydrostatic equilibrium can be obtained by solving the Tolman-Oppenheimer-Volkoff (TOV) equation~\citep{1939PhRv...55..364T,1939PhRv...55..374O} for the pressure $P$ and the enclosed mass $m$,
\begin{eqnarray}
\frac{dP(r)}{dr}& = &-\frac{Gm(r)\varepsilon(r)}{r^{2}}\frac{\Big[1+\frac{P(r)}{\varepsilon(r)}\Big]\Big[1+\frac{4\pi r^{3}P(r)}{m(r)}\Big]}
 {1-\frac{2Gm(r)}{r}} ,\\
\frac{dm(r)}{dr}& = &4\pi r^{2}\varepsilon(r).
\end{eqnarray}
$G$ is the gravitational constant. Starting with a central mass density $\varepsilon(r=0)=\varepsilon_c$, we integrate out until the pressure on the surface equals that corresponding to the density
of iron. This gives the stellar radius $R$, and the gravitational mass is then
\begin{eqnarray}
m(R) = 4\pi \int^R_0 dr r^2 \varepsilon(r)
\end{eqnarray}
For the description of the NS crust, we usually join the EOS $P(\varepsilon)$ with the NV EOSs of Negele and Vautherin in the medium-density regime~\citep{1973NuPhA.207..298N} and those of Baym-Pethick-Sutherland for the outer crust~\citep{1971ApJ...170..299B}.
After solving the TOV equations, we can obtain the maximum mass $M_{\rm TOV}$ and the mass-radius relation for comparison with astrophysical observations.

\subsection{Symmetry energy effects on neutron star structure}
\begin{figure*}
\vspace{-0.3cm}
{\centering
\resizebox*{0.48\textwidth}{0.3\textheight}
{\includegraphics{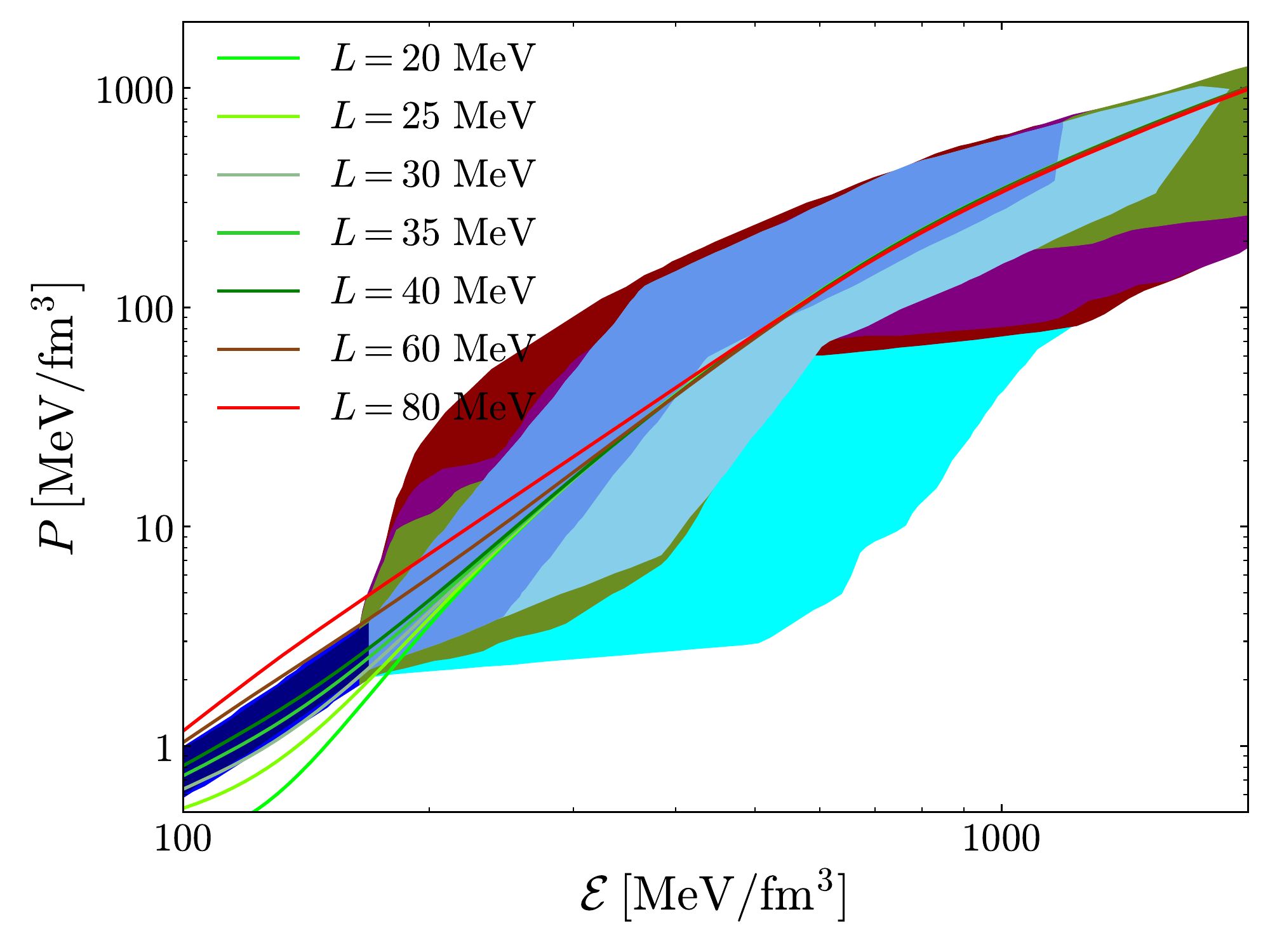}}
\resizebox*{0.48\textwidth}{0.3\textheight}
{\includegraphics{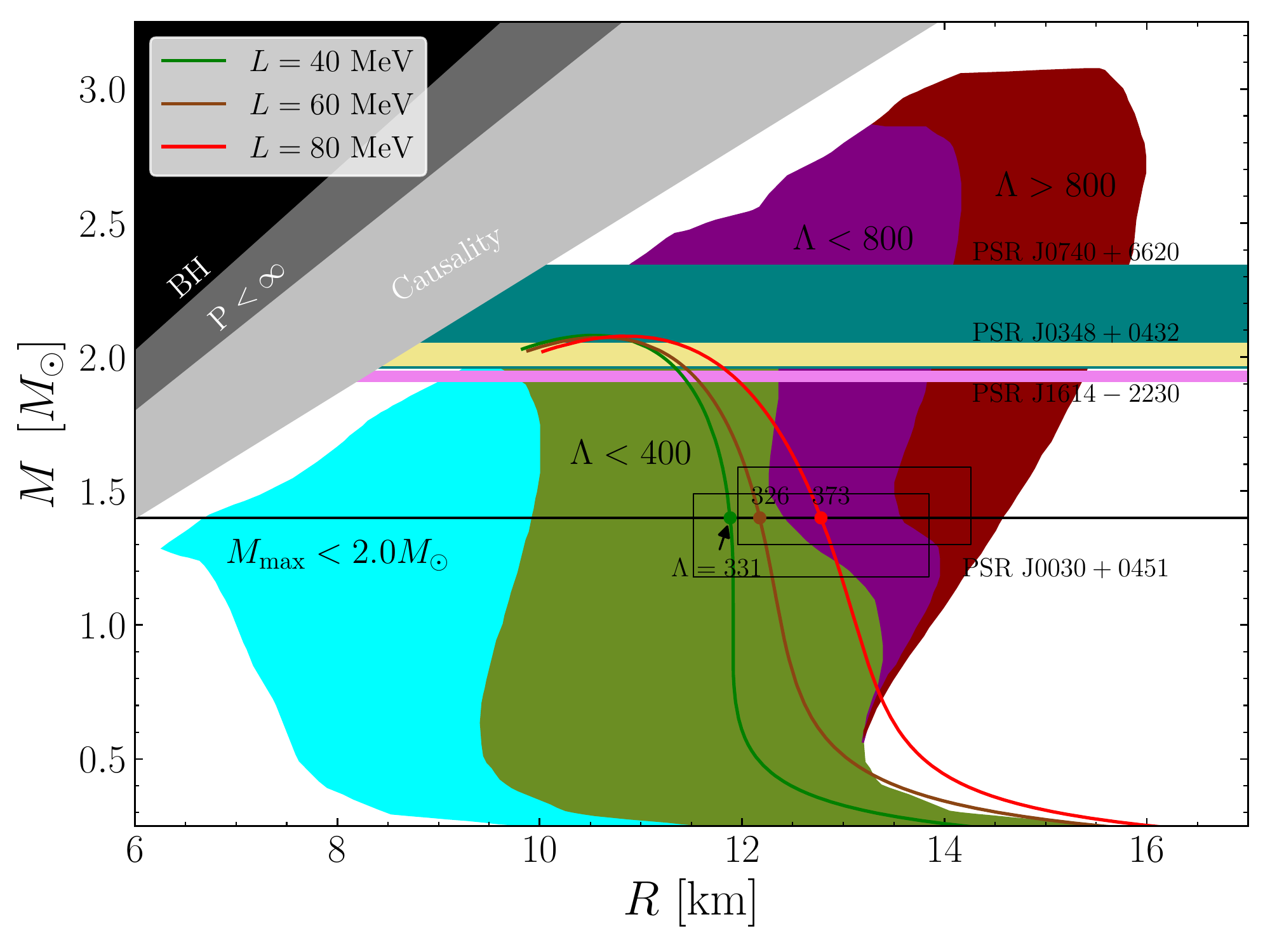}}}
\caption{ (Left) NS EOSs and (right) mass-radius relation within QMF with different values of symmetry energy slope $L$, with more $L$ cases shown in the left panel than in the right panel. The shaded region is the favored region from ab initio calculations at the subsaturation density in chiral effective field theory~\citep{2013ApJ...773...11H} and from~\citep{2018PhRvL.120q2703A}. They are causal and fulfill the two-solar-mass constraint of heavy pulsars ($M_{\rm TOV}>2 M_{\odot}$) and the tidal deformability constraint of binary merger event GW170817 ($\Lambda_{\rm 1.4}\le800$) for a $1.4 M_{\odot}$ star. Also shown are the latest NICER measurements from the pulse-profile modeling of the accretion hot spots of the isolated millisecond pulsar PSR J0030+0451~\citep{2019ApJ...887L..24M,2019ApJ...887L..21R}. The general constraints from the black hole limit, the Buchdahl limit and the causality limit are also included. The figure shows that the radius sensitively depends on the symmetry energy slope with the maximum mass only slightly modified. A smaller $L$ (softer symmetry energy) leads to a smaller radius. All cases of $L=20-80~\rm MeV$ lie within the $\Lambda \le800$ boundary~\citep{2017PhRvL.119p1101A} and fulfill the updated limit $\Lambda_{\rm 1.4}=190^{+390}_{-120}$~\citep{2018PhRvL.121p1101A} using the PhenomPNRT waveform model at a 90\% confidence level.}\label{fig:mr}
\vspace{-0.3cm}
\end{figure*}

Currently, the EOS of SNM ($\beta=0$) is constrained relatively well. Matter with nonzero isospin asymmetry remains unknown, largely due to the uncertainty in the symmetry energy. Conflicts remain for the symmetry energy (especially its slope) despite significant progress in constraining the symmetry energy around and below the nuclear matter saturation density~\citep{2014EPJA...50....9L,2016PrPNP..91..203B}. The symmetry energy slope characterizes the density dependence of the symmetry energy and largely dominates the ambiguity and stiffness of the EOS in NSs' high-density cores in the case of no strangeness phase transition.

Fig.~\ref{fig:mr} shows our EOSs and the corresponding mass-radius relation under different symmetry energy slopes $L$ in the range of $20-80$ MeV. The QMF parameters are fitted to reproduce the saturation properties in Table~\ref{tab:sat}, with the other five parameters ($\rho_0, E/A, K, E_{\rm sym}$) unchanged~\citep{2018ApJ...862...98Z}.
The TOV mass of the star hardly changes with changing $L$ and fulfills the recent observational constraints of three massive pulsars for which the masses are precisely measured~\citep{2010Natur.467.1081D,2013Sci...340..448A,2016ApJ...832..167F,2018ApJS..235...37A,2020NatAs...4...72C}.
There is a strong positive correlation between the slope parameter and the radius of a $1.4 M_{\odot}$ star~\citep[for more discussion, see, e.g.,][]{2018PhRvL.121f2701L,2018PhRvC..98f5804H}. However, a small dependence is found in~\citet{2018PhRvC..98f5804H}.
The cases of $L\approx30-60$ MeV in our QMF model may be more compatible with the neutron matter constraint~\citep{2013ApJ...773...11H}.
\citet{2020NatAs.tmp...42C} found that the radius of a $1.4 M_{\odot}$ NS is $R_{\rm 1.4}= 11.0^{+0.9}_{-0.6}~\rm km$ ($90 \%$ credible interval), assuming a description in terms of nuclear degrees of freedom remains valid up to $2\rho_0$.
The recent NICER measurements of PSR J0030+0451~\citep{2019ApJ...887L..24M,2019ApJ...887L..21R} might indicate $L \gtrsim  40 \rm MeV$.

The EOS governs not only the stable configuration of a single star but also the dynamics of NS mergers. During the inspiral phase, the influence of the EOS is evident on the tidal polarizability~\citep{2010PhRvD..82b4016P,2013PhRvD..88d4042R}.
In Fig.~\ref{fig:mr}, we also include the calculated results of the tidal deformability $\Lambda$ and the constraining region from binary merger event GW170817, namely, $\Lambda_{\rm 1.4}\le800$ for a $1.4 M_{\odot}$ star~\citep{2017PhRvL.119p1101A}.
The tidal deformability describes the magnitude of the induced mass quadrupole moment when
reacting to a certain external tidal field. It is zero in the black hole case.
The dimensionless tidal deformability $\Lambda$ is related to the compactness $M/R$ and the Love number $k_2$ through $\Lambda = \frac{2}{3}k_2(M/R)^{-5}$ (see more discussion later in Sec.~\ref{sec:binarytidal}).

\begin{figure*}
\vspace{-0.5cm}
{\centering
\resizebox*{0.48\textwidth}{0.28\textheight}
{\includegraphics{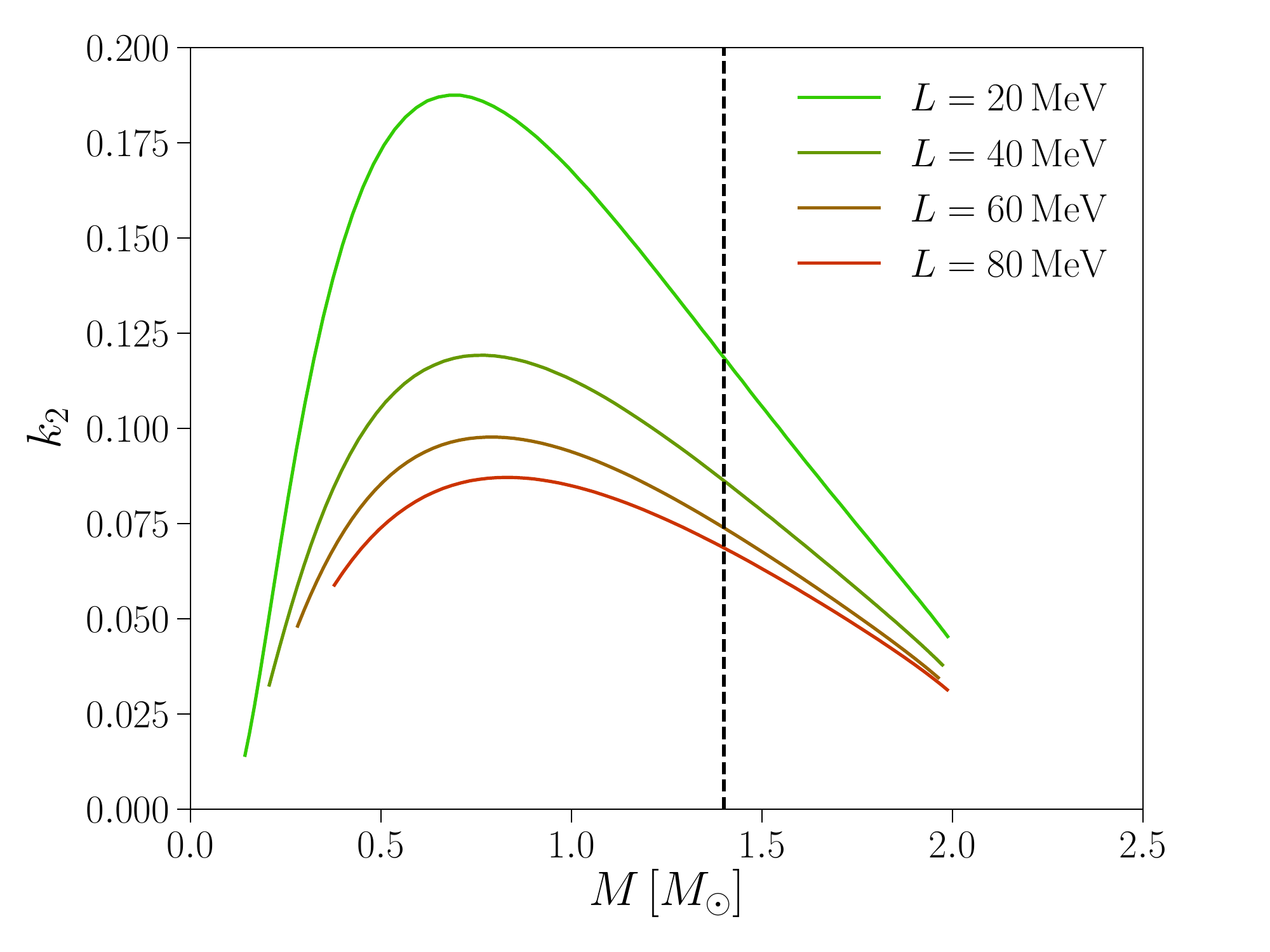}}}
{\centering
\resizebox*{0.48\textwidth}{0.28\textheight}
{\includegraphics{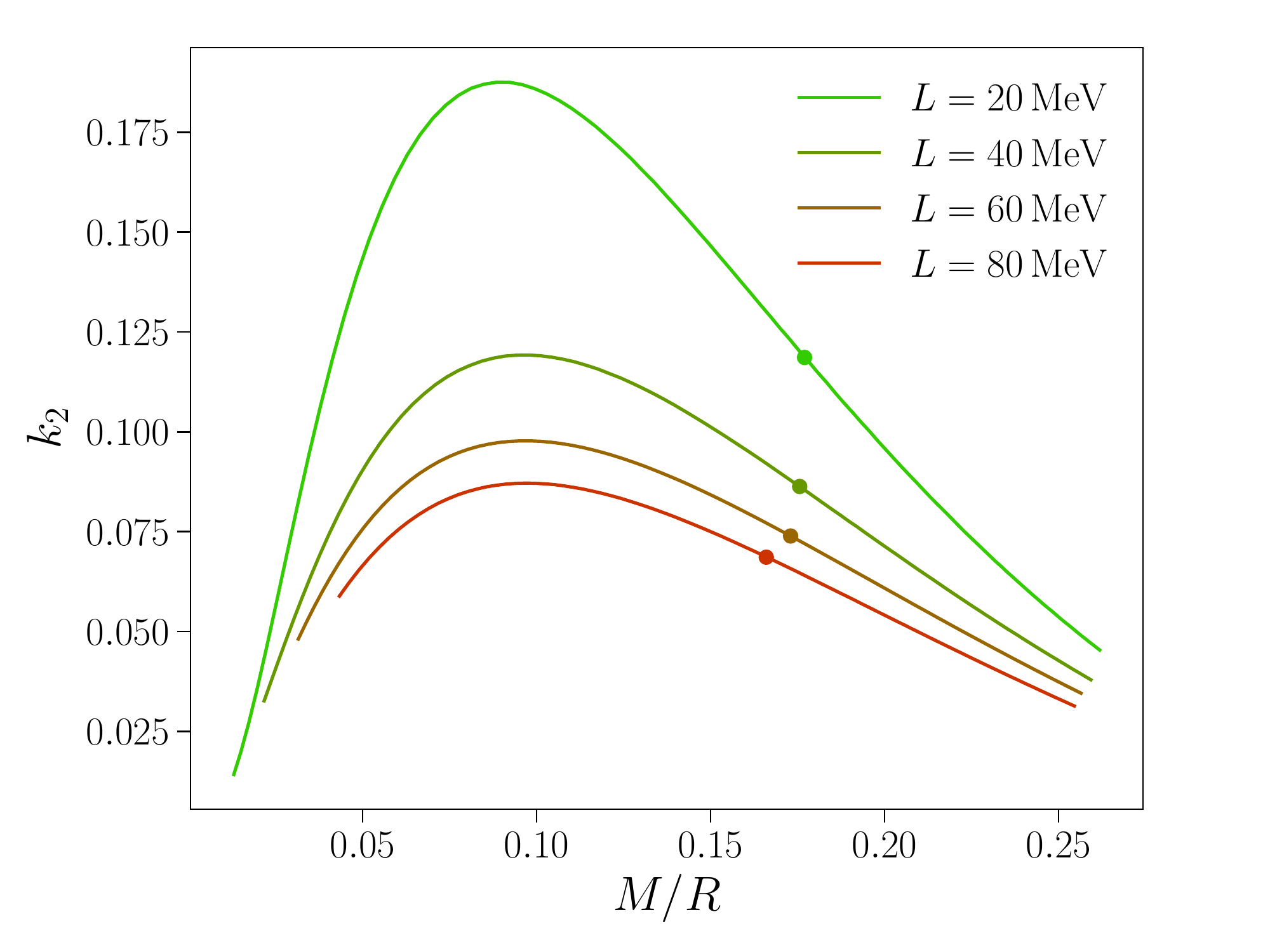}}}
\caption{Love numbers as a function of the mass (left) and compactness (right) for four EOSs with different values for the symmetry energy slope $L$ ($20$, $40$, $60$, $80$ MeV). $k_2$ first increases and then decreases with mass and compactness. The increase in $k_2$ (below $\sim$1.0$M_{\odot}$) is due to large radii and a large portion of soft crust matter.
The vertical line and colorful dots indicate $M = 1.4\,M_\odot$.
Taken from~\citet{2018ApJ...862...98Z}. }\label{fig:k2}
\vspace{-1cm}
\end{figure*}
\begin{figure*}
\vspace{-2cm}
{\centering
\resizebox*{0.48\textwidth}{0.28\textheight}
{\includegraphics{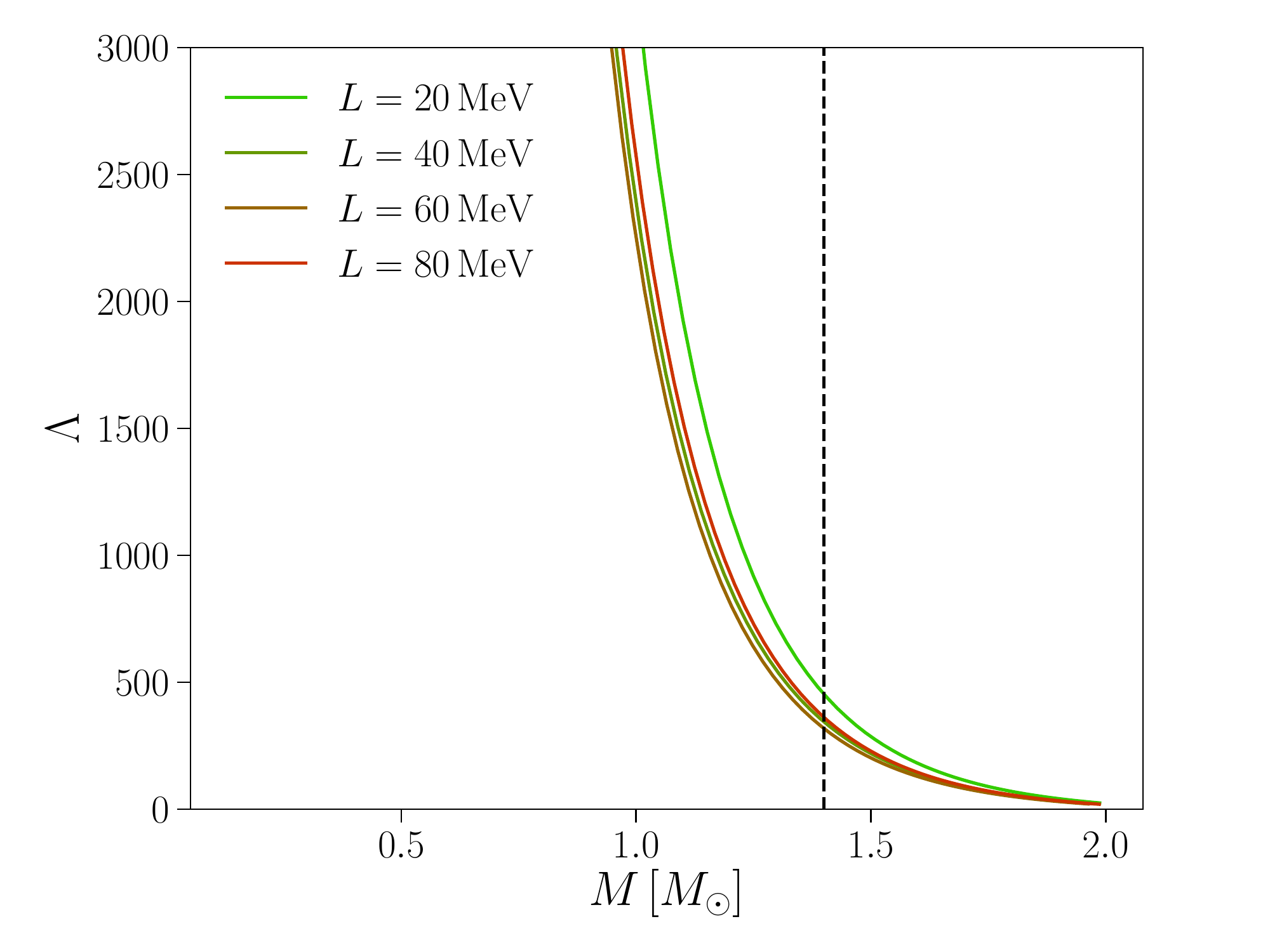}}}
{\centering
\resizebox*{0.48\textwidth}{0.28\textheight}
{\includegraphics{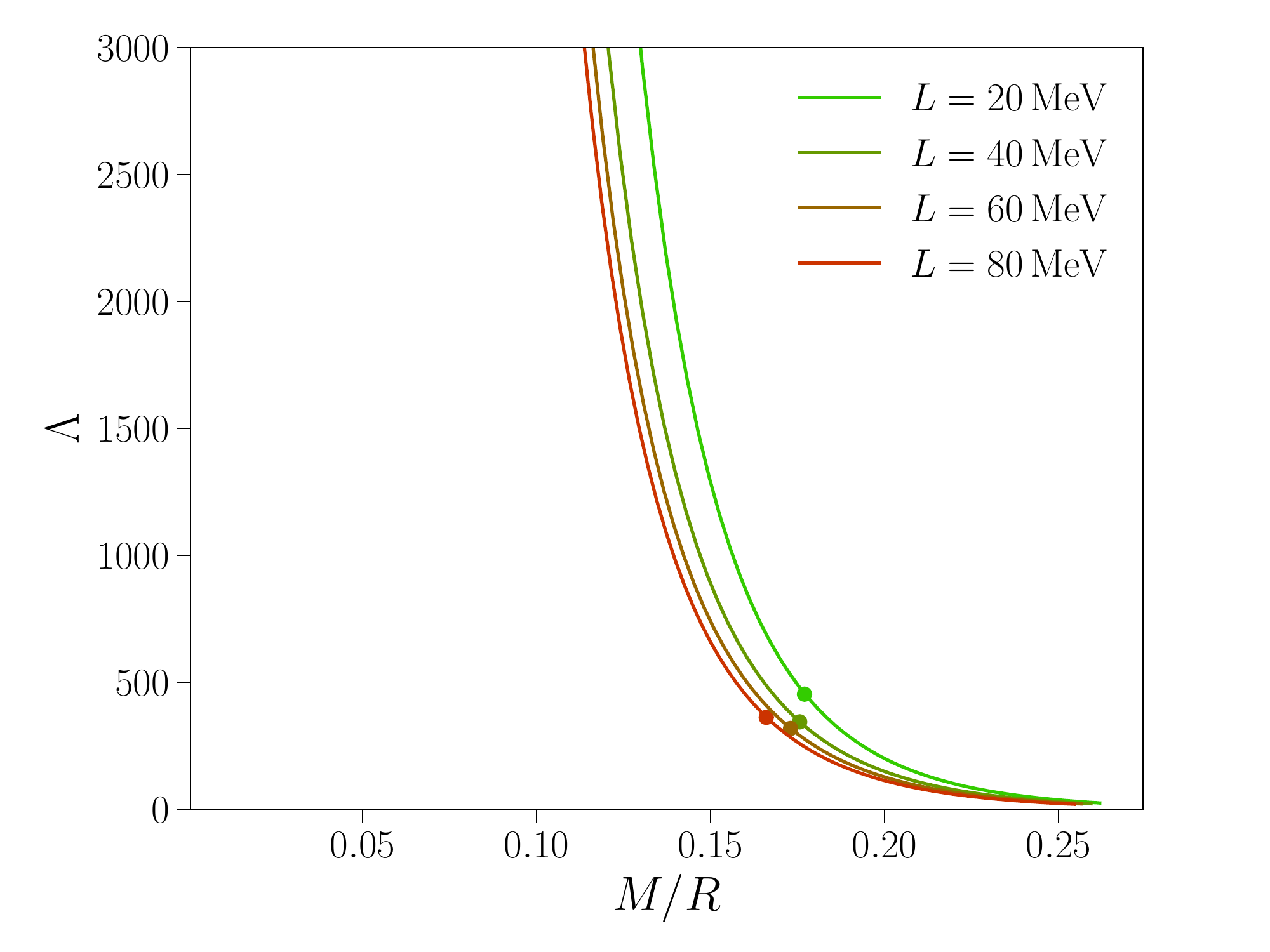}}}
\caption{Same as Fig.~\ref{fig:k2} but for the tidal deformabilities.
$\Lambda$ monotonously decreases with the mass and compactness. Similar to Fig~.\ref{fig:k2}, the large values of $\Lambda$ for small masses (below $\sim$1.0$M_{\odot}$) are due to large radii and a large portion of soft crust matter.
Taken from~\citet{2018ApJ...862...98Z}. }\label{fig:lambda}
\vspace{-1cm}
\end{figure*}

To study the effects of the symmetry energy slope $L$ in more detail, we present the resulting Love numbers (tidal deformabilities) as a function of the mass and the compactness for different $L$ in Fig.~\ref{fig:k2} (Fig.~\ref{fig:lambda}).
In Fig.~\ref{fig:k2}, $k_2$ first increases and then decreases with mass and compactness. In Fig.~\ref{fig:lambda}, $\Lambda$ monotonously decreases with the mass and compactness. The increase in $k_2$ and large values of $\Lambda$ for small masses (below $\sim$1.0$M_{\odot}$) are due to large radii and a large portion of soft crust matter. If no crust is considered (e.g., an EOS described by a pure polytropic function), $k_2$ still decreases monotonously with mass and compactness.
Further loud gravitational-wave measurements from merging binary NSs would provide data with good
precision for learning more about the slope parameter as well as the NS structure.
Moreover, the final fate of the merger, i.e., prompt or delayed collapse to a black hole or a single NS star, depends on the EOS, as well as the amount of ejected matter that undergoes nucleosynthesis of heavy elements. These discussions are presented in Sec.~\ref{sec:binary}.

\subsection{Rotating neutron star}\label{sec:rns}

NSs are usually rotating, and the rotational periods $P$ of rapidly rotating NSs (pulsars) could provide restrictions on the EOSs and their evolution processes when combined with the mass constraint.
When rapidly rotating, an NS is flattened by the centrifugal force, and the TOV equation, suitable for a static and spherically symmetric situation, cannot correctly describe the rotating stellar structure.
We assume NSs are steadily rotating and have an axisymmetric structure.
Based on the axial symmetry, the space-time metric used to model a rotating star can be expressed as
\begin{eqnarray}
ds^2 & = & -e^\nu dt^2 + e^\alpha dr^2 + e^\alpha r^2d\theta^2   \nonumber \\
&&+ e^\beta r^2 \sin^2\theta (d\phi - \omega dt)^2,
\end{eqnarray}
where $\nu, \alpha, \beta$ and $\omega$ is the function of $r,\theta$.
The matter inside the star is approximated by a perfect fluid, and the energy-momentum tensor is given by
\begin{eqnarray}
T^{\mu\nu}=(\varepsilon+p)u^{\mu}u^{\nu}-pg^{\mu\nu},
\end{eqnarray}
where $\varepsilon,p$ and $u^{\mu}$ are the energy density, pressure and four-velocity, respectively.
To solve Einstein¡¯s field equation for potentials $\nu, \alpha, \beta$ and $\omega$,
\citet{1989MNRAS.237..355K} transformed the Einstein equation from differential equations to integrals by using the Green function method.
In this form, the asymptotic flatness condition, which is the boundary condition of the Einstein equation, can be satisfied automatically.
This method for solving the Einstein equation is written as a standard code. This is the well-tested RNS code~\footnote{http://www.gravity.phys.uwm.edu/rns/}.
Using tabulated EOSs, the stationary and equilibrium sequences of rapidly rotating, relativistic stars can then be computed in general relativity~\citep[see more detail about the code in, e.g.,][]{1989MNRAS.237..355K,1994ApJ...422..227C,1995ApJ...444..306S}.

\begin{figure}
\vspace{-0.3cm}
{\centering
\resizebox*{0.85\textwidth}{0.4\textheight}
{\includegraphics{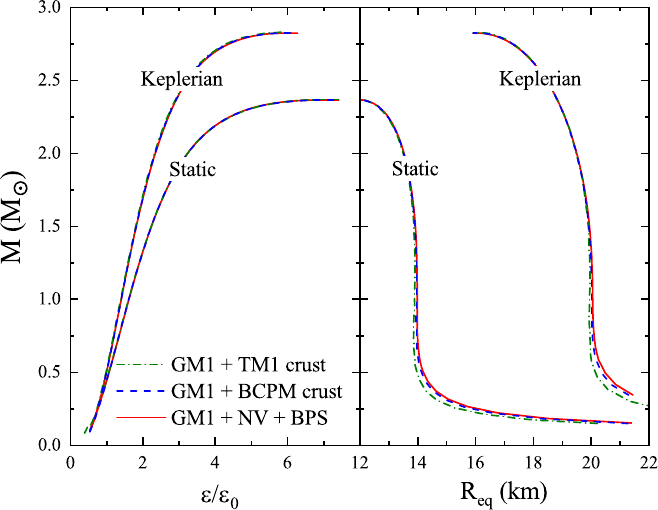}}}
\caption{ NSs' masses as a function of central energy density (left) and radius (right) for three cases of crust EOSs (TM1, BCPM, NV + BPS) matching one GM1 core EOS, with the detailed EOS matching data shown in Table~\ref{tab:sat}. The calculations are performed for both the static case and Keplerian rotating case. The maximum masses and radii, as well as the central densities, hardly depend on how the inner crusts are described for NSs heavier than $1.0 \, M_{\odot}$.}\label{fig:rnscrust}
\vspace{-0.3cm}
\end{figure}

The Keplerian (mass-shedding) frequency $f_{\rm K}$ is one of the most studied physical quantities for rotating stars. An EOS that predicts Kepler frequencies that are smaller than the observed rotational frequencies is to be rejected, as it is not compatible with observation.
An empirical formula was proposed in \citet{2004Sci...304..536L},
\begin{eqnarray}
f_{\rm K} = f_0 \left( \frac{M}{M_{\odot}}\right)^{\frac{1}{2}} \left( \frac{R}{10~\rm km}\right)^{-\frac{3}{2}},
\label{eq:fK}
\end{eqnarray}
where $M$ is the gravitational mass of the Keplerian configuration, $R$ is the radius of the nonrotating configuration of mass $M$, and $f_0$ is a constant that does not depend on the EOS.
An optimal prefactor $f_0 = 1080~\rm Hz$ was found in \citep{2009A&A...502..605H,2017PhRvD..96d3008W} for NSs as well as hybrid stars.
See more discussion in \citep{2009A&A...502..605H,2017PhRvD..96d3008W} regarding the justification of the functional form of Eq.~(\ref{eq:fK}) and its valid range.
The calculated highest spin frequencies $f_{\rm K}$ are all higher than 1000 Hz, while the current observed maximum is $f$ = 716 Hz~\citep{2006Sci...311.1901H} for PSR J1748-2446a in the globular cluster Terzan 5. A possible reason for this discrepancy is that the star fluid is suffering from $r$-mode instability~\citep{2001IJMPD..10..381A}. A simple estimation showed that this would lower the maximum frequency by $\sim37\%$, which might satisfactorily explain the observations to date.

Fig.~\ref{fig:rnscrust} shows the crust effects on the star's mass-radius relations in nonunified EOSs, where three widely used crust EOSs (TM1, BCPM, NV + BPS) are matched with one core EOS (GM1).
It is clear that for both the static case and Keplerian rotating case, the results hardly depend on how the inner crusts are described. This is true not only for the maximum mass and central densities but also for the radii. For less massive stars, crust-core matching has a slightly larger effect on the radii, and the TM1 curve deviates slightly from the other two due to the relatively larger difference in the crust-core interface for TM1 mentioned before. This deviation may be relevant only for NSs' masses smaller than $1.0 \, M_{\odot}$.

\begin{figure}
\vspace{0.3cm}
{\centering
\resizebox*{0.6\textwidth}{0.3\textheight}
{\includegraphics{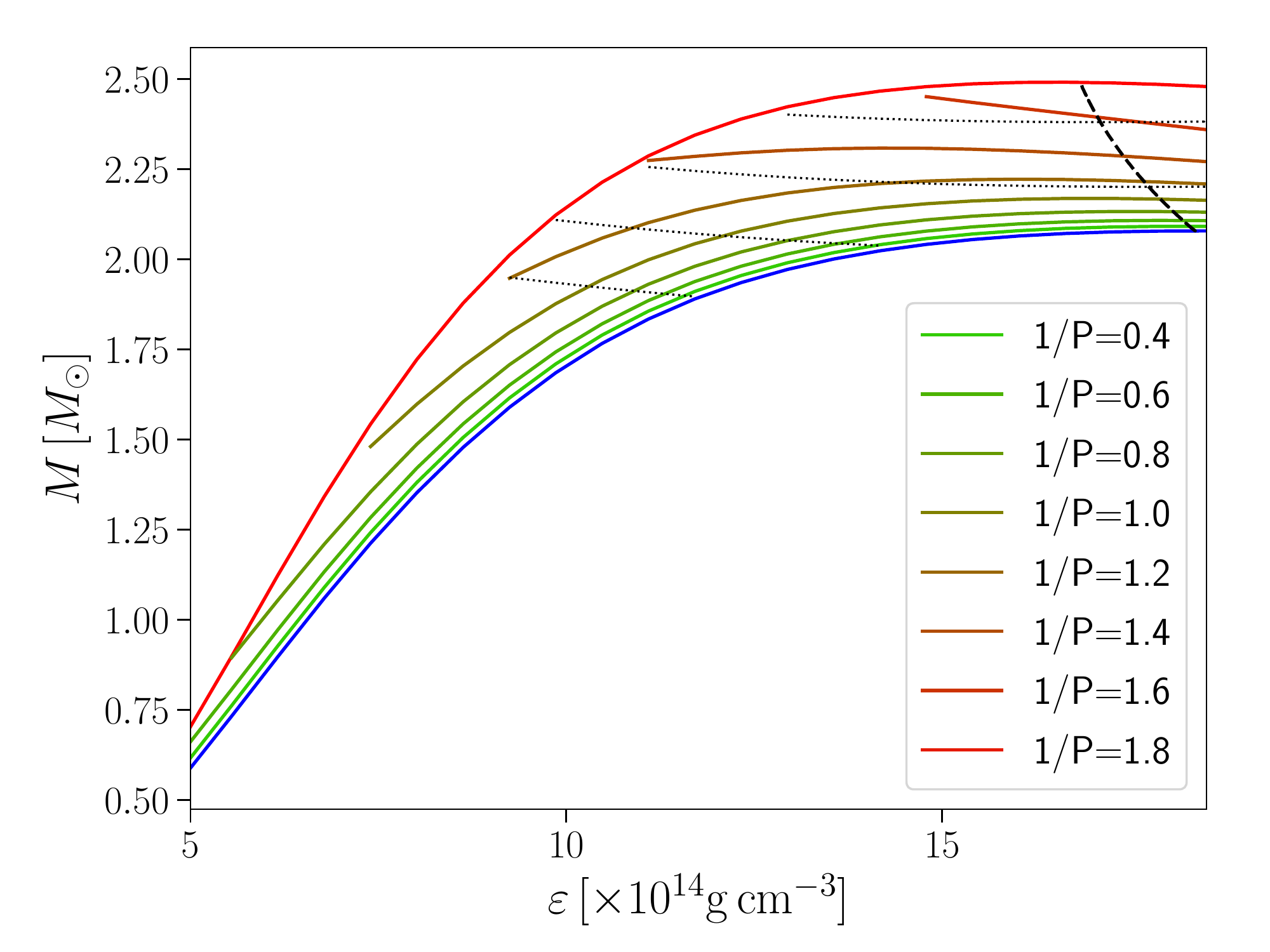}}
\caption{NSs' masses as a function of the central energy density with the QMF EOS at various fixed rotation frequencies $1/P$=0.4-1.8 kHz.
The lower blue curve is the static case, and the upper red curve corresponds to the Keplerian frequencies at different rotating cases.
The change in the maximum mass $M_{\rm crit}$ with frequency is indicated with a dashed black curve.
In static stars, QMF gives $M_{\rm TOV}= 2.08 M_{\odot}$ at a central density $\rho_c=6.92 \rho_0$ with a corresponding radius $10.5$ km.
At Keplerian frequency $f_{\rm K} = 1699$ Hz, the maximum mass and corresponding radius with QMF are $2.50 M_{\odot}$ and $14.0$ km at a central density $\rho_c=8.21 \rho_0$.
Curves with a fixed baryonic mass of $M_{\rm b} = 2.2, 2.4, 2.6, 2.8 M_{\odot}$ are also shown with nearly horizontal gray curves. }\label{fig:rns}}
\vspace{-0.3cm}
\end{figure}

Generally, rotation increases both the gravitational mass and the radius.
Based on the EOSs collected in Table~\ref{tab:sat}, rotation can increase the star's gravitational mass up to $\sim18-19\%$, and the star can be as massive as $\sim2.61 M_{\odot}$ in the APR case. Additionally, the star becomes flattened, and the corresponding circumferential radius is increased up to $\sim3-4$ km, i.e., $\sim29-36\%$. For lighter stars such as 1.4 $M_{\odot}$, the radius increase is more pronounced, reaching $\sim5-6$ km, i.e., $\sim41-43\%$. 
Additionally, rotation lowers the central density from $\sim7-10 \rho_0$ to $\sim6-9 \rho_0$, which is due to the effect of the centrifugal force, effectively stiffening the EOS.
We show in Fig.~\ref{fig:rns} the gravitational mass as a function of the central density at various fixed rotation frequencies based on the QMF EOS. In static stars, QMF18 gives $M_{\rm TOV}= 2.08 M_{\odot}$ at a central density $\rho_c=6.92 \rho_0$ with a corresponding radius of $10.5$ km.
At Keplerian frequency $f_{\rm K} = 1699$ Hz and with QMF18, the maximum mass is $2.50 M_{\odot}$ with a corresponding radius of $14.0$ km at a central density of $\rho_c=8.21 \rho_0$.
\begin{figure*}
\vspace{0.3cm}
{\centering
\resizebox*{0.48\textwidth}{0.28\textheight}
{\includegraphics{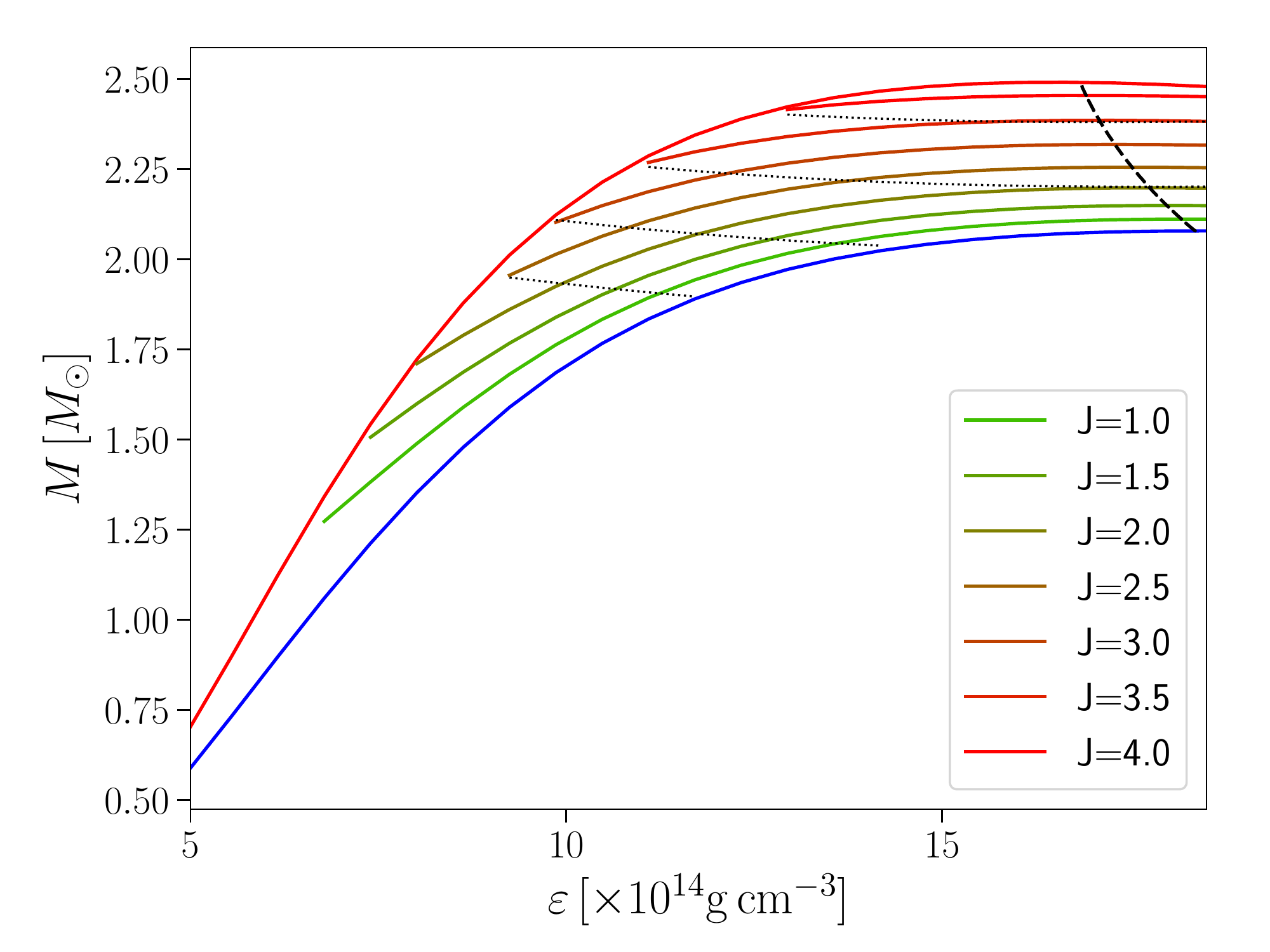}}}
{\centering
\resizebox*{0.48\textwidth}{0.3\textheight}
{\includegraphics{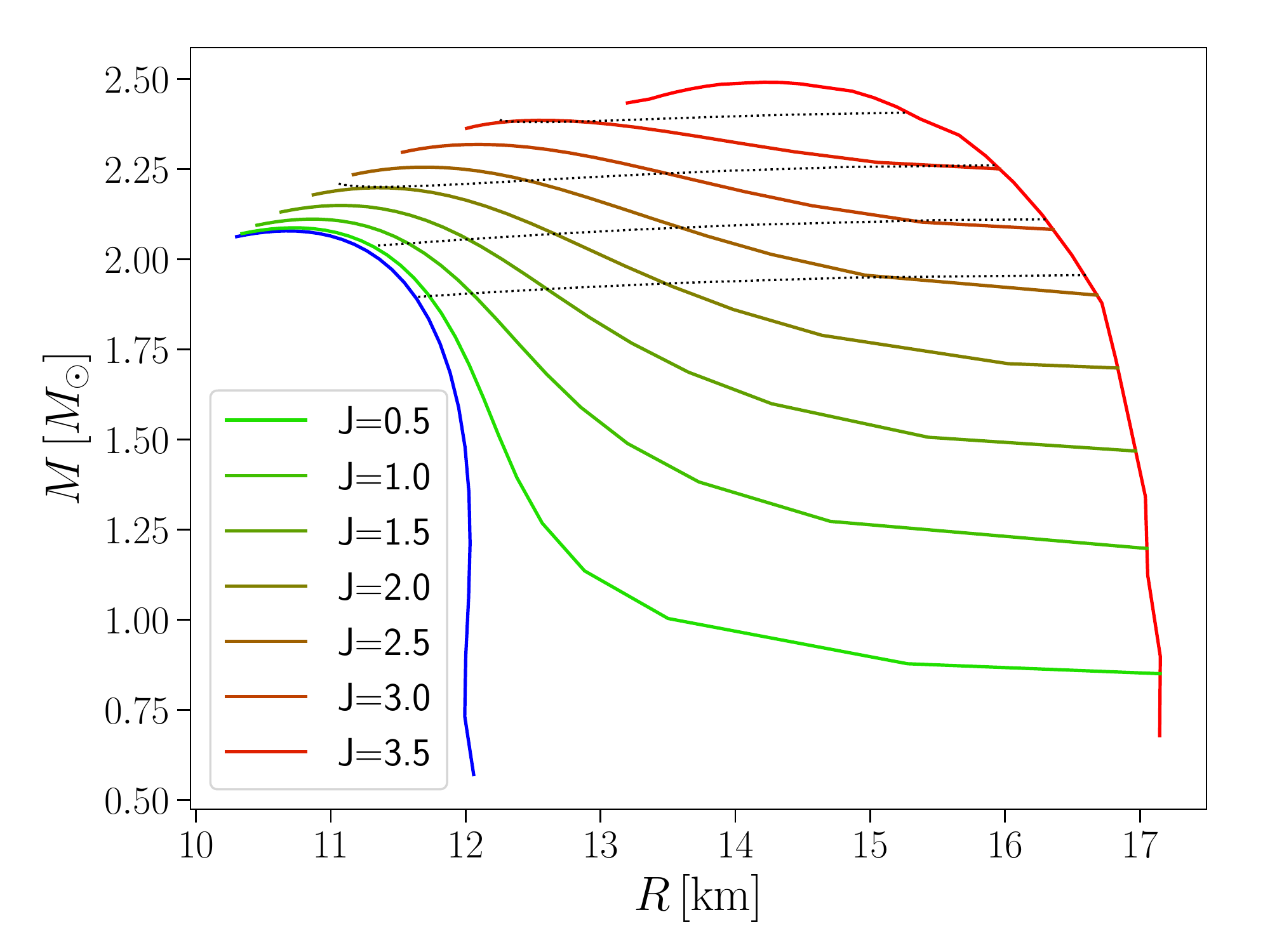}}}
\caption{ (Left) NS masses as a function of the central energy density and (right) mass-radius relations with the QMF EOS at various fixed angular momenta $J$.
The lower blue curve is the static case, and the upper red curve corresponds to the Keplerian frequencies in different rotating cases.
In the left panel, the change in the maximum mass $M_{\rm crit}$ with the angular momentum is indicated with a dashed black curve.
A spin-down star, losing angular momentum over its evolution, follows the lines with fixed baryonic mass $M_{\rm b}$, shown by the nearly horizontal gray curves for $M_{\rm b} = 2.2, 2.4, 2.6, 2.8 M_{\odot}$.
}\label{fig:rnsmr}
\vspace{-0.3cm}
\end{figure*}

One of the most interesting rotating stars is the so-called ``supramassive'' star, which exists only by virtue of rotation.
It is well known that the onset of the instability of the static sequence is determined by the condition $dM/d \rho_c=0$, i.e., the curve should stop at the maximum value of gravitational mass $M$.
In the rotating case, the above criteria have to be generalized, i.e., a stellar configuration is stable if its mass $M$ increases with increasing central density for a fixed angular momentum $J$. Therefore, the onset of the instability, which is called the secular axisymmetric instability, is expressed by
\begin{eqnarray}
\left(\frac{\partial M}{\partial \rho_c} \right)_{J}=0.
\end{eqnarray}
Since rotation increases the mass $M$ that a star of a given central density can support, the static configuration with the baryon mass $M_{\rm b} > M_{\rm b}^{\rm TOV}$ (the baryon mass of a TOV mass star) does not exist.
Such sequences are supramassive stars that are doomed to collapse as they lose energy and angular momentum during their spin-down, following the nearly horizontal line of fixed baryonic mass $M_{\rm b}$.

We show in Fig.~\ref{fig:rnsmr} the NS mass as a function of the central energy density as well as the mass-radius relations with the QMF EOS at various fixed angular momenta $J$.
The lower blue curve is the static case, and the upper red curve corresponds to the Keplerian frequencies in different rotating cases.
In the left panel, the change in the maximum mass $M_{\rm crit}$ with the angular momentum is indicated with a dashed black curve.
There may be a universal relation between $M_{\rm crit}/M_{\rm TOV}$ and $j/j_{\rm K}$~\citep[e.g.,][]{2016MNRAS.459..646B,2013MNRAS.433.1903U,2013Sci...341..365Y,2013PhRvD..88b3009Y,2014MNRAS.438L..71H,2014PhRvL.112l1101P} that does not depend on the specific choice of EOS or the $f_{\rm K}$ value,
\begin{eqnarray}
\frac{M_{\rm crit}}{M_{\rm TOV}} = 1 + a_2 \left( \frac{j}{j_{\rm K}}\right)^2 +a_4 \left( \frac{j}{j_{\rm K}}\right)^4
\end{eqnarray}
where $j=J/M^2$ is the dimensionless angular momentum
and the coefficients are $a_2 = 1.316\times 10^{-1}$ and $a_4 = 7.111\times 10^{-2}$~\citep{2016MNRAS.459..646B}.

Note that the above discussions focus only on the case of rigid rotation, while differential rotation can be much more efficient in increasing the maximum allowed mass.
In differentially rotating stars, the high-density inner part may rotate faster than the low-density outer part, so the inner part can be supported by rapid rotation without the equator having to exceed the Keplerian limit~\citep[e.g.,][]{2019PhRvD.100d3015Z}.
While both rigid and differential rotation follow axisymmetry, there are cases when a rotating NS breaks its axial symmetry if the rotational kinetic energy to gravitational binding energy
ratio, $T/|W|$, exceeds a critical value.
The abovementioned $r$-mode instability could also trigger NSs' motion with off-axis symmetry.
It is presently unclear whether such configurations of NSs can actually be realized in practice~\citep[e.g.,][]{2018PhRvD..97b3013Z}.
Overall, it is especially important to calculate models of rotating stars to better understand the observations of binary merger events (see details in Sec.~\ref{sec:binaryremnant}).

\section{EOS with exotic particles} \label{sec:exotic}

\subsection{Hyperon star and hyperon puzzle}

While around the saturation densities $\rho=\rho_0$, the matter inside an NS consists only of nucleons and leptons, at higher densities, several other species of particles may appear due to the fast increase in the baryon chemical potentials with density~\citep{2005PrPNP..54..193W,2016RvMP...88c5004G,2020arXiv200209223T}, just because their appearance is able to lower the ground state energy of the dense nuclear matter phase.
Among these new particles are strange baryons, namely, the $\Lambda, \Sigma^{0,\pm}, \Xi^{0,-}$ hyperons.
Other species (such as kaons and Delta isobars) might also appear in stellar matter, which the present paper does not cover.
Generally, the presence of one species of strange particle is found to push the onset of other species of strange particles to higher densities, even out of the physically
relevant density regime~\citep[e.g.,][]{2020PhLB..80235266M,2007ChPhy..16.1934L}.

It is necessary to generalize the QMF study of the nuclear EOS with the inclusion of hyperons~\citep[e.g.,][]{2002NuPhA.707..469S,2014PTEP.2014a3D02H,2014PhRvC..89b5802H,2016PhRvC..94d4308X,2017PhRvC..95e4310X,2017PhRvC..96e4304H}.
The density thresholds of hyperons are essentially determined by the masses and their interaction. The mass of $\Lambda (uds)$ is 1116 MeV. The masses of $\Sigma^+ (uus)$, $\Sigma^0 (uds)$, $\Sigma^- (dds)$ are 1189, 1193, and 1197 MeV, respectively. The masses of $\Xi^0 (uss)$ and $\Xi^- (dss)$ are 1315 and 1321 MeV, respectively.
From hypernuclei experiments in the laboratory~\citep{2015RPPh...78i6301F}, we know that $\Lambda$-nucleus and $\Lambda\Lambda$ interactions are attractive, while $\Sigma$-nucleus interactions are repulsive.
Additionally, the nature of the $\Xi$-nucleus interaction has been suggested to be attractive~\citep{2000PhRvC..61e4603K}.
Theoretically, any effective many-body theories should respect the available hypernuclei data before proceeding with other sophisticated studies~\citep{2016RvMP...88c5004G,2020arXiv200209223T}.
The adopted hyperon-meson couplings need to at least reproduce unambiguous hypernuclear data.

\begin{figure*}
\vspace{-0.3cm}
{\centering
\resizebox*{0.98\textwidth}{0.55\textheight}
{\includegraphics{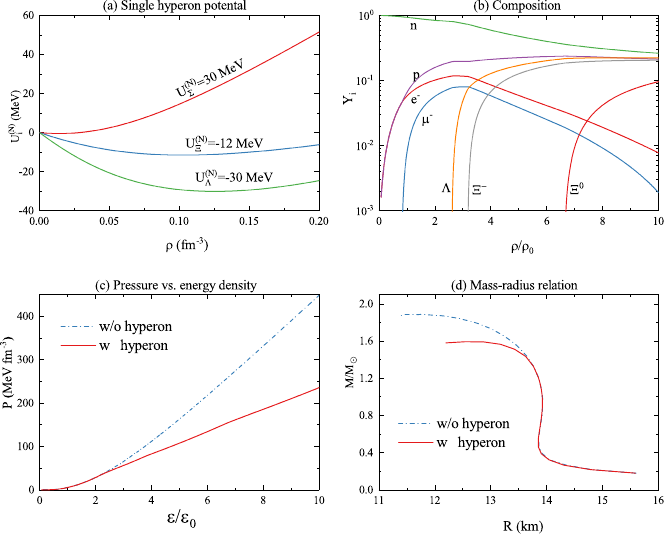}}
\caption{(a) Single hyperon potential, (b) fractions of leptons and baryons, (c) EOS, and (d) mass-radius relations for NSs with hyperons within QMF. The cases without hyperons in the star's core are also shown in the lower panels.
The maximum mass of QMF EOS without hyperons is slightly lower than $2M_{\odot}$ due to the absence of a high-order vector coupling term for effective nuclear interaction from an earlier work~\citep{2014PhRvC..89b5802H}.
When hyperons are included, the mass is largely reduced and well below the observational 2-solar-mass limit.
The hyperon puzzle is also present in many microscopic studies based on developed realistic baryon-baryon interactions~\citep[e.g.,][]{2015PhRvL.114i2301L,2011PhRvC..83b5804B,2016EPJA...52...21R,2017hspp.confj1002B}.
Adapted from~\citet{2014PhRvC..89b5802H}.} \label{fig:y}}
\vspace{-0.3cm}
\end{figure*}

At the mean-field level, the single $\Lambda$, $\Sigma$, $\Xi$ potential well depths in symmetric nuclear matter are $U^{(N)}_{\Lambda,\Sigma,\Xi} \sim-$30, 30, $-$14 MeV at the saturation density, respectively.
In the extended QMF model~\citep{2014PTEP.2014a3D02H,2014PhRvC..89b5802H}, we introduce different confining strengths for the $s$ quarks and the $u,d$ quarks in the corresponding Dirac equations (under the influence of the meson mean fields).
The confining strength of the $u,d$ quarks is constrained by finite nuclei properties, and that of the $s$ quarks is constrained by the well-established empirical value of $U^{(N)}_{\Lambda} \sim-$30 MeV.
The mass difference among baryons is generated by taking into account the spin correlations $E_B^* = \sum_ie_i^* + E^B_{\rm spin}$, and the spin correlations of the baryons are fixed by fitting the baryon masses in free space.
In addition, the spurious c.m. motion is removed through the usual square root method as
$M_B^* = \sqrt{E_B^{*2} - \langle p^2_{\rm cm} \rangle}$.
The contribution of the $\sigma$ meson is contained in the effective mass $M_{\Lambda,\Sigma,\Xi}^*$, and the $\omega$ and $\rho$ mesons couple to the baryons with the following coupling constants:
\begin{eqnarray}
&&g_{\omega N}=3g_\omega^q,\quad g_{\omega \Lambda}=cg_{\omega\Sigma}=2g_\omega^q,\quad
g_{\omega \Xi}=g_\omega^q \\
 & & g_{\rho N}=g_\rho^q,\quad g_{\rho \Lambda}=0,\quad g_{\rho \Sigma}=2g_\rho^q,\quad
 g_{\rho \Xi}=g_\rho^q
\end{eqnarray}
where a factor $c$ is introduced before $g_{\omega\Sigma}$ for a large $\Sigma\omega$ coupling, in contrast with the $\Lambda-\omega$ coupling, to simulate the additional repulsion on the $\Sigma-$nucleon channel, and $U_{\Sigma}^{(N)}$ = 30 MeV at the nuclear saturation density.
The basic parameters are the quark-meson couplings ($g^q_\sigma$, $g_\omega^q$, and $g_\rho^q$), the nonlinear self-coupling constants ($g_3$ and $c_3$), and the mass of the $\sigma$ meson ($m_\sigma$)~\citep[for more detail regarding the formalism and the parameters, see ] []{2000PhRvC..61d5205S,2014PhRvC..89b5802H}.
With such a parameter set, the saturation properties of nuclear matter can be described~\citep{2014PhRvC..89b5802H}.
The values of a single $\Xi$ hyperon in nuclear matter are obtained as $U^{(N)}_{\Xi}$ = $-12$ MeV, consistent with the BNL-E885 experiments~\citep{2000PhRvC..61e4603K}.
The density dependences of the single hyperon ($\Lambda,\Sigma,\Xi$) potentials are shown in Fig.~\ref{fig:y}(a).

Regarding the EOS of hyperonic matter, the baryon contributions can be obtained through the mean-field ansatz from the Lagrangian (including hyperons)~\citep{2000PhRvC..61d5205S,2014PhRvC..89b5802H}.
Electrons are again treated as a free ultrarelativistic gas, whereas the muons are relativistic, as in Eq. (22).
The total EOS can be calculated for a given composition of baryon components. This allows the determination of the chemical potentials of all species, which are the fundamental input for the equations of chemical equilibrium:
\begin{eqnarray}
&&\mu_n = \mu_{\Lambda} = \mu_{\Sigma^0} = \mu_{\Xi^0}  \\
&&\mu_e = \mu_{\mu} \\
&&\mu_n - \mu_e = \mu_p = \mu_{\Sigma^+} \\
&&\mu_n + \mu_e =\mu_{\Sigma^-} = \mu_{\Xi^-}
\end{eqnarray}
The above equations must be supplemented with two other conditions, i.e., charge neutrality and baryon number conservation. These are
\begin{eqnarray}
&&\rho_p + \rho_{\Sigma^+}  = \rho_e + \rho_{\mu} +\rho_{\Sigma^-} + \rho_{\Xi^-}, \\
&&\rho = \rho_n +\rho_p + \rho_{\Lambda} + \rho_{\Sigma} + \rho_{\Sigma^+} + \rho_{\Sigma^-} + \rho_{\Xi^-} + \rho_{\Xi^0} .
\end{eqnarray}
Finally, the actual detailed fraction $Y_i=\rho_i(\rho)$ of the dense matter is determined for each fixed baryon density $\rho$, as shown in Fig.~\ref{fig:y} (b).
In the low-density region (until $\rho <0.21$ fm$^{-3}$), the proton fraction $\rho_p/\rho$ is well below $1/9$, which fulfills the astrophysical observation that direct URCA cooling might not occur at too of low densities~\citep{2001PhR...354....1Y}.
With the properly chosen $\Lambda, \Sigma$ and $\Xi$ hyperon potentials, $\Lambda$ is the first hyperon appearing at $\sim2-3\rho_0$. Then, $\Xi^-$ hyperons appear at $\sim3\rho_0$ followed by $\Xi^0$ hyperons at $\sim7\rho_0$.
The fractions of hyperons increase with density. At densities above $\sim10\rho_0$, the fractions of $\Lambda$ and $\Xi^-$ are almost the same as the fractions of protons and neutrons. $\Sigma^{-}$, however, does not appear until very high density of $2.0$ fm$^{-3}$.

In Fig.~\ref{fig:y} (c), we show the pressure of beta-equilibrated matter as a function of the energy density. The solid curve represents the EOS including the hyperons, and the dot-dashed curve is the EOS without hyperons. The EOS becomes softer with the presence of strangeness freedom.
The NS properties are calculated by using the EOSs with/without hyperons obtained from the QMF model, and the NS mass-radius relations are plotted in Fig.~\ref{fig:y} (d).
It is found that the maximum mass of the NSs including hyperons is approximately $1.6~M_{\odot}$,
much lower than that without hyperons, which is below the observational limit.
This is the so-called hyperon puzzle, which is also found
in many microscopic studies based on developed realistic baryon-baryon interactions~\citep[e.g.,][]{2015PhRvL.114i2301L,2011PhRvC..83b5804B,2016EPJA...52...21R,2017hspp.confj1002B}.

Since hyperons are not present in nuclear matter, they cannot be determined from the nuclear matter properties. Moreover, the analysis of experimental data on hypernuclei shows that we cannot fix these parameters in a unique way.
How can a sufficiently stiff high-density EOS generate a heavy hyperon star with properly reproduced nuclear matter properties at the saturation density?
There may be three approaches forward:
\begin{enumerate}
\item \relax Three-body hyperon interactions can be introduced in microscopic studies or high-order meson fields in effective calculations. If they are repulsive, a stiff enough hyperon star EOS can be obtained by increasing the repulsion as the density increases, and there is a possibility of massive hyperon stars with a central density $>\sim5\rho_0$, \citep[e.g.,][]{2007NuPhA.792..341R,2014PhRvC..90d5805Y,2017EPJA...53..121H,2019EPJA...55..207L}. This is a natural solution based on the known importance of three-body nucleon forces in nuclear physics;
\item \relax Larger maximum masses can be produced through a transition to another phase of dense (quark) matter in the stellar core at high densities~\citep[e.g.,][]{2009ChPhC..33...61L,2015PhRvC..91c5803L,2008PhRvC..77f5807P}. This approach is presented in Sec.~\ref{sec:hybrid};
\item \relax A separate branch of pulsar-like objects can be introduced to account for the heavy ones, for example, QSs made of free quarks~\citep[e.g.,][]{2010MNRAS.402.2715L,2011RAA....11..482L,1986A&A...160..121H}. Unlike NSs, which are bound by gravity, QSs are bound by strong interactions; therefore, they have opposite M-R dependence. This is the so-called two-branch scenario~\citep[e.g.,][]{2017ApJ...846..163W,2018ApJ...852L..32D}, which is discussed in Sec.~\ref{sec:2branch}.
\end{enumerate}

\subsection{\label{sec:hybrid} Strange quark matter and hybrid stars}

The matter inside the NS core possesses densities ranging from a few times $\rho_0$ to one order of magnitude higher. At such densities, the hadronic matter might undergo a phase transition to quark matter, and a hybrid NS with a quark matter or mixed core can be formed.
However, the exact value of the transition density to quark matter is unknown and still a matter of recent debate not only in astrophysics but also within the theory of high energy heavy-ion collisions.
Additionally, it is not obvious whether the information on the nuclear EOS from high energy heavy-ion collisions can be related to the physics of NS interiors.
The possible quark-gluon plasma produced in heavy-ion collisions is expected to be characterized by low baryon density and high temperature, while the possible quark phase in NSs appears at high baryon density and low temperature.
Nevertheless, we must be careful that the transition cannot occur at too low of density (below the nuclear saturation density $\rho_0$) to maintain consistency with the current experimental data of heavy-ion collisions.

The possibility of the existence of strange quark matter in NS high-density cores is of special interest in the present era of GW astronomy~\citep[e.g.,][]{2019PhRvD.100j3022H,2019PhRvD..99h3014H,2019JPhG...46k4001A,2019MNRAS.484.4980A,2019PhRvL.122f1102B,2019PhRvL.122f1101M,2019ApJ...881...73W,2019arXiv190600826X,2019ApJ...877..139G,2020PhRvD.101d4019C,2020arXiv200300972F,2020arXiv200301259T,2020arXiv200207538N}.
Presently, we have no unified models to address the hadron phase and the quark phase, and it is still not clear whether the change in the hadron phase corresponding to that in the quark phase is a crossover or a first-order transition.
Here, we analyze a specific example in the context of a first-order transition (at transition pressure $\ptrans$) to express the experimental constraints in model-independent terms.
For the hadronic sector, we use the above QMF model. For the high-density quark phase, we utilize the CSS parametrization~\citep{2013PhRvD..88h3013A}, exploiting the fact that for a considerable class of microscopic quark matter models, the speed of sound is weakly density-dependent, e.g.,~\citep{2013A&A...551A..61Z,2015PhRvD..92h3002A,2016PhRvC..93d5812R}.
The present scheme can only discuss the transition that occurs at a sharp interface (Maxwell construction) between bulk hadronic matter and quark matter, i.e., the quark-hadron surface tension is high enough to disfavor mixed phases (Gibbs construction).
It has been shown that a strong first-order phase transition with a sharp interface is the most promising scenario to be tested or distinguished from pure hadronic matter by future observations~\citep{2019PhRvD.100j3022H,2019PhRvD..99h3014H,2020PhRvD.101d4019C}.
We tend to find that the observation of a two-solar-mass star and the accurate measurement of the typical NS radius constrain the CSS parameters, including the squared speed of sound in the high-density phase $c_{\rm QM}^2$, the hadron-quark phase transition density $\rho_{\rm trans}$, and the discontinuity in the energy density at the transition $\Delta\ep/\ep_{\rm trans}$, where $\ntrans \equiv \rho_{\rm NM}(\ptrans),~\etrans \equiv \ep_{\rm NM}(\ptrans)$.

\begin{figure*}
{\centering
\resizebox*{0.48\textwidth}{0.25\textheight}
{\includegraphics{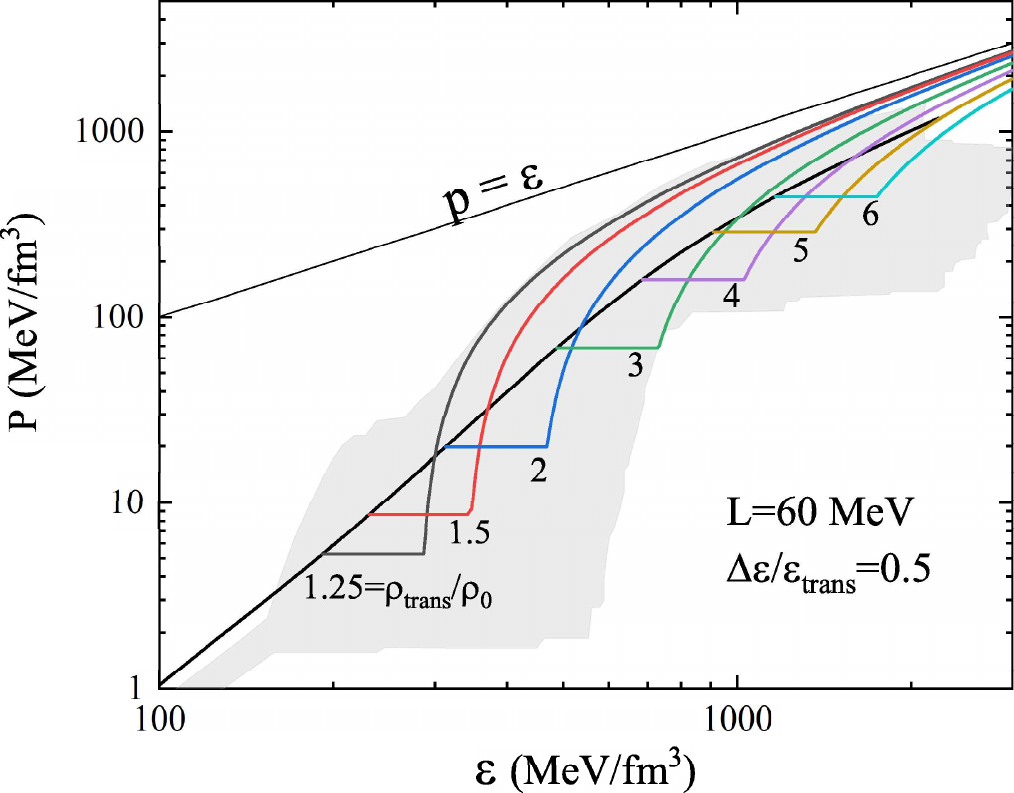}}}
{\centering
\resizebox*{0.48\textwidth}{0.25\textheight}
{\includegraphics{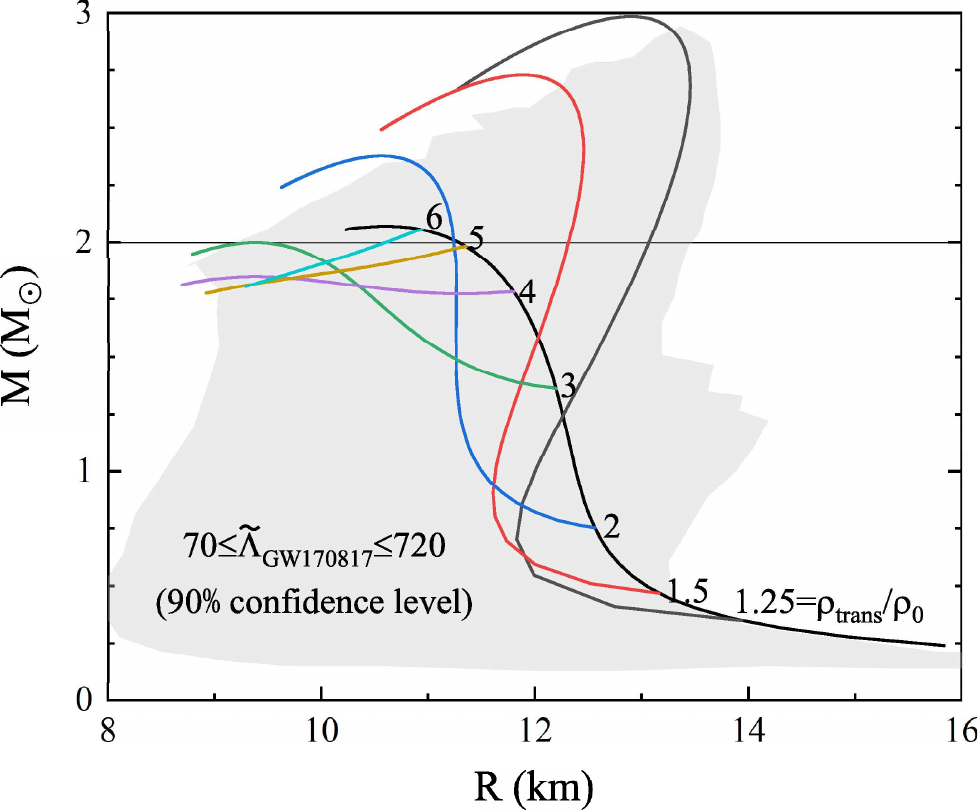}}}
\caption{\label{fig:L15} EOSs (left) and mass-radius relations (right) for hybrid stars (colorful curves) at a fixed discontinuity in energy density at the transition $\Delta\varepsilon/\varepsilon_{\rm trans}=0.5$ for different transition densities $\rho_{\rm trans}/\rho_0=1.25, 1.5, 2, 3, 4, 5, 6$, with the symmetry energy slope $60~\rm MeV$. The squared speed of sound is fixed at $c^2_{\rm QM}=1$ in the quark matter. The corresponding NS results within the QMF are shown for comparison (black curves).
The shaded region is the favored region from the maximal model~\citep{2019EPJA...55...97T}, with the underlying EOSs constrained at low densities from EFT, facilitating the complete allowed parameter space for the speed of sound above the saturation density $n_0$ and enforcing the LIGO/Virgo constraint from GW170817 ($70\leq~\tilde{\Lambda}\leq720$)~\citep{2019PhRvX...9a1001A}. }\label{fig:hybrideos}
\vspace{-0.3cm}
\end{figure*}

For a given nuclear matter EOS $\ep_{\rm NM}(P)$, the full CSS EOS is
\begin{equation}
\ep(P) = \left\{\!
\begin{array}{ll}
\ep_{\rm NM}(P) & P<\ptrans \\
\ep_{\rm NM}(\ptrans)+\Delta\ep+c_{\rm QM}^{-2} (P-\ptrans) & P>\ptrans
\end{array}
\right.\
 \nonumber \\
\label{eq:CSS_EoS}
\end{equation}

We perform the calculation by varying $\cQMsq$ from the causality limit ($\cQMsq=1$) to the conformal limit ($\cQMsq=1/3$, the value for systems with conformal symmetry that may be applicable to relativistic quarks).
It is worth mentioning that perturbative QCD calculations exhibit quark matter with $c_{\rm QM}^2$ values of approximately 0.2 to 0.3~\citep{2010PhRvD..81j5021K}.
We use units where $\hbar=c=1$.
In Fig~\ref{fig:hybrideos}, we show representative EOSs $P(\ep)$ for dense matter with a sharp first-order phase transition and the corresponding mass-radius relations. The two-solar-mass lower limit for maximum gravitational mass is explicitly indicated in the mass vs. radius plot. We include the curves with increasing transition density from $1.25\rho_0$ to $6\rho_0$ at a fixed energy density discontinuity and speed of sound in quark matter, and the nuclear matter EOS is chosen to be the QMF model result with the symmetry energy slope $L=60$ MeV. We mention here that $L\sim30-60$ MeV is the preferred range within QMF as indicated by the ab initio calculations (shown in Fig.~\ref{fig:mr}).
We see that a lower transition density (pressure), therefore a stiffer EOS, leads to a heavier hybrid star. The smallest hybrid star is typically the heaviest.

\begin{figure*}
{\centering
\resizebox*{0.9\textwidth}{0.4\textheight}
{\includegraphics{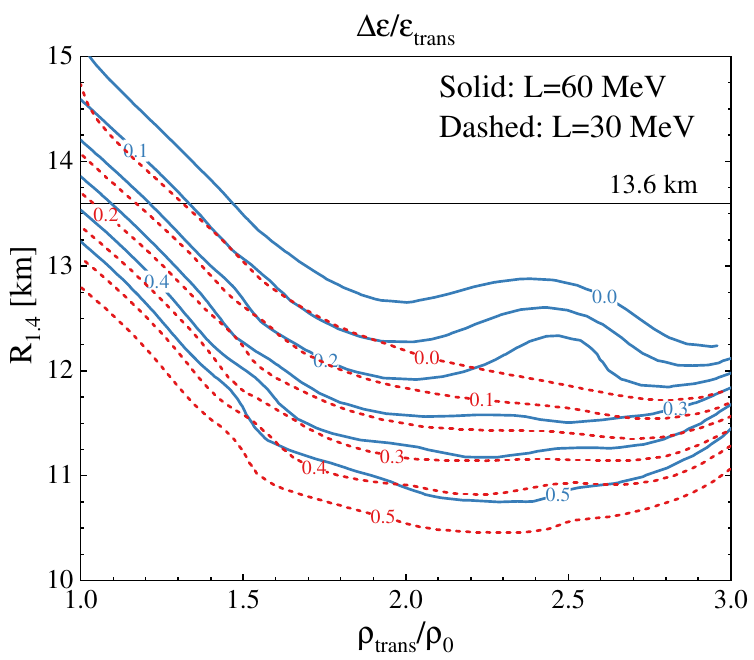}}}
\vskip+4mm\
{\centering
\resizebox*{0.9\textwidth}{0.4\textheight}
{\includegraphics{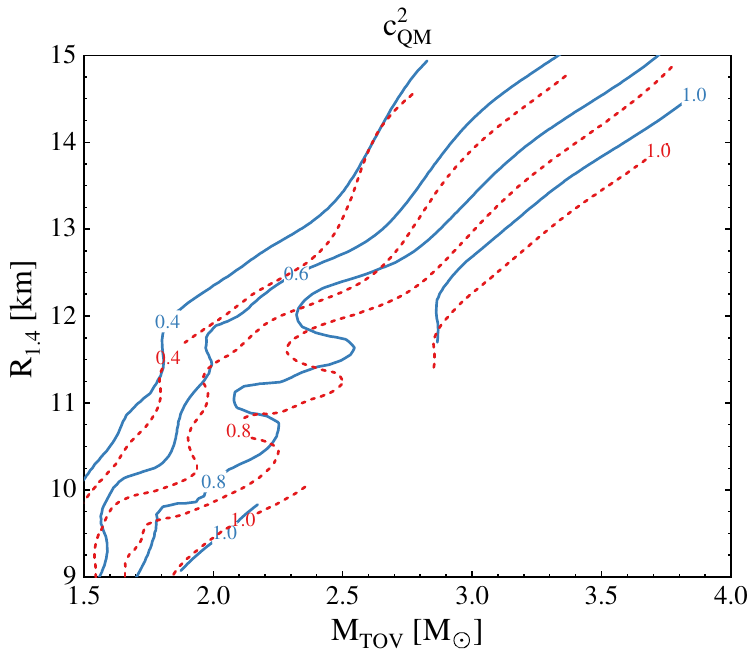}}}
\caption{Upper: Radius of a $1.4 M_{\odot}$ star vs. the transition density, with the energy density discontinuity explicitly indicated; Lower: Radius of a $1.4 M_{\odot}$ star vs. the maximum mass, with the sound speed explicitly indicated.
The results are shown for two kinds of symmetry energy slopes $60~\rm MeV$ (solid lines) and $L=30$ MeV (dashed lines). The horizontal line represents a conservative upper limit of $13.6$ km for a $1.4 M_{\odot}$ star with or without a phase transition~\citep[e.g.,][]{2017ApJ...850L..34B,2018PhRvL.120q2703A,2018PhRvL.120z1103M,2016PhRvC..94c5804F,2018PhRvL.121p1101A,2019PhRvD..99j3009M,2018ApJ...860..139B,2018PhRvL.121i1102D}. The vertical line in the lower panel represents the two-solar-mass constraint. There are cases when no $1.4 M_{\odot}$ star is possible, shown by breaks in the curves. Taken from Miao et al.~\citep{Miao2020}. }\label{fig:R14}
\end{figure*}

Systematically, we carry out calculations for the mass-radius of hybrid stars spanning the whole parameter space of the speed of sound in a domain with a transition density up to $\rho_{\rm trans}=6\rho_0$ and an energy density discontinuity up to $\Delta\varepsilon=1.5\,\varepsilon_{\rm trans}$. The calculations are performed using two values for the symmetry energy slope parameter of $L=30$ MeV and $L=60$ MeV, and the results are shown in Fig.~\ref{fig:R14}. 

Figure~\ref{fig:R14} displays the correlation of the radius of a $1.4 \,M_{\odot}$ hybrid star $R_{1.4}$ with the transition density $\rho_{\rm trans}/\rho_0$ (upper panels) and the maximum mass $M_{\rm TOV}$ (lower panels). In general, there exists an anticorrelation between $R_{\rm 1.4}$ and $\rho_{\rm trans}/\rho_0$ and a correlation between $R_{\rm 1.4}$ and $M_{\rm TOV}$.
A conservative upper limit of $13.6$ km for $R_{\rm 1.4}$ can be obtained with different analyses~\citep[e.g.,][]{2017ApJ...850L..34B,2018PhRvL.120q2703A,2018PhRvL.120z1103M,2016PhRvC..94c5804F,2018PhRvL.121p1101A,2019PhRvD..99j3009M,2018ApJ...860..139B,2018PhRvL.121i1102D}.
In the upper panel, the upper limit of $13.6$ km for $R_{\rm 1.4}$ corresponds to $\approx1.3\rho_0$ for $L=30$ MeV and $\approx1.5\rho_0$ for $L=60$ MeV. The possible onset of a first-order phase transition below such densities might be strongly disfavored.
From the mass measurement of heavy pulsars, we can put lower limits on $R_{\rm 1.4}$ by making use of the $R_{\rm 1.4}-M_{\rm TOV}$ correlation in the lower panel.
The two-solar-mass constraint leads to a lower limit of $\approx 9.6$ km, in good concurrence with other analyses in the literature based on X-ray observations or LIGO/Virgo measurements~\citep[e.g.,][]{2017ApJ...850L..34B, 2018PhRvL.120z1103M,2018PhRvL.121p1101A,2019PhRvD..99j3009M}.
An upper limit on the maximum mass can also be indicated from $R_{\rm 1.4}<13.6$ km of $ M_{\rm TOV} < 3.6 M_{\odot}$.

We conclude this section by further discussing the following aspects:
\begin{itemize}
\item \relax $Submillisecond~rotation$: It is commonly believed that only self-bound stable QSs may rotate rapidly with a submillisecond period~\citep{1990MPLA....5.2197G}. However, it is suggested that pulsars rotating with approximately half a millisecond period could also be interpreted as hybrid stars~\citep{2003A&A...408..675B}, with NSs containing a metastable deconfined quark phase at their centers. This conclusion does not depend on the quark matter EOS models. Therefore, very rapidly rotating pulsars may be interpreted as either QSs or NSs with deconfined quark matter interiors.
\item \relax $Mixed~phase$: In the present work, we adopt the simple Maxwell construction. The Gibbs construction provides a realistic model of the phase transition between the hadronic and quark phases inside the star~\citep{1992PhRvD..46.1274G}, yielding a range of baryon densities where both phases coexist, which provides an EOS containing a pure hadronic phase, a mixed phase, and a pure QM region~\citep[e.g.,][]{2015PhRvC..91c5803L,2007PhRvD..76l3015M}. The pressure is the same in the two phases to ensure mechanical stability, while the chemical potentials of the different species are related to each other, satisfying chemical and beta stability. Both the hadron and quark phases are separately charged while preserving total charge neutrality~\citep{1992PhRvD..46.1274G}. As a consequence, the pressure is a monotonically increasing function of the density. The realization of the mixed phase depends on the nuclear surface tension, which is currently an unknown parameter~\citep{2019PhRvD..99j3017X}. The Gibbs treatment is the zero surface tension limit of the calculations, including finite-size effects. It was demonstrated that the influence of different constructs on the maximum mass value is rather small~\citep{2007PhRvD..76l3015M}.
\end{itemize}

\subsection{\label{sec:2branch} Quark stars and two-branch scenario}

We now turn to the description of the bulk properties of uniform quark matter.
The strange quark matter is composed of up ($u$), down ($d$) and strange ($s$) quarks with charge neutrality maintained by the inclusion of electrons (hereafter muons as well if present):
\begin{equation}
\frac{2}{3}\rho_u-\frac{1}{3}\rho_d-\frac{1}{3}\rho_s-\rho_e=0,  \label{eq:Chargeneut}
\end{equation}
The baryon number conservation,
\begin{equation}
\frac{1}{3}\left(\rho_u + \rho_d + \rho_s\right) =\rho, \label{eq:Baryonconserv}
\end{equation}
is also satisfied with $n$ being the baryon number density.
Due to the weak interactions between quarks and leptons,
\begin{eqnarray}
&& d \rightarrow u + e + \tilde{\nu}_e\ ,~u + e \rightarrow d + \nu_e;\nonumber \\
&& s \rightarrow u + e + \tilde{\nu}_e\ ,~u + e \rightarrow s + \nu_e;\nonumber \\
&& s + u \leftrightarrow d + u\ , \nonumber
\end{eqnarray}
The $\beta$-stable conditions $\mu_s = \mu_d = \mu_u + \mu_e $ should be fulfilled in neutrino-free matter.
The energy density and pressure include both contributions from quarks and leptons, and those of leptons can be easily calculated by the model of an ideal Fermi gas such as in the NS matter case.

In the density regime achieved inside compact stars, the dense matter properties cannot be calculated directly from the first principle lattice QCD or perturbative QCD. The latter is only applicable at ultrahigh densities beyond the range of compact stars.
The current theoretical description of quark matter is based on phenomenological models~\citep[e.g.,][]{2010MNRAS.402.2715L,2011RAA....11..482L} and burdened with large uncertainties.
In the following, we consider the nonperturbative contributions from perturbative QCD~\citep{2014ApJ...789..127K}.
For simplicity, we use the pQCD thermodynamic potential density to the order of $\alpha_\mathrm{s}$~\citep{2005PhRvD..71j5014F},
\begin{eqnarray}
\Omega^\mathrm{pt} = \Omega_0 + \Omega_1 \alpha_\mathrm{s}\ , \label{eq:omegapt_pQCD}
\end{eqnarray}
with
\begin{equation}
\Omega_1 =\sum_{i=u,d,s} \frac{g_i m_i^4}{12\pi^3}
               \left\{ \left[ 6 \ln\left(\frac{\bar{\Lambda}}{m_i}\right) + 4 \right]\left[u_i v_i - \ln(u_i+v_i)\right]
               + 3\left[u_i v_i - \ln(u_i+v_i)\right]^2 - 2 v_i^4 \right\},
\label{eq:omega1}
\end{equation}
where $u_i \equiv \mu_i/m_i$ and $v_i \equiv \sqrt{u_i^2-1}$. The coupling constant $\alpha_\mathrm{s}$ and quark masses $m_i$ run with the energy scale and can be determined by~\citep{2005PhRvD..71j5014F},
\begin{eqnarray}
\alpha_\mathrm{s}(\bar{\Lambda})
  &=& \frac{1}{\beta_0 L}   \left(1- \frac{\beta_1\ln{L}}{\beta_0^2 L}\right),
\label{eq:alpha} \\
m_i(\bar{\Lambda})
  &=& \hat{m}_i \alpha_\mathrm{s}^{\frac{\gamma_0}{\beta_0}}
      \left[ 1 + \left(\frac{\gamma_1}{\beta_0}-\frac{\beta_1\gamma_0}{\beta_0^2}\right) \alpha_\mathrm{s} \right].
\label{eq:mi}
\end{eqnarray}
Here, $L\equiv \ln\left( \frac{\bar{\Lambda}^2}{\Lambda_{\overline{\mathrm{MS}}}^2}\right)$, and we take the $\overline{\mathrm{MS}}$ renormalization point $\Lambda_{\overline{\mathrm{MS}}} = 376.9$ MeV based on the latest results for the strong coupling constant~\citep{2014ChPhC..38i0001O}. Following Eq.~(\ref{eq:mi}), the invariant quark masses are $\hat{m}_u= 3.8$ MeV, $\hat{m}_d = 8$ MeV, and $\hat{m}_s = 158$ MeV.
The parameters for the $\beta$-function and $\gamma$-function are $\beta_0=\frac{1}{4\pi}(11-\frac{2}{3}N_\mathrm{f})$, $\beta_1=\frac{1}{16 \pi^2} (102-\frac {38}{3} N_\mathrm{f})$,
$\gamma_0=1/\pi$, and $\gamma_1=\frac{1}{16\pi^2} (\frac{202}{3} - \frac{20}{9}N_\mathrm{f})$~\citep{1997PhLB..405..327V} (The formulas are for arbitrary $N_\mathrm{f}$, and in this study, $N_\mathrm{f}=3$). It is not clear how the renormalization scale evolves with the chemical potentials of quarks, and we use $\bar{\Lambda} = \frac{C_1}{3} \sum_i\mu_i$, with $C_1=1-4$~\citep{2014ApJ...789..127K}.

\begin{figure*}
{\centering
\resizebox*{0.95\textwidth}{0.39\textheight}
{\includegraphics{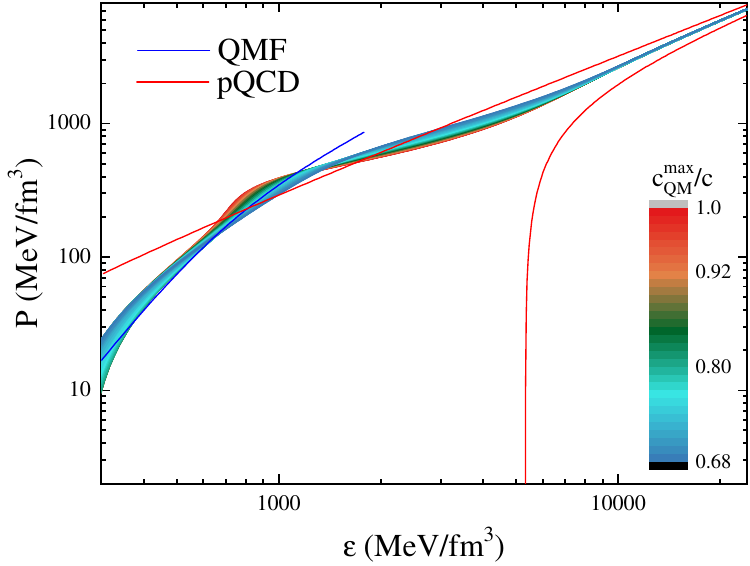}}}
{\centering
\resizebox*{0.95\textwidth}{0.39\textheight}
{\includegraphics{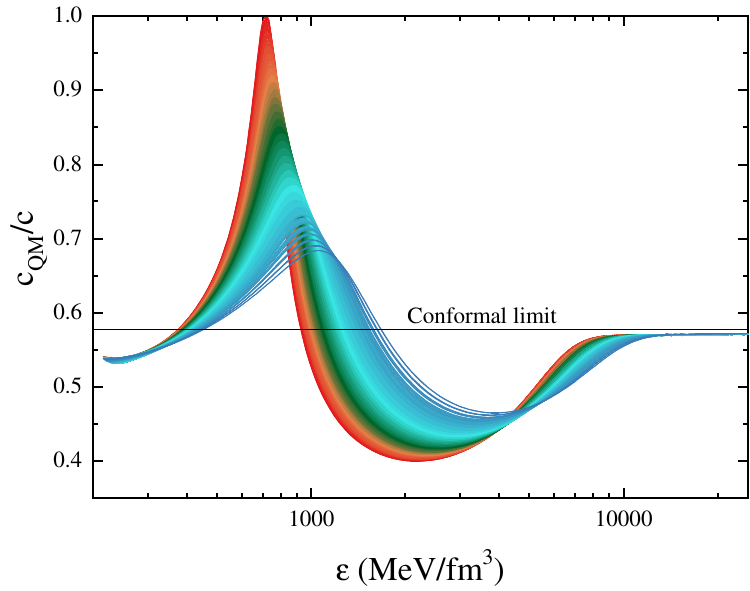}}}
\caption{EOSs (left) and sound speed $c_{\rm QM}$ (right) of SQM. The EOSs are generated in the perturbation model, fulfilling the available astrophysical constraints of mass~\citep{2010Natur.467.1081D,2013Sci...340..448A,2016ApJ...832..167F,2018ApJS..235...37A,2020NatAs...4...72C}, radius~\citep{2019ApJ...887L..24M,2019ApJ...887L..21R} and tidal deformability~\citep{2019PhRvX...9a1001A} for QSs. They are compared with the results of the perturbative QCD (red curve) without nonperturbative corrections.
The EOS of the nuclear matter obtained with the QMF (blue curve) is shown in the left panel.
The maximum sound speeds $c_{\rm QM}^\mathrm{max}$ are explicitly indicated.
The horizontal lines in the $c_{\rm QM}$ plot show the conformal limit.
To ensure a large mass for QSs above two solar mass, the sound speed is necessarily large, $c_{\rm QM}/c>0.68$. Taken from~\citep{2019arXiv190600826X}. }\label{fig:cs}
\end{figure*}

We also introduce the bag mechanism to account for the energy difference between the physical vacuum and perturbative vacuum, and the bag parameter is  dynamically scaled~\citep[e.g.,][]{2002PhLB..526...19B,2004PhRvD..70d3010M}.
The total thermodynamic potential density for strange quark matter can be written as ~\citep{2019PhRvD..99j3017X},
\begin{eqnarray}
\Omega  &=& \Omega^\mathrm{pt} + B  \nonumber \\
&\equiv& \Omega^\mathrm{pt} + B_\mathrm{QCD} + (B_0 - B_\mathrm{QCD})\exp{\left[-\left( \frac{\sum_i\mu_i-930}{\Delta\mu}\right)^4\right]}  \nonumber \\
\label{eq:omega_pQCD}
\end{eqnarray}
where we take $B_0=40,~50~\rm MeV/fm^3$~\citep{1975PhRvD..12.2060D}
for the calculations and $\Delta\mu=\infty$ indicates no medium effect for the bag parameter.
If $\alpha_\mathrm{S}$ and $m_{u,d,s}$ run with the energy scale as reported by the particle data group~\citep{2014ChPhC..38i0001O}, the maximum mass of QSs does not reach $\sim$2$M_{\odot}$. In such cases,
the dynamic rescaling of the bag constant with finite $\Delta \mu$ is essential, which basically originates from nonperturbative effects such as chiral symmetry breaking and color superconductivity~\citep{2008RvMP...80.1455A,2018RPPh...81e6902B,2005PhR...407..205B}.
$B_{\rm QCD} =400~\rm MeV/fm^3$ is demanded by the dynamic equilibrium condition at the critical temperature of the deconfinement phase transition and is obtained by equating the pressures of the QGP ($-B_{\rm QCD} + 37\pi^2T^4/90$) and pion gas ($\pi^2T^4/30$) at $T = T_c~(\sim$ 170 MeV).

At given chemical potentials $\mu_i$, the pressure $P$, particle number density $\rho_i$, and energy density $\varepsilon$ are determined by:
\begin{eqnarray}
P &=& -\Omega, \label{eq:pressure_pQCD}\\
\rho_i &=& \frac{g_i}{6\pi^2} \left(\mu_i^2-m_i^2\right)^3
          -\frac{\partial \Omega_1}{\partial \mu_i}
           \alpha_\mathrm{s} + \rho_0, \label{eq:ni_pQCD}\\
\varepsilon &=& \Omega + \sum_i \mu_i \rho_i. \label{eq:E_pQCD}
\end{eqnarray}
The common term for the particle number density in Eq.~(\ref{eq:ni_pQCD}) is obtained with
\begin{eqnarray}
\rho_0   &=& -\frac{C_1}{3}\sum_i \left( \frac{\partial \Omega_0}{\partial m_i}
                          +\frac{\partial \Omega_1}{\partial m_i}\alpha_\mathrm{s}\right) \frac{\mbox{d} m_i}{\mbox{d} \bar{\Lambda}} \nonumber \\
                          && + \frac{C_1}{3} \frac{\partial \Omega_1}{\partial\bar{\Lambda}} \alpha_\mathrm{s}
                          +\frac{C_1}{3} \Omega_1 \frac{\mbox{d} \alpha_\mathrm{s}}{\mbox{d} \bar{\Lambda}}
                          -\frac{\partial B}{\partial \mu_i}.  \label{eq:n0}
\end{eqnarray}

In Fig.~\ref{fig:cs}, we show the EOSs generated in the perturbation model, which fulfill the available astrophysical constraints of mass~\citep{2010Natur.467.1081D,2013Sci...340..448A,2016ApJ...832..167F,2018ApJS..235...37A,2020NatAs...4...72C}, radius~\citep{2019ApJ...887L..24M,2019ApJ...887L..21R} and tidal deformability~\citep{2019PhRvX...9a1001A} for QSs.
They are compared with the results of pQCD without nonperturbative corrections, taking $C_1=1-4$ and $B=0$.
The EOSs for the nuclear matter obtained with the QMF models are also shown.
\citep{2015PhRvL.114c1103B} pointed out that if the two-solar-mass constraint is combined with the hadronic matter EOS below and around the nuclear saturation density, $c_{\rm QM}$ might first increase then decrease after reaching a maximum (maybe even up to $0.9c$) and finally approach the conformal limit $c/\sqrt{3}$ from below.
This peculiar shape resembles the analysis of the case of crossover EOS~\citep{2019ApJ...885...42B}.
To ensure a large mass for QSs above two solar mass, the obtained peak value ($c_{\rm QM}^\mathrm{max}$) ranges from 0.68$c$ to $c$, similar to previous results for NSs~\citep{2014ApJ...789..127K,2018MNRAS.478.1377A,2018ApJ...860..149T,2019PhRvL.122l2701M}.

\begin{figure*}
{\centering
\resizebox*{0.98\textwidth}{0.45\textheight}
{\includegraphics{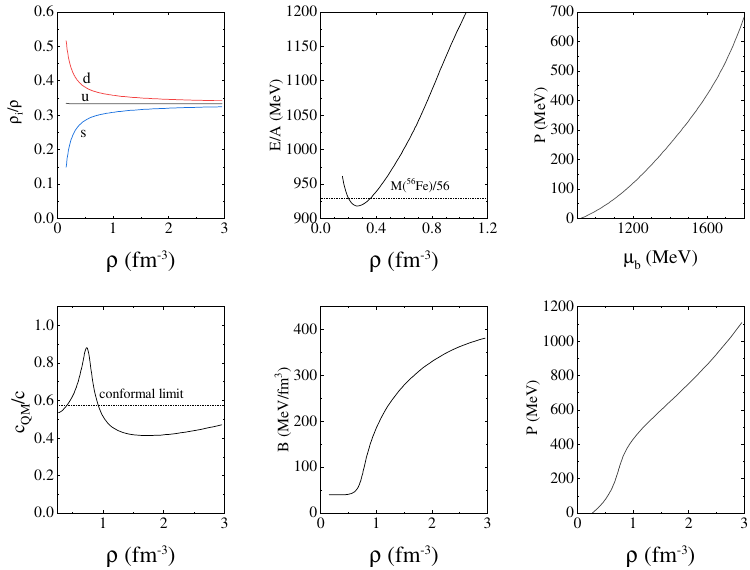}}}
\caption{Various properties of betastable strange quark matter, including the quark fractions, the binding energy, the pressure, the sound velocity, and the scaled bag parameter, which are plotted as a function of the baryon density or chemical potential. The calculations are performed based on the perturbation model using the parameters $C_1=3.5,\ B_0=40$\ MeV/fm$^3$, and \ $\Delta\mu=800$\ MeV.
The horizontal line in the $E/A$ plot indicates that of the most stable $^{56}\rm Fe$ nucleus.
The horizontal line in the $c_{\rm QM}$ plot shows the conformal limit.
}\label{fig:qseos}
\vspace{-0.3cm}
\end{figure*}
\begin{figure*}
{\centering
\resizebox*{0.9\textwidth}{0.4\textheight}
{\includegraphics{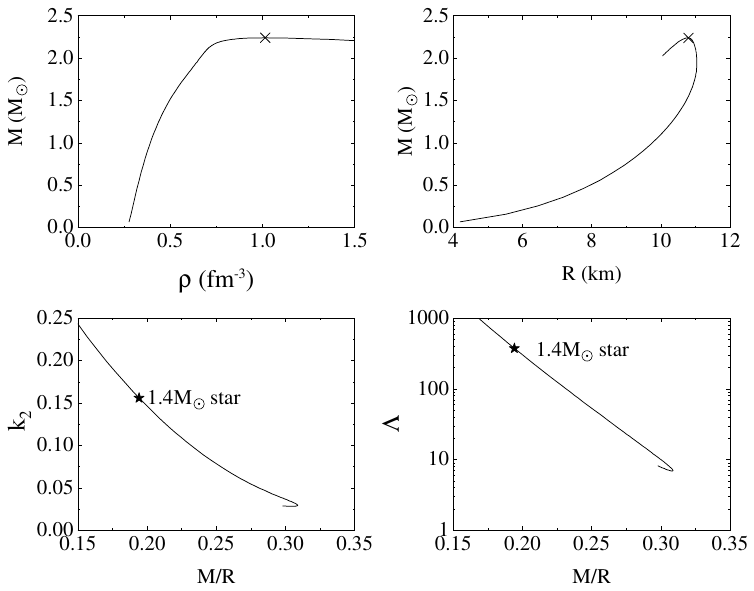}}}
\caption{Various properties of QSs based on the EOS in Fig.~\ref{fig:qseos}, including the gravitational mass, the Love number, and the tidal deformability, which are plotted as a function of central density, radius, or compactness.
The crosses in the upper two panels show where the maximum mass is reached.
The locations of a $1.4 M_{\odot}$ star are explicitly indicated in the lower two panels. }\label{fig:qsstar}
\vspace{-0.3cm}
\end{figure*}

We show in Fig.~\ref{fig:qseos} the various properties of strange quark matter based on the perturbation model using the exemplary parameters of $C_1=3.5,\ B_0=40$\ MeV/fm$^3$, and \ $\Delta\mu=800$\ MeV.
The composition, binding energy, pressure, sound velocity, and scaled bag parameter are shown as functions of the baryon density or chemical potential.
Note that in the binding energy plot, the condition that the strange quark matter be the absolute stable strong-interaction system, requiring that at $P=0, E/A \le M(^{56}\rm Fe/56)$ =930 MeV, is fulfilled. Fig.~\ref{fig:qsstar} shows the predicted properties of the QSs, including the mass, Love number, and tidal deformability.

Previously, although the quark star EOS models could reach 2 solar mass, they required a too small surface density (not much larger than the nuclear saturation density) in some cases, and a larger maximum mass meant an even smaller surface density (because of the anticorrelation between the two~\citep{2017ApJ...844...41L}), for example, the CDDM2 model~\citep{2016PhRvD..94h3010L} and the PMQS3 model~\citep{2017ApJ...844...41L}. Then, the radius (and the tidal deformability) exceeded the observational values~\citep{2017PhRvL.119p1101A,2019ApJ...887L..24M,2019ApJ...887L..21R,2019PhRvX...9a1001A}.
There models were not welcomed by particle physicists studying hadrons (for which experiments have been established studying the nonperturbative effects) because in such a density realm, the quarks are thought to be very dilute and are very possibly confined inside hadrons. In the real world, we do obtain nuclear matter rather than quark matter around the nuclear saturation density.
The present perturbative model with an in-medium bag can achieve both a reasonable surface density and a maximum mass as large as $2.2 M_{\odot}$. The predicted properties of dense matter ($c_{\rm QM}, \Gamma$) and quark stars ($R, M$), as well as the EOS of Fig.~\ref{fig:qseos}, are collected in Table \ref{tab:qseos}.

\begin{table*}
\begin{center}
\caption{QS~EOS~with~proper~sound~velocity~behavior and the predicted properties of dense matter ($c_{\rm QM}, \Gamma$) and quark stars ($R, M$).
The calculations are done based on the perturbation model using the parameters $C_1=3.5,\ B_0=40$\ MeV/fm$^3$, and \ $\Delta\mu=800$\ MeV. See Sec.~\ref{sec:2branch} for details.}
\setlength{\tabcolsep}{5pt}
\renewcommand{\arraystretch}{1.6}
\begin{tabular}{c|c|c|c|c|c|c}
\hline\hline
 $\rho$~(fm$^{-3}$) & $\varepsilon$~(MeV/fm$^3$) & $P$~(MeV/fm$^3$) & $c_{\rm QM}/c$ &$\Gamma$ & $R$~(km) &  $M/M_{\odot}$\\ \hline
0.277&	254.68&	3.22&	0.5365&	23.053&	3.834&	0.0530\\
0.298&	274.56&	8.98&	0.5400&	9.2107&	6.241&	0.2350\\
0.332&	306.51&	18.42&	0.5481&	5.2979&	8.078&	0.5320\\
0.366&	340.66&	28.88&	0.5590&	3.9983&	9.123&	0.7994\\
0.401&	376.61&	40.38&	0.5725&	3.3848&	9.778&	1.0253\\
0.436&	413.89&	52.94&	0.5886&	3.0548&	10.205&	1.2125\\
0.471&	452.00&	66.56&	0.6073&	2.8730&	10.488&	1.3669\\
0.506&	490.42&	81.22&	0.6288&	2.7826&	10.675&	1.4945\\
0.539&	528.59&	96.90&	0.6535&	2.7564&	10.796&	1.6004\\
0.571&	566.01&	113.56&	0.6816&	2.7802&	10.871&	1.6890\\
0.601&	602.20&	131.15&	0.7134&	2.8460&	10.914&	1.7638\\
0.629&	636.77&	149.60&	0.7487&	2.9465&	10.933&	1.8274\\
0.655&	669.49&	168.87&	0.7865&	3.0713&	10.936&	1.8821\\
0.663&	679.97&	175.46&	0.7994&	3.1152&	10.934&	1.8986\\
0.686&	710.22&	195.71&	0.8364&	3.2383&	10.923&	1.9439\\
0.708&	739.02&	216.62&	0.8665&	3.3126&	10.904&	1.9837\\
0.728&	767.11&	238.17&	0.8818&	3.2818&	10.879&	2.0186\\
0.749&	795.72&	260.33&	0.8741&	3.0995&	10.851&	2.0495\\
0.771&	826.60&	283.12&	0.8405&	2.7687&	10.819&	2.0766\\
0.795&	862.04&	306.60&	0.7856&	2.3523&	10.785&	2.1000\\
0.824&	904.94&	330.86&	0.7197&	1.9348&	10.749&	2.1199\\
0.859&	958.68&	356.09&	0.6532&	1.5756&	10.710&	2.1360\\
0.904&	1027.09&	382.51&	0.5930&	1.2958&	10.668&	2.1483\\
0.959&	1114.25&	410.42&	0.5421&	1.0915&	10.623&	2.1565\\
1.054&	1266.67&	450.58&	0.4896&	0.9136&	10.558&	2.1609\\
 \hline\hline
\end{tabular}\label{tab:qseos}
\end{center}
\end{table*}

We conclude this section by further discussing the following aspects:
\begin{itemize}
\item \relax $QS~vs.~hybrid~stars$: Although it is known that the degree of freedom is hadronic around the nuclear saturation density, the QCD phase state for cold, dense matter of intermediate densities is unfortunately unknown, and a great deal of effort is being applied in the communities of astrophysics, nuclear physics, and particle physics due to the crucial importance of this aspect. One key point is still not clear: Does the matter go through a phase transition from hadron matter to quark matter at some densities (which is relevant to compact star physics) or is quark matter the absolute ground state of strongly interacting matter (the conjecture of Bodmer-Witten~\citep{1971PhRvD...4.1601B,1984PhRvD..30..272W})?. Therefore, there remains the problem of how to verify QSs or distinguish them from NSs or hybrid stars~\citep{2005ApJ...629..969A,2019ApJ...887..151W}. For a fixed gravitational mass, hybrid stars are characterized by a smaller radius than their hadronic counterparts and could be as compact as QSs for masses above $1.0 M_{\odot}$.
The similarity of the sound speed of the hadron-quark mixed phase with that of the pure quark matter in the intermediate density region of $\sim3-8\rho_0$, a particular shape with a peak, further complicates the distinguishing of QS from hybrid stars~\citep{2019arXiv190600826X}.
\item \relax $Two~branch~scenario$: Because of the tension of a low tidal deformability ($190^{+390}_{-120}$~\citep{2018PhRvL.121p1101A}) and a high maximum mass ($2.14^{+0.20}_{-0.18}M_{\odot}$ for the presently heaviest pulsar~\citep{2020NatAs...4...72C} and $\le2.35M_{\odot}$ based on the numerical simulation studies on NS binary mergers~\citep{2018ApJ...852L..25R,2018PhRvD..97b1501R,2019PhRvD.100b3015S})
for a certain EOS in the NS model, binary QSs have been proposed as the possible scenario for the GW170817 event~\citep{2018PhRvD..97h3015Z,2018RAA....18...24L}. A binary QS merger for some binary configurations could eject amounts of matter (comparable to the binary NS case) to account for the electromagnetic observations in the optical/infrared/UV bands (namely, kilonova)~\citep{2009PhRvL.103a1101B}. A magnetar with QS EOS is preferred as the post-merger remnant for explaining some groups of short gamma-ray burst (SGRB) observations~\citep{2016PhRvD..94h3010L,2017ApJ...844...41L}. It has been suggested that because of this discrepancy (if confirmed), small and large stars of the same mass could coexist as hadronic and quark matter stars~\citep{2017ApJ...846..163W,2018ApJ...852L..32D}.
\item \relax $Comments~on~the~maximum~mass~of~NS/QS$: Various microscopic calculations of NS matter (without strangeness) indicate a possible upper limit of $\sim2.3-2.4M_{\odot}$ for the NS maximum mass, for example, Brueckner theory calculations~\citep{2016EPJA...52...21R} and quantum Monte Carlo calculations~\citep{2020arXiv200101374G}.
Exotic particles, if they are present, usually lower the limit as a result of the extra degrees of freedom during the phase transition (while the appearance of quarks might increase the limit in the case of crossover~\citep{2019arXiv191202312W,2019ApJ...885...42B}). The quark-hadron crossover EOS gives a maximum mass of $2.35 M_{\odot}$~\citep{2019ApJ...885...42B}.
The bound of $M_{\rm TOV} \lesssim 2.3-2.4 M_{\odot}$ may be applicable to QSs. For example, the maximum mass of QSs is $2.18 M_{\odot}$ ($2.32 M_{\odot}$ with color-flavor-locked superfluity~\citep{1999NuPhB.537..443A}) within the MIT bag model~\citep{2018PhRvD..97h3015Z}.
The present perturbation model yields $2.24 M_{\odot}$ with the peak sound speed ($c_{\rm QM}^\mathrm{max}$) approaching the speed of sound (we expect an increase in the value including the uncertain superfluity of $\sim0.1 M_{\odot}$).
These high theoretical limits on the maximum mass are higher than (but close to) the observational bound of pulsars of approximately $2.14 M_{\odot}$~\citep{2020NatAs...4...72C}.
A looser upper limit based on extreme causal EOSs may be in the range of $M_{\rm TOV} < 3.6-4.8M_{\odot}$~\citep{2019EPJA...55...97T,1996ApJ...470L..61K,1974PhRvL..32..324R,1976Natur.259..377B}.
The observations of accreting black holes, on the other hand, revealed a paucity of sources with masses below $5M_{\odot}$~\citep[e.g.,][]{1998ApJ...499..367B,2010ApJ...725.1918O,2011ApJ...741..103F,2012ApJ...757...36K}. Presently, binary mergers involving one or two companions have masses that fall into the so-called mass gap range ($3-5 M_{\odot}$) that are hard to distinguish~\citep[e.g.,][]{2019arXiv190407789W,2020PhRvL.124g1101T,2020arXiv200101761T}.
\end{itemize}

\section{Neutron star binary}\label{sec:binary}

The gravitational waves (GWs) detected from binary neutron star (BNS) merger event GW170817 \citep{2017ApJ...850L..39A}, as well as its electromagnetic (EM) counterparts \citep{2017ApJ...851L..16A}, announced the beginning of the multimessenger astronomy era. In addition to hinting at the origin of SGRB \citep{2017PhRvL.119p1101A,1992ApJ...395L..83N} and revealing the site of r-process nucleosynthesis \citep{2017ApJ...848L..12A,1989Natur.340..126E}, our knowledge of the EOS of cold dense matter at supranuclear densities has been greatly enriched. In the past year, various studies have been performed to constrain the EOS of dense matter, either by putting constraints on observable characteristics of NSs~\citep[i.e., radius or tidal deformability; see e.g.,][]{2018PhRvL.120q2703A,2018PhRvL.120z1103M,2018PhRvL.121p1101A} or by connecting the constraint with model parameters in nuclear physics~\citep[i.e., symmetry energy slope or neutron skin parameter; see e.g.,][]{2018ApJ...862...98Z,2018PhRvL.120q2702F}), which could be tested by nuclear physics experiments. Some studies also go beyond the conventional NS scenario and put constraints on compact star models involving strong interaction phase transitions \citep[e.g.,][]{2018PhRvL.120z1103M,2018PhRvD..97h3015Z,2020PhRvD.101d4019C,2019PhRvD..99j3009M}.

Those constraints mainly come from 3 aspects from a BNS merger event. First, during the inspiral stage, unlike binary black hole (BBH) mergers, deformation is induced for each NS due to the tidal field of the companion, providing additional dissipation of the orbital energy and angular momentum and hence accelerating coalescence \citep{2017ApJ...850L..39A}. This deformation therefore leaves a detectable signature in the GW signal of the late inspiral stage, from which we can learn about the tidal deformability of the NS EOSs \citep{2008PhRvD..77b1502F}. Second, the detection of SGRB hints at a delayed collapse to a BH for the merger remnant \citep{2015ApJ...808..186L,2014ApJ...788L...8M}.
This interpretation of the SGRB observation provides information on the maximum mass of a nonrotating configuration for the NS EOS (namely, $M_\mathrm{TOV}$). For instance, the EOS should not be too stiff; otherwise, the remnant supramassive NS lives much longer~\citep{2018ApJ...860...57A} in the magnetar central engine model \citep{2006Sci...311.1127D,2008MNRAS.385.1455M}. However, if the EOS is too soft, the merger might result in a prompt collapse to a BH. In such occasions, the magnetic field might not be enhanced sufficiently (by a differentially rotating NS remnant) and thus not able to launch a jet \citep{2018PhRvD..97b1501R}. Third, the features of the transient optical/infrared/UV observations (namely, the kilonova) powered by the radio activity of the neutron-rich elements in the ejecta depend directly on the mass, velocity and electron fraction in the ejecta, which is related to the properties of the EOS for the merging NS.

In this section, we briefly review the information we have learned about the EOS of NSs from the BNS merger events GW170817 and GRB170817A as well as AT2017gfo.

\subsection{GW170817 and tidal deformability} \label{sec:binarytidal}

The finite size effects of NSs alter the late inspiral GW signal compared with that of the BBH case \citep{2008PhRvD..77b1502F,2008ApJ...677.1216H}. Through the leading order, the GW observations constrained a combination of the tidal deformability for each NS in the binary ($\Lambda_1$ and $\Lambda_2$) \citep{2017ApJ...850L..39A}.

\begin{equation}
\tilde{\Lambda}=\frac{16}{13}\frac{(12q+1)\Lambda_1+(12+q)q^4\Lambda_2}{(1+q)^5},
\end{equation}
in which $q=M_2/M_1$ is the mass ratio of the binary. The dimensionless tidal deformability of each star is
\begin{equation}
\Lambda=\frac{2}{3}k_2(\frac{R}{M})^5,
\end{equation}
where $k_2$ is the tidal Love number describing the fraction between the induced quadrupole moment of a star and the external tidal field and $R$ and $M$ are the radius and mass of the star, respectively. On the other hand, $k_2$ can be obtained for a given EOS for any given mass and hence can be tested with the observation of GW170817.

Practically, the tidal deformability is fitted to the GW observation together with other parameters \citep{2019PhRvX...9a1001A}. For instance, in the Taylor F2 post-Newtonian aligned-spin model, 13 parameters need to be fitted, including 7 extrinsic parameters (sky location, distance of the source, polarization angle, inclination angle and coalescence phase and time) and 6 intrinsic parameters (mass of each NS, tidal deformability of each NS and the aligned spin of each NS). Therefore, the uncertainties in the estimation of other parameters weakly correlate with the determination of the tidal deformability. Hence, the constraint on the tidal deformability is normally made with certain assumptions.

For instance, as seen in \citep{2017ApJ...850L..39A}, the assumption in the spin parameter of the NS could significantly affect the interpretation of the mass of each NS, thus further affecting the constraint on the tidal deformability. Through the assumption that the $\Lambda$ of each NS vary independently, the first constraint on $\Lambda_1$ and $\Lambda_2$ could be obtained under different spin priors. The result favors the softer EOS, i.e., the EOS
that predicts more compact stars. Another analysis assumes a uniform prior for $\tilde{\Lambda}$, which sets an upper limit of 800 on $\tilde{\Lambda}$ in the low-spin case and 700 in the high spin case. Alternatively, through the expansion of $\Lambda(M)$ around a certain $M$, constraints can be directly placed on the tidal deformability of a certain mass star. This constraint is $\Lambda(1.4)\le1400$ in the high-spin case and $\Lambda(1.4)\le800$ in the low-spin case.

Follow-up analysis further improves these constraints under more assumptions. For example, in \citep{2018PhRvL.121p1101A}, instead of assuming an independent and uniform prior for the tidal deformability of each star, the EOS of each star in the merging binary is assumed to be the same. Consequently, the area of the 90\% confidence region in the $\Lambda_1$-$\Lambda_2$ parameter space shrinks by a factor of 3. This also improves the determination of $\Lambda(1.4)$ to $190^{+390}_{-120}$. In \citep{2019PhRvX...9a1001A}, a lower cut-off frequency of 23\,Hz is used instead of the 30\,Hz in the original analysis. Although the tidal effects mainly affect the GW signals above several hundred Hz, a lower frequency cut off allows for the better determination of other parameters, hence improving the measurement of the tidal deformability. In \citep{2018PhRvL.121i1102D}, it was pointed out that under the assumption that two stars in a binary system have a common EOS, there is an approximate relation between the tidal deformability of each star, i.e., $\Lambda_1/\Lambda_2=q^6$, where $q$ is the mass ratio. With the aid of this relation, once the assumption in the mass ratio of the binary is made, the tidal deformability can be further constrained. In \citep{2018PhRvL.121i1102D}, the improved analysis shows that $\tilde{\Lambda}$ is $222^{+420}_{-138}$ for a uniform mass prior and $245^{+453}_{-151}$ for a mass prior inferred from observed double neutron star systems and $233^{+448}_{-144}$ for a mass prior informed by all galactic neutron star masses within the 90\% credibility level.

We can directly test existing EOS models by simply calculating the tidal deformability and comparing it with the observational constraint. Nevertheless, more insight could be gained regarding the EOS of neutron-rich matter by more systematically studying the impact of the EOS (i.e., $p$ as a function of $\varepsilon$, where $p$ and $\varepsilon$ are the pressure and rest mass density of the matter) on the tidal deformability. Such interpretations are presented in \citep[e.g.,][]{2018PhRvL.120q2703A,2018PhRvL.120z1103M,2018PhRvL.121p1101A}. For instance, in \citep{2018PhRvL.121p1101A}, the logarithm of the adiabatic index of the EOSs is treated as a polynomial of the pressure for the high density EOS, namely, $\Gamma=\Gamma(p;\gamma_i)$ and $\gamma_i=(\gamma_0,\gamma_1,\gamma_2,\gamma_3)$ are free parameters. For densities below half the nuclear saturation density, the EOS is connected to the SLy EOS \citep{2001A&A...380..151D}. The sampling of the EOS then consists of uniformly sampling $\gamma_i$ in certain intervals. For each of the EOS samples, the mass radius relation and tidal deformability could be theoretically obtained and constrained by the observation of both the tidal deformability and 2 solar mass pulsars \citep{2010Natur.467.1081D,2013Sci...340..448A,2016ApJ...832..167F,2018ApJS..235...37A,2020NatAs...4...72C}. The constraint on the neutron star radius is $R=11.9^{+1.4}_{-1.4}$\,km for the merging binary of GW170817. Similar analysis can be found in, e.g., \citep{2018PhRvL.120q2703A}, which shows that the radius of a 1.4 solar mass NS is in the range of $[9.9,13.8]$\,km. However, it is worth noting that such analysis might be affected by the choice of EOS priors. In \citep{2018PhRvL.120z1103M}, it was pointed out that when the prior for possible twin star (for which there is a hadron-quark phase transition inside the star) branch EOSs is considered, the radius becomes less constrained, i.e., $R_{1.4}\in[8.53,13.74]\,$km.

\begin{figure*}
\vspace{-0.5cm}
{\centering
\resizebox*{0.48\textwidth}{0.3\textheight}
{\includegraphics{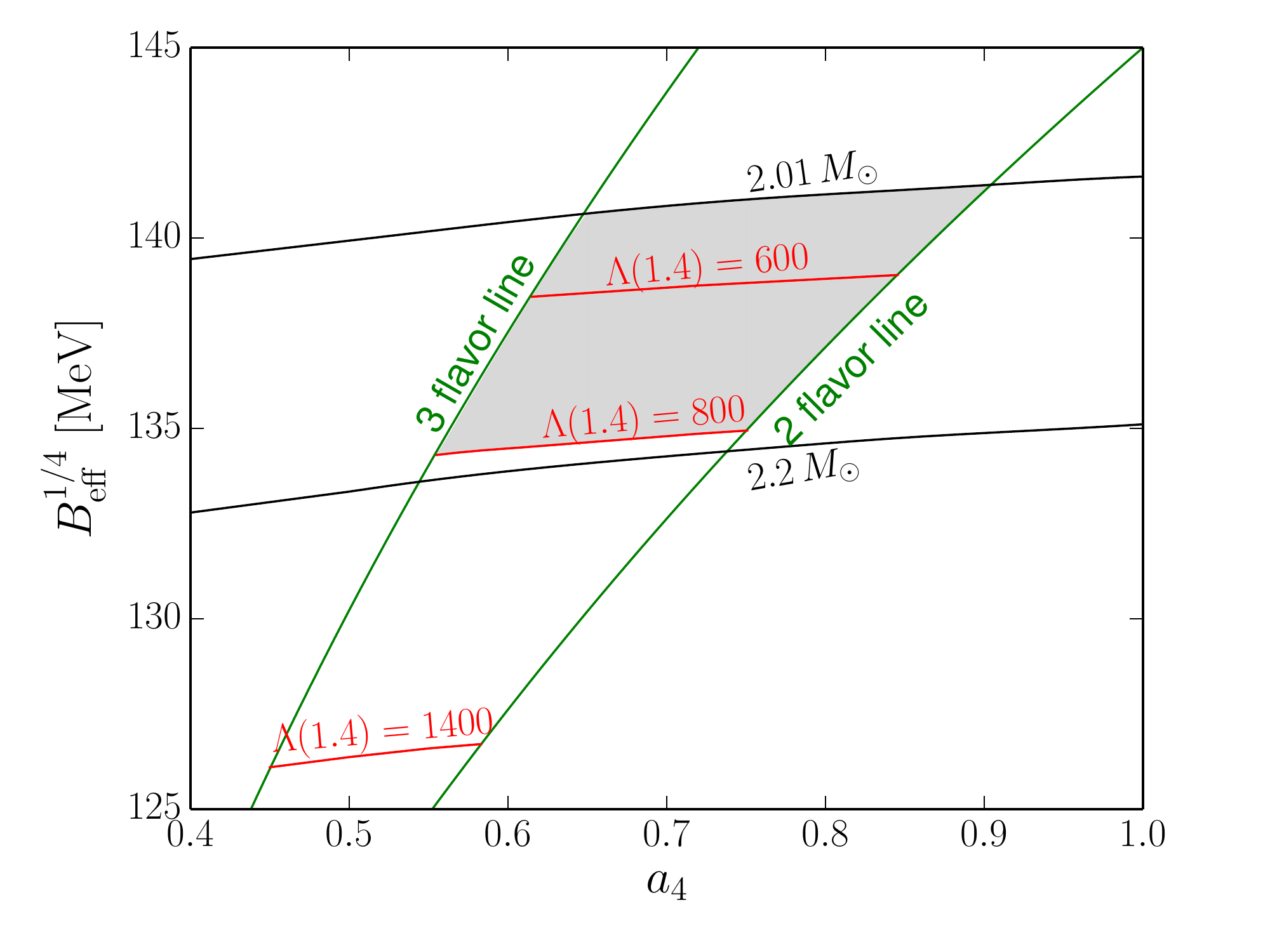}}}
{\centering
\resizebox*{0.48\textwidth}{0.3\textheight}
{\includegraphics{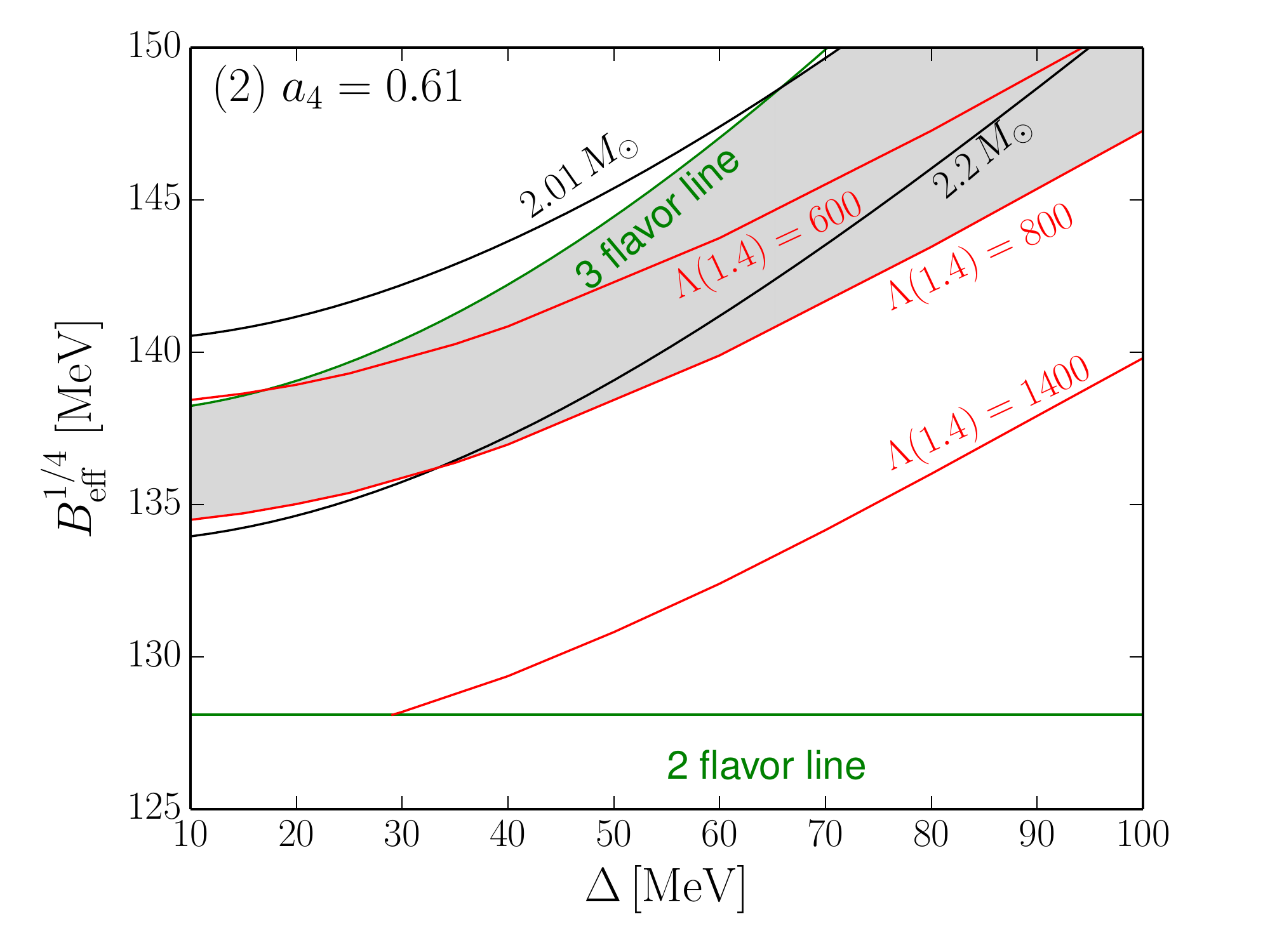}}}
\caption{Parameter space for QS EOS models within the MIT bag model ($B_{\rm eff},a_4,\Delta$) for normal QSs (left) and superfluid QSs (right) obtained by combining the GW170817 constraint on $\Lambda$(1.4), the two-solar-mass constraint on $M_\mathrm{TOV}$ and the stability window for quark matter.
With the constraint of the "2 flavor'' line, we ensure that normal atomic nuclei do not decay into nonstrange quark matter. With the constraint of the "3 flavor'' line, we ensure that strange quark matter is more stable than normal nuclear matter, namely, Bodmer-Witten's conjecture~\citep{1971PhRvD...4.1601B,1984PhRvD..30..272W}.
The perturbative QCD correction parameter $a_4$ characterizes the degree of the quark interaction correction, with $a_4 = 1$ corresponding to no QCD corrections (Fermi gas approximation).
$a_4 = 0.61$ is chosen to be close to the calculated result with different choices of the renormalization scale \citep{2001PhRvD..63l1702F}.
The effective bag constant ($B_\mathrm{eff}$) also includes a phenomenological representation of nonperturbative QCD effects.
Due to the strong correlation between $M_\mathrm{TOV}$ and $\Lambda(1.4)$, a lower bound can be inferred for $\Lambda(1.4)$ from the two-solar-mass limit, namely, $\sim$510 ($\sim$380) MeV for normal (superfluid) QSs.
Taken from~\citet{2018PhRvD..97h3015Z}. }\label{fig:cfl}
\vspace{-0.3cm}
\end{figure*}

In addition to systematic studies on parameterized EOS priors, phenomenological models can be applied to interpret tidal deformability constraints. According to \citep{2018PhRvL.120q2702F}, a better upper limit for neutron skin effects is obtained compared with that of the experiment done by PREX, and better results could be achieved with future GW observations and terrestrial nuclear physics experiments. Both the symmetry energy parameter and the symmetry energy slope are better constrained with respect to previous nuclear physics studies \citep{2018ApJ...862...98Z}. Under the assumption that GW170817 originates from a binary quark star (BQS) merger, the quark interaction parameters are studied in \citep{2018PhRvD..97h3015Z}.
Fig.~\ref{fig:cfl} shows the results of both normal and superfluid QSs.
It is worth noting that due to the finite surface density of QSs, a surface correction needs to be taken into account when calculating $\Lambda$ of a QS \citep{2009PhRvD..80h4035D}. Therefore, a QS could have a larger TOV maximum mass without violating the tidal deformability constraint compared with those of NSs \citep{2019EPJA...55...60L}.

\subsection{GRB170817A and merger remnant} \label{sec:binaryremnant}

\begin{figure*}
{\centering
\resizebox*{0.95\textwidth}{0.4\textheight}
{\includegraphics{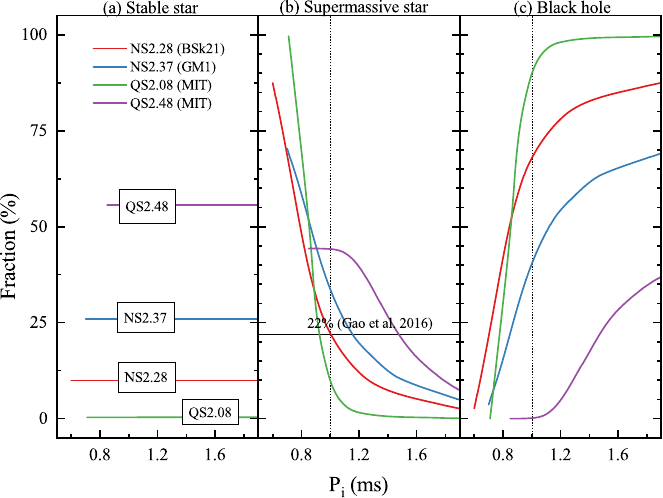}}}
\caption{Postmerger product fractions for (a) stable star, (b) supermassive star and (c) black hole for the NS and QS EOS models, labelled with the star type plus the corresponding $M_{\rm TOV}$: Unified BSk21 (red line labelled NS2.28), nonunified GM1 (blue line labelled NS2.37), and MIT model (green and purple lines labelled QS2.08 and QS2.48). In panel (b), the observed 22\% constraint for supermassive stars from~\citet{2016PhRvD..93d4065G} is shown by the horizontal line for comparison. The vertical dotted line in the same panel is for a typical initial period of 1 ms. Taken from~\citet{2017ApJ...844...41L}.}\label{fig:pmqs}
\vspace{-0.3cm}
\end{figure*}

BNS mergers have long been proposed as the central engine of SGRBs \citep{1992ApJ...395L..83N}. This suggestion has been verified by GRB170817A detected by Fermi/GBM and INTEGRAL/SPI-ACS, which accompanies the detection of GW170817. According to the time of the merger implied from the chirp signal, there is a $1.74\pm0.05\,$s delay for the onset of the SGRB \citep{2017ApJ...848L..12A}. The detection of GRB170817A not only helps the determination of the location of the source, which allows for abundant follow-up observations in other bands, but also provides useful information about the post-merger evolution of the merger event, thus providing constraints on the EOSs \citep[e.g.,][]{2016PhRvD..93d4065G,2018ApJ...858...74M,2020FrPhy..1524603G}.
Exemplary fractions of the outcome of the binary are shown in Fig.~\ref{fig:pmqs} using both NS and QS EOSs. The dependence on the EOS, as well as the initial period, is evident~\citep{2017ApJ...844...41L}. It is found that the fraction of stable star (panel (a)) is determined by the static maximum mass $M_{\rm TOV}$. The fractions of supermassive star (panel (b)) and black hole (panel (c)) are further sensitive to the initial period since the fast-rotating configurations of the star have to be taken into account for them~\citep{2016PhRvD..94h3010L}.

Depending on the TOV maximum mass of the NS EOS and the total mass of the merging binary, there could be 4 different outcomes after the merger:
\begin{itemize}
\item if the total mass of the binary system ($M_\mathrm{tot}$) is much larger than $M_\mathrm{TOV}$, the direct formation of a black hole (BH) on a dynamic time scale, namely, prompt collapse, occurs. The total binary mass, above which prompt collapse can occur, is denoted as the threshold mass ($M_\mathrm{thres}$);
\item if $M_\mathrm{tot}$ is smaller than $M_\mathrm{thres}$ but larger than the maximum mass that can be reached by uniformly rotating NSs (denoted $M_\mathrm{supra}$), then a short-lived NS could exist as a remnant supported by differential rotation \footnote{Such NSs are called hypermassive NSs (HMNS). NSs that can be supported by only uniform rotation are called supramassive NSs (SMNS)}. The differential rotation dissipates due to magnetorotational instabilities as well as viscosity within a timescale of $\sim$100\,ms, and then the NS collapses to a BH;
\item if $M_\mathrm{tot}$ is smaller than $M_\mathrm{supra}$ but larger than $M_\mathrm{TOV}$, the remnant is a long-lived supramassive NS. The uniformly rotating NS could still lose angular momentum by magnetic dipole radiation, but it takes a much longer time to sufficiently reduce the angular momentum to induce the collapse to a BH;
\item if $M_\mathrm{tot}$ is smaller than $M_\mathrm{TOV}$, a stable NS remnant exists.
\end{itemize}

The GW observations do not provide any hint to which scenario applies to the case of GW170817 due to the lack of post-merger GW observations \citep{2017ApJ...850L..39A,2017ApJ...848L..13A}. The electromagnetic counterparts, on the other hand, can indicate what happens after the merger of the two NSs.

A very robust interpretation is that scenario 1) should be excluded due to the SGRB detected. According to \citet{2018PhRvD..97b1501R}, in the BH central engine model for SGRBs, enhancement in the magnetic field of the merger remnant due to the differential rotation of the hypermassive NS is essential for jet formation. A prompt collapse results in a magnetic field that is too weak to explain the SGRB observations. Hence, the detection of GRB170817A directly implies $M_\mathrm{tot}<M_\mathrm{thres}$ for the case of GW170817. The total mass of the binary can be measured by the inspiral GW signal, which places a constraint on $M_\mathrm{thres}$.

For a given EOS model, the threshold mass can be determined by performing numerical simulations with different total binary masses. This can be extremely time-consuming for full 3-dimensional general relativistic simulations. In \citet{2017ApJ...850L..34B}, the so-called smooth particle hydrodynamics (SPH) method as well as a conformally flat assumption for the spacetime metric is used to reduce the simulation time to make it plausible. According to the results, the threshold mass is related to $M_\mathrm{TOV}$ and $R_\mathrm{TOV}$ (i.e., the radius of the TOV maximum mass configuration) as
\begin{equation}
M_\mathrm{thres}=(-3.38\frac{M_\mathrm{TOV}}{R_\mathrm{TOV}}+2.43)M_\mathrm{TOV}.
\end{equation}
Alternatively, the results can be fitted with similar accuracy in terms of the radius of a 1.6 solar mass star ($R_{1.6}$) as
\begin{equation}
M_\mathrm{thres}=(-3.606\frac{M_\mathrm{TOV}}{R_{1.6}}+2.38)M_\mathrm{TOV}.
\end{equation}
Note that this reveals a quadratic relation between $M_\mathrm{thres}$ and $M_\mathrm{TOV}$ once $R_\mathrm{TOV}$ (or $R_{1.6}$) is fixed. Particularly, since the coefficient of the $M_\mathrm{TOV}^2$ term is negative, there exists a maximum value for $M_\mathrm{thres}$. This maximum possible value of $M_\mathrm{thres}$ must be larger than $M_\mathrm{tot}$ in the case of GW170817; otherwise, there is no parameter space to prevent a prompt collapse. In other words, any choice of $R_\mathrm{TOV}$ (or $R_{1.6}$) that results in a maximum possible value $M_\mathrm{thres}$ smaller than 2.74 solar mass should be excluded by the observations. This requires $R_\mathrm{TOV}$ to be larger than 9.26\,km and $R_{1.6}$ to be larger than 10.30\,km.

For NS EOS models, scenario 4) can also be excluded. Scenario 4) requires a very large $M_\mathrm{TOV}$, which results in a large tidal deformability, hence violating the tidal deformability constraint \footnote{Note that this possibility still remains for QSs, as discussed in \citep{2019EPJA...55...60L}}. Distinguishing between scenarios 2) and 3) could indicate more on the TOV maximum mass of NSs; however, this is quite model dependent. Under different SGRB central engine model assumptions, totally opposite conclusions could be drawn. If the SGRB originates from a BH central engine, as assumed in \citep{2018PhRvD..97b1501R}, the delay collapse has to occur within 1.7\,s after the merger. Therefore, scenario 2) most likely occurred for the case of GW170817. As shown by previous studies \citep{2016MNRAS.459..646B}, the ratio between $M_\mathrm{supra}$ and $M_\mathrm{TOV}$ is almost universal for various NS EOS models, and the value is approximately 1.2. Combining the total mass of the merging binary, an upper limit \citep[approximately 2.15-2.25$\,M_\odot$ according to different studies, e.g.,][]{2018ApJ...852L..25R,2018PhRvD..97b1501R,2017ApJ...850L..19M,2017PhRvD..96l3012S} could then be obtained for the TOV maximum mass. In addition, it has been pointed out that the merger remnant might collapse to a BH with angular momentum smaller than that of the Keplerian limit due to angular momentum transfer by post-merger GW emission, neutrino and viscous effects, which leads to a different constraint on the TOV maximum mass. With this in mind, \citet{2019PhRvD.100b3015S} performed an analysis by considering conservation laws of baryonic mass, energy and angular momentum, and the constraint on the TOV maximum mass is found to be $2.10\,M_\odot<M_\mathrm{TOV}<2.35\,M_\odot$. It is worth noting that this constraint is valid only under the BH central engine assumption. In fact, it has been pointed out that the multimessenger observation of GW170817 is consistent in a magnetar central engine model \citep{2018ApJ...861L..12L} and could even be favored by an X-ray activity detected very long time after the merger \citep{2019MNRAS.483.1912P}. In such a magnetar central engine model, in contrast, scenario 3) is favored, and hence, the upper limit of $M_\mathrm{TOV}$ mentioned before becomes a lower limit instead.

\subsection{AT2017gfo and ejecta properties}
It has long been suggested that the BNS merger is an important site for the production of heavy elements in the Universe \citep{1989Natur.340..126E}. R-process nucleosynthesis is expected to occur in the neutron-rich matter ejected during the merger. The radioactive decay of such neutron-rich isotopes could power a transient in optical/UV/IR, i.e., a kilonova \citep{1998ApJ...507L..59L,2010MNRAS.406.2650M}. Such a transient event (AT2017gfo) was detected several hours after the merger time of GW170817 \citep{2017ApJ...848L..12A}, the luminosity, spectrum and light curve of which are consistent with the prediction of the kilonova model. Such a kilonova detection not only enriches our knowledge about the abundance of heavy elements in the Universe but also greatly increases our understanding of NS EOSs.

The observational properties, for example, the peak luminosity and peak time, of the kilonova are closely related to the ejecta properties (c.f. \citep{2017Natur.551...80K}). The abundance of lanthanides (atomic numbers from 58 to 71) is strongly related to the electron fraction ($Y_e$) of the ejecta. On the other hand, the opacity of the ejecta is mainly determined by the lanthanides in it, and is hence related to the electron fraction of the ejecta. The overall luminosity is related to the amount of radioactive heavy elements and thus the total mass of the ejecta. The ejecta is expanding at a certain velocity and becomes translucent after a period of time. Therefore, the characteristic duration of a kilonova is related to the velocity of the ejecta. To summarize, the mass, velocity and electron fraction of the ejecta are key parameters for understanding the observations of AT2017gfo.

The ejecta during a BNS merger mainly consists of two components. The first component is the so-called dynamic ejecta, which is normally more neutron-rich (lower $Y_e$). Part of the dynamic ejecta is due to the tidal torque during the inspiral \citep{1997A&A...319..122R,1999A&A...341..499R}; hence, it has a lower temperature and very low $Y_e$ (smaller than $0.1-0.2$) and is spatially distributed in the equatorial plane of the binary. Another part of the dynamic ejecta results from shock during coalescence (also called shock-driven ejecta) \citep{2007A&A...467..395O,2013PhRvD..87b4001H}. Due to the higher temperature at coalescence, this part of the dynamic ejecta normally has a slightly higher electron fraction ($Y_e>0.25$) \citep{2014ApJ...789L..39W,2016PhRvD..93l4046S} and can expand in the polar direction. In addition to the dynamic ejecta, the neutrino emissions from the remnant before collapsing to BH as well as the viscosity could further drive more ejecta (wind-driven ejecta) from the disc surrounding the remnant \citep{2008MNRAS.390..781M,2009ApJ...690.1681D,2013MNRAS.435..502F}. Due to neutrino irradiation, this part of the ejecta has a broader distribution of $Y_e$, which could be as high as 0.5 \citep{2014MNRAS.441.3444M,2014MNRAS.443.3134P,2015ApJ...813....2M}. Clearly, the amount of wind-driven ejecta is dependent on the lifetime of the remnant NS. For instance, in the case of a prompt collapse, the wind-driven ejecta could be significantly suppressed.

The kilonova observation following GW170817 has shown clear evidence of two distinct ejecta components \citep{2017ApJ...848L..17C,2017Natur.551...80K,2017PASJ...69..102T} \footnote{Note that there are studies arguing a model with 3 components (c.f. \citep{2017ApJ...850L..37P}).}, an early rising ($\sim2$ days after the merger) and bluer component (which indicates a lower opacity and higher velocity) and a more extended redder component. The required amount of ejecta accounting for the "blue" component is approximately $0.01M_\odot$ with a relatively larger electron fraction $Y_e>0.25$ and velocity $v^{\mathrm{blue}}\sim0.2-0.3c$. For the red component observed at later times, in total, approximately $0.05M_\odot$ lanthanide-rich ($Y_e<0.25$) ejecta is needed, with a lower velocity of $v^{\mathrm red}\sim0.1-0.2c$. The inferred properties can be used to constrain the EOS of the merged NSs, although this constraint is quite model dependent.

One property we can use to constrain NS EOS models is the mass of the ejecta, as it is related to the properties of the merging binary. As summarized in \citep{2018ApJ...852L..29R}, stiffer EOS models (for which the tidal deformability is larger) typically have a smaller amount of tidal-induced dynamic ejecta than softer EOS models. However, softer EOSs normally eject more dynamic ejecta overall because of a more violent coalescence and hence eject more shock-driven ejecta. The amount of wind-driven ejecta depends on the lifetime of the merger remnant before collapsing to a BH, which is determined again by the $M_\mathrm{TOV}$ of the NS EOS. Ideally, the details of the ejecta property for BNS mergers with different EOS models could be determined through extensive numerical simulations. However, to fully resolve the wind-driven ejecta, a very long-term post-merger simulation with the implementation of viscosity and neutrino cooling is required, which is normally extremely time consuming and not affordable in practice. Nevertheless, conservative estimations can still be made. In \citep{2018ApJ...869..130R}, it has been found that the total dynamic ejecta plus all the mass in the disc surrounding the remnant has a positive correlation with the $\tilde{\Lambda}$ parameter of the binary.
As not all the matter in the remnant disc is ejected, the dynamic ejecta plus the remnant disc mass has to be larger than the inferred mass of the ejecta to explain the observation of AT2017gfo, which is approximately 0.05 solar mass. This sets up a lower limit for the dynamic ejecta mass plus the remnant disc mass and hence a lower limit for the binary tidal deformability. In \citep{2018ApJ...869..130R}, this conservative constraint is $\tilde{\Lambda}>400$. However, in the more systematic study of \citep{2019ApJ...876L..31K}, which employs a set of more general parameterized EOS and considers unequal-mass binaries, a contradiction was found. In other words, it was shown that a binary system with $\tilde{\Lambda}<400$ could still be consistent with the luminosity of AT2017gfo in terms of the ejecta mass. Therefore, this lower limit must be considered with caution.

Another implication of the observation of AT2017gfo is the fate of the merger remnant (the 4 scenarios mentioned in the previous subsection). This observed "blue" component of the ejecta clearly rules out the possibility of a prompt collapse, in which case there is a negligible amount of high $Y_e$ shock-driven ejecta and wind-driven ejecta. In such a case, the kilonova observation should be dominated by the tidal ejecta and thus should be red. Distinguishing whether the remnant is long-lived is very uncertain and model dependent. In \citep{2017ApJ...850L..19M}, it was suggested that if a long-lived SMNS is produced, then a significant amount of the rotational kinetic energy of the SMNS is injected into either a collimated relativistic jet or the ejecta within tens of seconds. This extra energy injection is considered to be inconsistent with the observations of GRB170817A and AT2017gfo. Therefore, the authors of \citep{2017ApJ...850L..19M} believe an HMNS or very short-lived SMNS is produced in the remnant and put a similar upper limit on $M_\mathrm{TOV}$ of 2.17\,$M_\odot$. Nevertheless, in \citep{2018ApJ...861L..12L}, it was shown that with a long-lived SMNS as the merger remnant, both the early and late emission of AT2017gfo can be explained without requiring an unrealistically low opacity and high ejecta mass. Similar arguments can also be found in \citep{2017PhRvD..96l3012S}. A long-lived SMNS remnant is believed to be able to provide strong neutrino emissions to reduce the lanthanide contamination in our line of sight as well as to produce enough ejecta with relatively high speed in the post-merger phase. Particularly, a temporal feature observed 155\,d after the merger in the X-ray afterglow provides a more direct hint supporting a long-lived remnant. Considering the possibility of different merger outcomes, the allowed range for the TOV maximum mass for NSs could actually be larger~\citep{2019arXiv191206369A}.

\subsection{NS vs QS in the multimessenger era}

We have summarized some of the many studies on NS EOSs in light of GW170817 and its EM counterparts. Nevertheless, it is worth noting that the phase diagram of strong interactions at supranuclear densities is not yet clearly understood due to the nonperturbative nature of QCD at low energy scales. Apart from conventional NS models, other models involving strong interaction phase transitions are suggested, e.g., twin stars or strange stars \citep{2013PhRvD..88h3013A,2014PhRvD..89d3014D,1986ApJ...310..261A}. As is summarized below, a binary quark star (BQS) scenario could be totally consistent with the observation of GW170817 and its EM counterparts. Due to the self-bound nature, QSs have a very large surface density. This leads to many significant differences between QSs and NSs. For example, when supported by uniform rotation, QSs can reach an even higher maximum mass with respect to their TOV maximum mass ($40\%$ more) than NSs ($20\%$ more) \citep{2000A&A...363.1005G}. The finite surface density leads to a correction when calculating the tidal deformability \citep{2010PhRvD..82b4016P}. QSs could reach a much higher $T/|W|$ ratio when rotating, which could lead to more significant GW radiation in the post-merger phase \citep{2018PhRvD..97h3015Z}. As a result, the above analysis on NS models should not be directly applied to QSs. It is interesting and useful to understand the constraints on QS models from what we have learned from GW170817 and its EM counterparts and how to distinguish between NS/QS models in the multimessenger era.

Qualitatively, the tidal deformability measurement constrains the stiffness of QSs, similar to the case of NSs. Stiffer EOSs normally reach higher $M_\mathrm{TOV}$ but also have larger tidal deformability than a softer EOS due to the larger size of the star described by a stiffer EOS.
There is an overall positive correlation between $M_\mathrm{TOV}$ and $\Lambda$ in NS models \citep{2018PhRvL.120q2703A}. Investigating QS properties based on the MIT bag model reveals a similar relation between $M_\mathrm{TOV}$ and $\Lambda(1.4)$ \citep{2018PhRvD..97b3013Z}. However, the quantitative results are quite different. In \citep{2018PhRvL.120q2703A}, creating NS EOS models stiff enough to reach $M_\mathrm{TOV}\sim$2.8\,$M_\odot$ with $\Lambda(1.4)\le800$ was found to be impossible. In \citep{2019EPJA...55...60L}, it was shown that a self-bound strange star model can be stiff enough to reach $M_\mathrm{TOV}>3\,M_\odot$ without violating the tidal deformability constraint.

It was suggested that a BQS merger should result in a clean environment with little or no hadronic dynamic ejecta \citep{1991ApJ...375..209H}. It is not easy to verify this argument with numerical simulations, as the density discontinuity on the QS surface is difficult to handle numerically. Nevertheless, in \citep{2009PhRvL.103a1101B}, several BQS merger simulations were performed with the SPH method, and it was shown that with some binary configurations, a BQS merger could eject a comparable amount of (quark) matter in the case of BNS mergers. According to \citep{1985PhRvD..32.1273A,1986PhRvD..34.2947M}, under certain conditions (i.e., if the baryon number of an ejected quark nugget is smaller than a critical value), quark matter could efficiently evaporate into normal nucleon matter and contribute to the kilonova observation \citep{2018RAA....18...24L,2017IJMPS..4560042P}. In addition, as uniformly rotating QSs can reach a higher maximum mass, the post-merger remnant is likely to be longer lived than the case of the BNS merger. It has been found that a magnetar with QS EOS as the post-merger remnant is better for understanding the internal X-ray plateau observations following SGRBs \citep{2016PhRvD..94h3010L}. In addition, both differentially rotating and uniformly rotating triaxial QSs are found to be sufficient GW emitters \citep{2018PhRvD..97h3015Z,2019PhRvD.100d3015Z}, which could be tested by future GW observations.
\begin{figure}
\begin{center}
		\includegraphics[height=60mm]{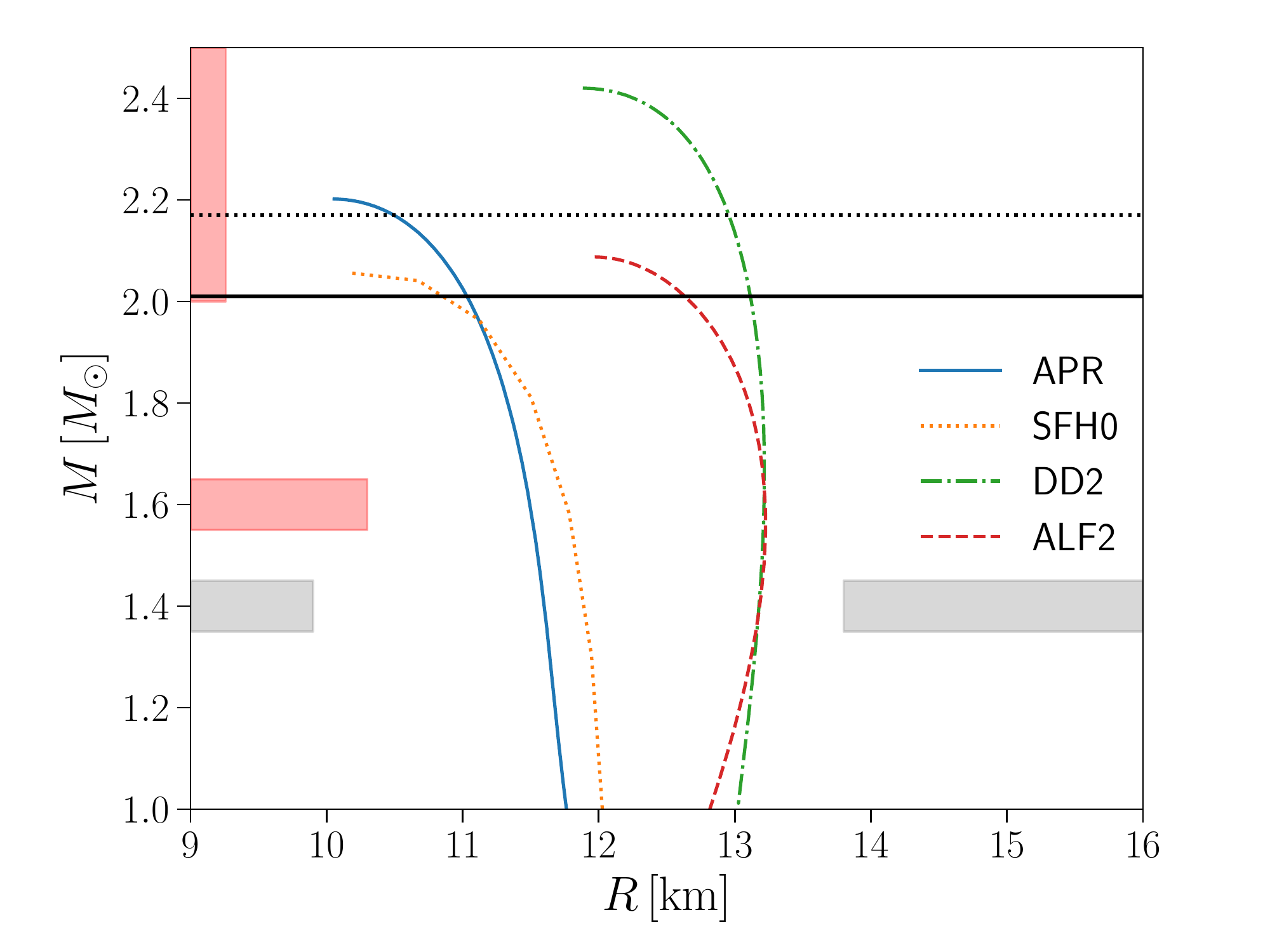}
\end{center}
\caption{The combined constraints on NS EOSs from the multimessenger observations of GW170817, GRB170817A and AT2017gfo. The gray shaded regions are excluded by the tidal deformability measurement of GW170817 \citep{2017ApJ...848L..13A,2018PhRvL.120q2703A}. The red shaded regions are forbidden because the GRB170817A and AT2017gfo observations exclude a prompt collapse after merger \citep{2017ApJ...850L..34B}. The solid horizontal line is the 2.01 solar mass lower limit for $M_\mathrm{TOV}$ according to the observation of the massive pulsar \citep{2013Sci...340..448A}, and the dotted horizontal line is the 2.17 solar mass constraint. Note that if the SGRB is powered by a BH central engine (or magnetar central engine), the dotted horizontal line is an upper limit (or lower limit). The M-R relation of several commonly used NS EOSs is shown in the figure. }
	\label{fig:constraint}
	\vspace{-0.3cm}
\end{figure}

Another interesting possibility is a BNS merger that leads to a conversion of the merger remnant to a QS. In such a case, the inspiral GW signal and dynamic ejecta properties should be exactly the same as the case of a normal BNS merger, whereas the post-merger behavior could be quite different. On the one hand, if the phase transition occurs partially inside the star (i.e., only the high-density core part of the remnant), a softening of the EOS occurs, hence reducing the lifetime of the merger remnant as well as shifting the $f_2$ peak in the post-merger GW signal to a higher frequency compared with those of the purely nucleonic merger remnant case \citep{2019PhRvL.122f1102B,2019ApJ...881...73W}. On the other hand, if the entire star could be converted to a QS after merger, which results in a stiffening of the EOS, the lifetime of the remnant is longer, and the $f_2$ frequency is smaller \citep{2019ApJ...881..122D}.
In this scenario, the time delay between the merger and the SGRB is believed to be the time needed for converting the surface of the remnant to quark matter, which significantly reduces the baryon load in the environment, thus helping the formation of a collimated jet \citep{2016PhRvD..93j3001D}.

Overall, the possibilities of a BQS merger or BNS merger with a QS remnant are consistent with the multimessenger observations of GW170817. However, current knowledge of the details of mergers involving QSs is quite limited. More simulations need to be performed for a better and more complete understanding in the future. With the help of numerical results and more future observations, whether QS could be involved or formed in a merger event could be distinguished, particularly according to the post-merger GW signals.

\subsection{Conclusion}
To summarize, the multimessenger observation of BNS merger GW170817 has greatly increased our knowledge about the EOS of dense matter. The most robust constraint is from the tidal deformability encoded in the GW signal during the inspiral. Such a tidal deformability measurement could translate into a constraint on the radius of the NS at a given mass.
The EM counterparts contain large amounts of information on the EOS models, the most reliable of which is to exclude the prompt collapse scenario. This provides independent constraints on the neutron star radius for a given mass. Other constraints on the lifetime of the remnant NS, however, depend on the central engine model of the SGRB and are not reliable. The constraints could be totally opposite in the BH central engine model and magnetar model, and the current observations could not reliably rule out either possibility for GW170817. We summarized all the constraints mentioned above, which is shown in Fig. \ref{fig:constraint}.
Nevertheless, GW170817 is just the beginning of the multimessenger era. As an increasing number of GW signals and EM counterparts from BNS mergers are detected in the future, our knowledge of NS and even QS EOS models will be enriched. In particular, if the time delay between the merger and the collapse to BH could be more robustly determined by future observations (either by post-merger GW signals or by more extensive EM counterpart observations), the TOV maximum mass can be crucially inferred, leading to a much better understanding of the EOS of dense neutron-rich matter.

\section{Other opportunities from compact objects}

\subsection{Neutron star cooling}

With the ever-increasing accuracy of observational instruments, more details of the signals emitted by NSs can be quantitatively monitored.
Apart from the measurements of NS masses, radii and tidal deformabilities, high-density NS models can be confronted with the surface temperatures of isolated NSs of known or estimated ages and the thermal photon luminosity of the X-ray transients in quiescence with an estimated time-averaged accretion rate on the NS~\citep{2007PhR...442..109L,2001PhR...354....1Y,2006ARNPS..56..327P,2006NuPhA.777..497P,2015SSRv..191..239P}.
The NS EOS determines the stellar structure, as well as the effective masses and superfluid gaps of baryons, and is therefore crucial for the heat capacity and neutrino emission rate~\citep[e.g.,][]{2001PhR...354....1Y,2016PhRvC..93a5803L,2020arXiv200103859S}.

\begin{figure}
\begin{center}
\includegraphics[width=0.7\textwidth]{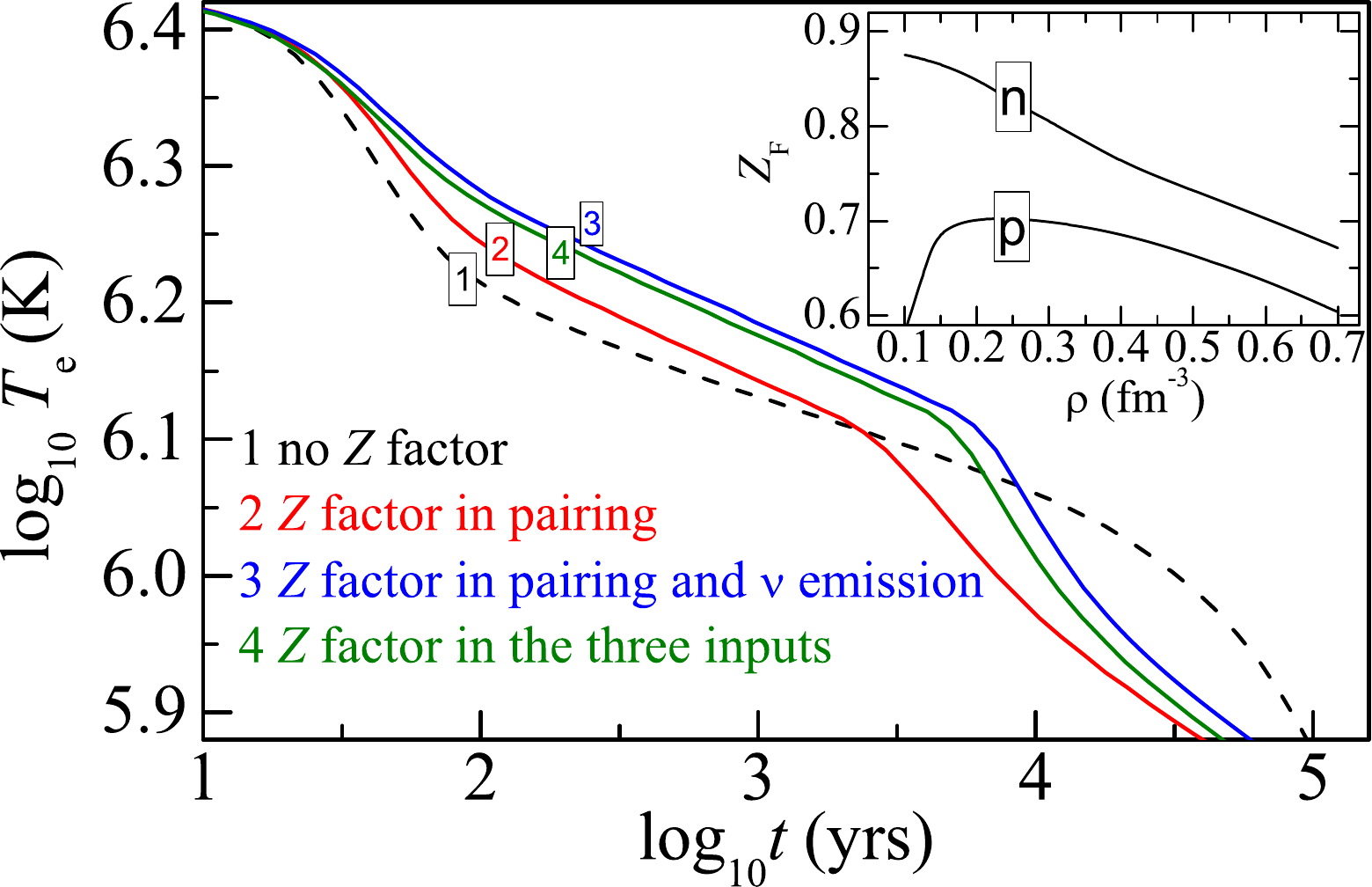}
\caption{Cooling curves of a canonically isolated NS within the minimal cooling paradigm~\citep{2004ApJS..155..623P}, without including fast neutrino emissions, charged meson condensate, hyperons, or confinement quarks in canonical NSs. The stellar structure is built with the APR EOS~\citep{1998PhRvC..58.1804A}. The calculations are carried out in three cases for a comparison: without any $Z$ factors (the Fermi surface depletion due to the SRC), with $Z$ factors only in superfluidity, with $Z$ factors both in superfluidity and neutrino emission and with $Z$ factors. Dong et al.~\citep{2016ApJ...817....6D} found that the SRC effect reduces the neutrino emissivity for the DU, MU, NNB and PBF processes, as well as the heat capacity of the stellar interior~\citep{2016ApJ...817....6D}.}\label{fig:cooling}
\end{center}
	\vspace{-0.3cm}
\end{figure}

In the Newtonian framework, the energy balance equation for NS cooling is written as~\citep{2006NuPhA.777..497P},
\begin{equation}
\frac{dE_{\rm{th}}}{dt}=C_{v}\frac{dT}{dt}=-L_{\nu }-L_{\gamma}+H,\label{AA}
\end{equation}
where $T$ and $C_{v}$ are the stellar internal temperature and the total heat capacity, respectively. The loss of the thermal energy $E_{\rm{th}}$ occurs through neutrino emission (total luminosity
$L_{\nu }$) and photon emission (total luminosity $L_{\gamma }$). $H$ represents all possible energy sources to heat the star, such as the decay of the magnetic field energy stored in stars.
Current simulations of thermal evolution are usually based on a general relativistic formulation, and some robust codes have already been established~\footnote{http://www.astroscu.unam.mx/neutrones/
NSCool/},
which comprises all the relevant cooling reactions: direct URCA (DU) ($n\rightarrow p+l+\overline{\nu _{l}}, p+l\rightarrow n+\nu _{l}$), modified URCA (MU) ($n+N\rightarrow p+N+l+\overline{\nu _{l}},p+N+l\rightarrow n+N+\nu _{l}$), nucleon-nucleon bremsstrahlung (NNB) ($N+N\rightarrow N+N+\nu +\overline{\nu}$), Cooper pair breaking and formation (PBF) processes ($N+N\rightarrow \left[ NN\right] +\nu +\overline{\nu}$).
Exemplary cooling curves of a $1.4 M_{\odot}$ NS~\citep{2016ApJ...817....6D} are displayed in Fig.~\ref{fig:cooling}.

In the quiescent state of X-ray transients, the accreted matter sinks gradually in the interior of the NS and undergoes a series of nuclear reactions~\citep{2018MNRAS.475.5010F}.
These reactions release some heat, which propagates into the whole NS, inwardly heating the core and outwardly emitted in the form of photons at the surface. This is the so-called deep crustal heating. The heating curves of X-ray transients can be derived, relating the $L_{\gamma}$ in quiescence to the estimated time-averaged accretion rate $\dot{M}$~\citep{2003A&A...407..265Y}.
The relevance of the pasta phase, which is beyond the neutron drip density, to explaining some X-ray transients (if confirmed) might be regarded as smoking-gun evidence of the NS model for pulsar-like objects~\citep[e.g.,][]{2017ApJ...839...95D} and disfavors the alternative QS model.

Above all, a reliable theory for NS cooling, in combination with accurate observations, is indispensable for gaining important information about the stellar interior.
The complexity of NS systems, such as anisotropic magnetic fields and the compositions
of the stellar core and envelope, is not controlled well theoretically, currently rendering the task of providing reliable and quantitative predictions extremely difficult, and considerable effort might be achieved in the future.

\subsection{Pulsar glitch and glitch crisis}

A glitch (sudden spin-up)~\footnote{An ``antiglitch'', i.e., an abrupt spin-down, has also been detected~\citep{2013Natur.497..591A}.} is a common phenomenon in pulsar observations and was discovered during pulsar timing studies in the Vela pulsar~\citep{1969Natur.222..228R}.
Since then, the number of known glitches has greatly increased, with more than 555 glitches now known in more than 190 pulsars.
The observed glitches are collected by the Jodrell Bank Observatory~\footnote{http://www.jb.man.ac.uk/pulsar/glitches/gTable.html} and the ATNF Pulsar Catalogue~\footnote{http://www.atnf.csiro.au/research/pulsar/psrcat/glitchTbl.html}.
During glitches, the pulsar spin frequency $\Omega$ suddenly increases over a very short time and then usually relaxes to its preglitch rate over a longer time.
The glitch size, often defined as the relative increase in the spin frequencies during glitches, $\Delta \Omega_g/\Omega$, has a bimodal distribution ranging from $\sim$ $10^{-10}$ to $10^{-5}$ with peaks at $\sim$ $10^{-9}$ and $10^{-6}$~\citep{2011MNRAS.414.1679E,2013MNRAS.429..688Y}.
The glitches in young pulsars, including magnetars, are generally large~\citep{2018arXiv180104332M}.
However, the youngest pulsars, e.g., the Crab pulsar, tend to have more frequent and smaller glitches.
Various authors have used observed glitch properties as a probe to investigate the pulsar inner structure, i.e., the EOS of dense matter~\citep{2015IJMPD..2430008H}.

\begin{figure}
{\centering
\resizebox*{0.8\textwidth}{0.4\textheight}
{\includegraphics{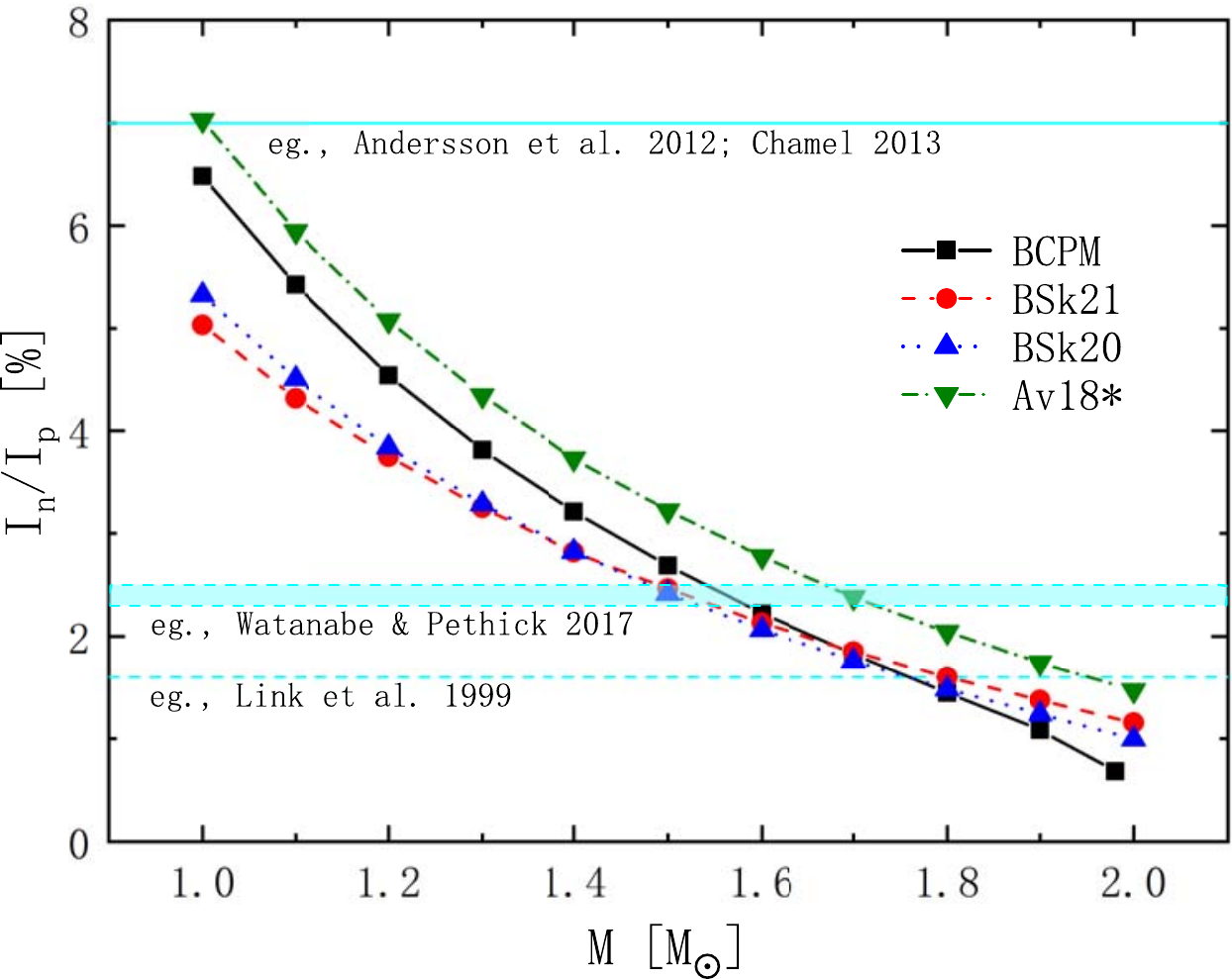}}
\caption{Fractional moments of inertia as a function of the stellar mass for both the charged component $I_p$ and the (crustal) superfluid component $I_n$, with four cases of NS EOSs (BCPM, BSk21, BSk20, Av18*).
Av18* indicates the nonunifed EOS of "Av18 + NV + BPS''. Taken from~\citet{2018IAUS..337..360L}. }\label{fig:glitch}}
	\vspace{-0.3cm}
\end{figure}

The physical mechanism behind glitches is suggested to be a sudden decrease in the NS moment of inertia, which could result from the coupling and decoupling between the (observed) charged component (rotating slower; labelled as index $p$) and the superfluid component (rotating faster; labelled as index $n$)~\citep{1975Natur.256...25A}.
The fractional moment of intertia $I_n/I_p$ is related to the glitch activity $A_g=(1/T)(\sum\Delta\Omega_p)/\Omega_p$ as follows,
\begin{equation}
2\tau_c A_g
\lesssim \frac{I_n}{I_p} \ ,
\label{eq:Ag}
\end{equation}
where $T$ is the total data span for glitch monitoring and $\sum(\Delta\Omega_g)$ is the sum of all observed glitch frequency jumps. $\tau_c = - { \Omega_p / 2 \dot{\Omega_p}}$ is the characteristic age of the pulsar.
The glitch activity $A_g$ can be estimated for systems that have exhibited at least two glitches of similar magnitude, like the Vela pulsar.
The glitch observations from the Vela pulsar place a constraint on the fractional moment of inertia, , which is $I_n/I_p\gtrsim  1.6\%$~\citep{1999PhRvL..83.3362L,2012PhRvL.109x1103A}.
It has been argued that ``entrainment'' of the neutron superfluid by the crystalline structure of the crust greatly reduces its mobility, increasing the lower limit from $1.6\%$ to $\sim7\%$ and
making it very difficult for the nuclear EOSs to fulfill with a normal $M > 1.0 M_{\odot}$ NS~\citep{2016ApJS..223...16L,2012PhRvL.109x1103A,2013PhRvL.110a1101C}.
This is clearly shown in Fig.~\ref{fig:glitch}.
Consequently, the unpinning of the crustal superfluid is insufficient to account for large glitches. This is the so-called glitch crisis problem. Other mechanisms, e.g., the unpinning of core superfluid neutrons, may be required.
However, the mobility of superfluid neturons are related to the effective neutron mass, which has been discussed actively in the literature, see e.g.,~\citep{2017PhRvL.119f2701W}.
According to the calculation in \citet{2017PhRvL.119f2701W}, the constraint for the fractional moment of inertia is $I_c/I \ge 2.5-2.3\%$
Then, an NS of $M \lesssim 1.55M_{\odot}$ NS would be acceptable, and there is no need to invoke the core superfluid. However, this is an open problem, and more detailed work has to be done.

At present, we are still far from a thorough understanding of the general picture of glitches; for example, is there a connection between the stellar interior and the magnetosphere of a star? How can various types of post-glitch behavior be explained? Is there an alternative model besides the superfluid model?
The original starquake model~\citep{1969Natur.224..872B} suggested that the change in the moment of inertia was due to relaxation in the NS (solid) crust to the current equilibrium oblateness but has difficulty explaining the large glitches observed from the Vela pulsar. In a solid star model, the whole body of the star, rather than only the crust, is in a solid state. In such cases, the glitch amplitude could be explained~\citep{2014MNRAS.443.2705Z,2018MNRAS.476.3303L}.
It is a challenge to quantitatively describe glitch behaviors since the related physical processes are complicated. There has been progress in the determination of NS properties in the literature~\citep[e.g.,][]{2015SciA....1E0578H,2017NatAs...1E.134P,2019NatAs...3.1143A,2019NatAs...3.1122G}.

\section{Summary and perspectives}

Although NSs were anticipated in the early thirties and discovered as pulsars in the late sixties, the state of their liquid interiors remains unclear due to a lack of a good understanding of QCD at low energy scales.
The current and upcoming multimessenger observatories~\citep[e.g.,][]{2016arXiv161006892W,2016ApJ...832...92O,2018SCPMA..61c1011L,2018IMMag..19..112L,2019arXiv190303035R,2019SCPMA..6229503W} will continue improving the detection of pulsars together with the precise measurements of their masses and radii.
Laboratory experiments will provide an emerging understanding of nuclear matter EOS and the transition to deconfined quark matter.
Hopefully, the high density behavior of the NS EOS can be determined soon, shedding light on many unsolved problems in nuclear physics and high-energy astrophysics~\citep[e.g.,][]{2017ApJ...850...30L,2017MNRAS.472.2403Z,2018arXiv181202002Y,2019PhRvD..99b3018L,2020ApJ...892....4D,2020ApJ...891..148S}.

Since the compact star EOS is such a demanding problem, it is necessary to combine efforts from different communities and discuss mutual interests and problems~\citep[e.g.,][]{2020NatAs.tmp...42C,2018ApJ...859...90Z,2020arXiv200211355D}. Additionally, it is important to establish new quantitative results testable by experiments/observations.
In this work, the microscopic physics of dense matter are modelled within the QMF, which connects the structure of a nucleon to the EOS of infinite nuclear matter, with a wide range of experimental and observational data available for use.
The parameter space of the QMF has been constrained well for describing NSs, following the present robust measurements of mass, radius, and tidal deformability. The pure NS maximum mass is approximately $2.1 M_{\odot}$, with a satisfying reproduction of the nuclear matter properties around the saturation density. The results have a modest dependence on the model parameters.
Based on the available hyperonuclei data, the hyperon puzzle is present, and we need to understand better how hyperon three-body interaction plays a role to understand more clearly whether hyperons are relevant in NSs (especially the heavy ones).
The CSS parametrization allows us to handle the high-density cores of NSs in a model-independent way.
After demonstrating how the NSs' mass and radius depend on the CSS parameters for the phase transition of deconfined quark matter, we find a safe upper limit for the hybrid star maximum mass at approximately $3.6 M_{\odot}$ based on the extreme causality EOS, similar to previous studies.
In particular, the NS/QS maximum mass is predicted to be approximately $2.3-2.4 M_{\odot}$ from various model calculations, as well as analysis on the merger remnant of GW170817.
Therefore, if more massive pulsars above $2.4 M_{\odot}$ or fewer massive black holes below $5 M_{\odot}$ are found, what is their nature? Is there a mass gap between NSs and black holes and why? Such problems are mysteries to be solved.
The detailed feature of the sound speed in quark matter is explored in a perturbative model, and an enhancement in the sound speed is necessary to fulfill the two-solar-mass constraint of pulsars, located at intermediate densities, indicating that the pair of quarks starts to play a nontrivial role at such densities. Quark superfluids should not be ignored in the quark matter relevant to compact stars.
This is also consistent with a lower bound for tidal deformability that is more consistent with the GW170817-like event in another MIT model than in the case without superfluid.
In the future, even if we understand the stiffness of the EOS, a further challenge is the particle degree of freedom in cold, dense matter. This could help us understand the nonperturbative properties of low-energy QCD (or the parameters of an effective model).
This is work for the future.

\section*{Acknowledgements}
We are grateful to the members of the XMU neutron star group.
We are thankful to the anonymous referee for his or her beneficial comments and Hong Shen, Jirina Stone, Anthony Thomas for insightful discussions.
This work was supported by the National Natural Science Foundation of China (Grant Nos. 11873040, 11775276, 11705163, and 11775119), the Youth Innovation Promotion Association of the Chinese Academy of Sciences, the Natural Science Foundation of Tianjin, the China Scholarship Council (Grant No. 201906205013), and the Ningbo Natural Science Foundation (Grant No. 2019A610066).

\end{document}